\documentclass[
  twocolumn,
  english,
  aps,
  prapplied,
  longbibliography,
  superscriptaddress,
  amsmath,
  amssymb,
  floatfix,
]{revtex4-2}
\usepackage[utf8]{inputenc}
\usepackage{microtype}
\usepackage{graphicx}
\usepackage{amsmath}
\usepackage{amssymb}
\usepackage{mathtools}
\usepackage{qcircuit}
\usepackage[colorlinks=true,urlcolor=blue,citecolor=blue,linkcolor=blue]{hyperref}
\usepackage{natbib}
\usepackage{braket}
\DeclareMathOperator{\Tr}{Tr}
\DeclarePairedDelimiter{\abs}{\lvert}{\rvert}

\DeclareMathOperator*{\argmin}{arg\,min}
\usepackage{cleveref}
\crefname{equation}{Eq.}{Eqs.}
\Crefname{equation}{Eq.}{Eqs.}
\crefname{section}{Sec.}{Secs.}
\Crefname{section}{Sec.}{Secs.}
\crefname{figure}{Fig.}{Figs.}
\Crefname{figure}{Fig.}{Figs.}
\crefname{appendix}{Appendix.}{Appendices.}
\crefname{appendix}{Appendix.}{Appendices.}
\usepackage{sidecap}
\usepackage{makecell}
\usepackage{listings}
\usepackage[linesnumbered,lined,commentsnumbered]{algorithm2e}
\Crefname{algocf}{Algorithm}{Algorithms}
\usepackage{xcolor}
\definecolor{C0}{HTML}{8a3ffc}
\definecolor{C0light}{HTML}{c49ffd}
\definecolor{C1}{HTML}{0072c3}
\definecolor{C1light}{HTML}{cce3f3}
\definecolor{C2}{HTML}{198038}
\usepackage{tikz}
\usetikzlibrary{fadings,shapes.arrows,shadows,positioning,fit}
\usepackage{cancel}
\usepackage{multirow}
\usepackage{siunitx}
\newcommand{\pa}{p_\text{a}}
\newcommand{\pc}{p_\text{c}}
\newcommand{\ps}{p_\text{s}}
\newcommand{\psp}{p_\text{sp}}
\newcommand{\fa}{f_\text{a}}
\newcommand{\fc}{f_\text{c}}
\newcommand{\fs}{f_\text{s}}
\newcommand{\fsp}{f_\text{sp}}
\newcommand{\fspfast}{f_\text{sp}^\text{fast}}
\newcommand{\fspslow}{f_\text{sp}^\text{slow}}
\newcommand{\pspfast}{p_\text{sp}^\text{fast}}
\newcommand{\pspslow}{p_\text{sp}^\text{slow}}
\newcommand{\hatpsp}{\hat{p}_\text{sp}}

\begin{document}

\title{Mitigating errors in state preparation and measurement with
noncomputational states}

\date{\today}
\author{Conrad J. Haupt}
\affiliation{IBM Quantum, IBM Research Europe -- Zurich, 8803
R\"{u}schlikon, Switzerland}
\affiliation{Laboratory of Theoretical Physics of Nanosystems,
{\'E}cole Polytechnique F{\'e}d{\'e}rale de Lausanne, 1015 Lausanne,
Switzerland}
\author{Almudena Carrera Vazquez}
\affiliation{IBM Quantum, IBM Research Europe -- Zurich, 8803
R\"{u}schlikon, Switzerland}
\author{Laurin E. Fischer}
\affiliation{IBM Quantum, IBM Research Europe -- Zurich, 8803
R\"{u}schlikon, Switzerland}
\affiliation{Theory and Simulation of Materials, {\'E}cole
Polytechnique F{\'e}d{\'e}rale de Lausanne, 1015 Lausanne, Switzerland}
\author{Stefan Woerner}
\affiliation{IBM Quantum, IBM Research Europe -- Zurich, 8803
R\"{u}schlikon, Switzerland}
\author{Daniel J. Egger}
\email{deg@zurich.ibm.com}
\affiliation{IBM Quantum, IBM Research Europe -- Zurich, 8803
R\"{u}schlikon, Switzerland}

\begin{abstract}
  Error mitigation has enabled quantum computing applications with
  over one hundred qubits and deep circuits.
  Many error mitigation methods are
  noise-aware, relying on a faithful characterization of
  the noise channels of the hardware.
  However, fundamental limitations lead to unlearnable degrees of
  freedom of the underlying noise models when considering qubits.
  Here, we show how to leverage non-computational states as an
  additional resource to learn state-preparation errors in
  superconducting qubits.
  This allows one to fully constrain the noise models.
  We can thus independently and accurately mitigate state-preparation
  errors, gate errors and measurement errors.
  Our proposed method is also applicable to dynamic circuits with
  mid-circuit measurements.
  This work opens the door to improved error mitigation for
  measurements, both at the end of the circuit and mid-circuit.
\end{abstract}

\maketitle{}

\section{Introduction}

Progress in quantum computing hardware and error mitigation has
enabled utility-scale experiments~\cite{kim2023evidence,
  fischer2024dynamical, fuller2025improved}.
At this scale, a brute-force classical simulation of the underlying
quantum circuits is no longer possible.
Moving forward, error mitigation may enable a quantum advantage
before the onset of full fault-tolerance~\cite{zimboras2025myths}.
In the fault-tolerant regime error mitigation will remain relevant by
enhancing performance and reducing residual logical errors beyond the
capabilities of error correction alone~\cite{Piveteau2021,
  aharonov2025importance}.
Importantly, quantum computing is not only made of sequences of unitary gates.
Mid-circuit measurements (MCMs) offer a powerful computational extension.
They form the bedrock of dynamic circuits
and enable entanglement distribution~\cite{Baumer2024a}, improved
algorithmic execution~\cite{Corcoles2021, Baumer2024, Baumer2025a},
and circuit cutting~\cite{Piveteau2024, Brenner2023, Carrera2024,
  Mitarai2021, Singh2024}.

Many error mitigation methods rely on accurate noise learning
experiments which also help characterize quantum processors at
scale~\cite{van2023probabilistic, Mckay2023, kim2023evidence}.
Gate noise is commonly characterized by repeating a gate layer to
amplify its noise, e.g., in \emph{cycle
  benchmarking}~\cite{erhard2019characterizing,carignan-dugas2023}.
A common assumption is to express the gate noise models as Pauli
channels, which is in practice justified by applying Pauli twirling
or randomized compiling~\cite{twirling_bennet, twirling_knill,
  wallman2016noise}.
Pauli noise learning is made scalable by \textit{sparsifying} the
noise generators~\cite{van2023probabilistic}.
This is typically done by dropping the noise generators that do not
match with the physical qubit couplings.
In this framework, state-preparation and final measurement (SPAM)
errors can be characterized with a model-free twirled readout circuit
in a technique popularized as \emph{twirled readout error extinction}
(TREX)~\cite{VanDenBerg2022}.
The SPAM errors on expectation value estimators are then mitigated in
post-processing.
By contrast, learning the noise models of MCMs requires extending cycle
benchmarking~\cite{hines2025pauli, zhang2025generalized}.
Here, readout-induced leakage can be characterized~\cite{hazra2025benchmarking}
and probabilistic error cancellation has been generalized to dynamic
circuits~\cite{gupta2024probabilistic, koh2024readout, Hashim2025}. Crucially,
when separating SPAM errors from gate errors there are fundamentally
non-learnable degrees of freedom.
These degrees of freedom in the noise model are rigorously understood
as the cut space of a cycle graph~\cite{chen2023learnability}.
The same happens when separating state from measurement errors.
These non-learnable degrees of freedom can be circumvented by imposing symmetry
assumptions on the noise models~\cite{van2023probabilistic} which may be
violated in real experiments.
In some cases this comes with the cost of additional ancilla qubits and
assumed ideal CNOT gates, implemented with error
mitigation~\cite{Yu2025}.
For successful noise-model based error mitigation it can be crucial
to go beyond these naive symmetry assumptions~\cite{fischer2024dynamical}.

Recent work thus focuses on producing self-consistent noise
characterization protocols that learn gate, state-preparation, and
measurements noise.
This approach results in a holistic noise model up to unlearnable
\emph{gauge} degrees of freedom such that any measured observable in
circuits composed of the given state-preparation, gate, and
measurement layers are insensitive to the gauge~\cite{chen2024efficient}.
While this approach is sufficient for noise-learning based error
mitigation of unitary circuits~\cite{Chen2025}, it does not reveal
the ground truth for all parameters of the noise model, such as
state-preparation errors.
This can limit certain uses of the noise model, such as combining
noise models of different layers to form models for gate structures
that have not been learned.
Furthermore, its applicability to dynamic circuits with measurements
and classical feed-forward still needs to be investigated.
Other methods for separating state-preparation from measurement noise
come with trade-offs, such as additional qubits~\cite{huggins2021,koczor2021},
two-qubit gates to perform noise learning~\cite{Yu2025, laflamme2022}, or
repeated measurements during mitigation~\cite{santos2025}.

Here, we show how non-computational states help learn
state-preparation errors and establish a ground truth for a noise
model of mid-circuit measurements.
By using non-computational states, we can overcome the existing no-go
theorems for noise learning~\cite{chen2023learnability, chen2024efficient}.
We demonstrate this in superconducting qubit hardware~\cite{Krantz2019},
and anticipate the method may be
applicable to other hardware platforms where non-computational states are
addressable.

In Sec.~\ref{sec:exp},  we start by demonstrating the importance of
characterizing state-preparation errors separately from measurement errors.
In Sec.~\ref{sec:noise_learning}, we propose a protocol to fix a
ground truth in the noise model of MCMs.
Finally, in Sec.~\ref{sec:sim} we present numerical results of the
proposed protocol.
We discuss and conclude our work in Sec.~\ref{sec:conclusion}.

\begin{figure*}[htb]
  \centering
  \includegraphics[width=\textwidth]{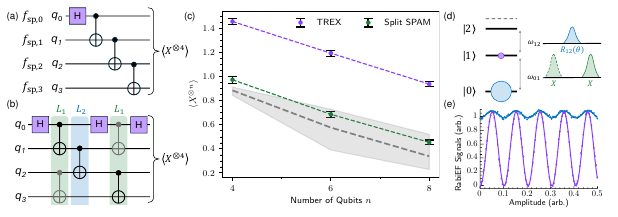}
  \caption{
    \textbf{Splitting SPAM errors with non-computational states.}
    (a) GHZ state preparation circuit.
    (b) A circuit inspired from (a) but with two unique CNOT layers
    $L_1$ and $L_2$ to facilitate gate noise learning.
    (c) SPAM error mitigation on the circuit in (b) with TREX (purple
    data) and a split mitigation of SPAM errors (green data).
    The error bars, discussed in
    Appendix~\ref{appendix:error_propagation}, represent $\pm 2\sigma$.
    The shaded area represents possible values of $\langle X^{\otimes
          n}\rangle$ based on the CNOT noise model, see
    Appendix~\ref{appendix:cnot_noise}.
    The gray dashed line represents a symmetric splitting of the
    conjugate pairs of fidelities in the learned CNOT noise.
    (d) The first three levels of a superconducting transmon and the
    RabiEF pulse sequence.
    The circles indicate a thermal state, with populations
    corresponding to their sizes.
    (e) A RabiEF experiment on IBM Quantum hardware.
    The markers are hardware measurements and the curves are fits of
    $a\sin^2(b\theta) + c$ to the data.
    Here, $\theta$ is proportional to the pulse amplitude on the x-axis.
    Blue and purple are for the no-pi ($\cancel{\pi}$) and pi ($\pi$)
    RabiEF signals, respectively.
  }
  \label{fig:experiment}
\end{figure*}

\section{Motivation\label{sec:exp}}

We now present an experiment, done on an IBM quantum Eagle device,
showing that jointly mitigating state-preparation errors and
measurement errors fails to error mitigate global observables of up
to eight qubits.
A GHZ state preparation circuit with a ladder structure of CNOT
gates, shown in Fig.~\ref{fig:experiment}(a), turns the single-qubit
$Z_0$ operator into a global $X^{\otimes n}$ operator.
Hence, this $+1$ stabilizer is subject to measurement errors from all
qubits, while only being affected by state-preparation errors from
the first qubit.
This makes the observable $O=X^{\otimes n}$ maximally sensitive to
the separation of SPAM errors.
We would thus like to study the circuit of Fig.~\ref{fig:experiment}(a).
However, to account for gate noise we would need to learn the noise
of the $n-1$ distinct layers of CNOT
gates~\cite{van2023probabilistic, Mckay2023}.
To reduce this overhead we convert the circuit in
Fig.~\ref{fig:experiment}(a) to a similar circuit built from only two
distinct layers of CNOT gates, see Fig.~\ref{fig:experiment}(b),
which features the same behavior for the $X^{\otimes n}$ observable.

We mitigate the circuit in Fig.~\ref{fig:experiment}(b) with
TREX~\cite{VanDenBerg2022} to illustrate how inaccurate noise models can
result in unphysical mitigated results.
We start by measuring the raw, unmitigated
expectation value $\langle O\rangle_\text{raw}$.
All two-qubit gates are Pauli twirled so that the effective CNOT
noise is a Pauli noise channel.
We repeat this for $16n^2$ twirling configurations and collect $128$
shots per configuration.

Next, we correct SPAM errors in $\langle O\rangle_\text{raw}$ with
TREX by normalizing it by a correction factor $\langle Z^\star\rangle$.
This factor is found by measuring $Z^{\otimes{}n}$ on the all-zero
initial state, with measurements twirled with $X$ gates and an
assumed ideal prepared state.
Under ideal state preparation the TREX correction factor is $\langle
  Z^\star\rangle=\prod_{i=0}^{n-1}f_{\text{m},i}$, where
$f_{\text{m},i}$ denotes the final measurement fidelity and subscript
$i$ denotes the corresponding qubit.
However, in the presence of state-preparation errors the measured
TREX mitigator becomes $\langle
  Z^\star\rangle=\prod_{i=0}^{n-1}f_{\text{sp},i}f_{\text{m},i}$, where
$f_{\text{sp},i}$ are the state-preparation fidelities.
Normalizing $\langle O\rangle_\text{raw}$ by $\langle Z^\star\rangle$
can thus cause TREX to overcompensate when the set of qubits whose
state-preparation errors affect  the observable (in our case only
qubit 0) differs from the support of the observable (in our case all
qubits), see, for example, Refs.~\cite{Carrera2024, Chen2025}.

Here, we measure $\langle Z^\star\rangle$ after a passive reset into
the thermal state with $5000$ shots per each of the $16$ randomizations.
We observe that TREX indeed over-corrects the expectation value
$\langle X^{\otimes n} \rangle$ which we attribute to the factor
$\prod_{i\neq 0}f_{\text{sp}, i} \neq 1$, see the purple curve in
Fig.~\ref{fig:experiment}(e).
For our circuits with four and six qubits this even results in
unphysical expectation values.
This calls for a SPAM error mitigation that accounts for
state-preparation errors independently of measurement errors.

\section{Noise learning with excited states\label{sec:noise_learning}}

Jointly mitigating SPAM errors can lead to non-physical expectation values.
We now demonstrate how to leverage non-computational states to
separate state-preparation from mid-circuit measurement errors.
The resulting learned noise model is applicable to advanced error mitigation
methods such as Probabilistic Error Cancellation
(PEC)~\cite{van2023probabilistic, gupta2024probabilistic}.
Our work is based around the idea that well-controlled initial states, such as
thermal states, can serve as a resource to this end.

\subsection{Thermal states and passive reset}

Learning the state-preparation error precisely necessitates stable
state preparation in the computational basis.
Therefore, we employ a simple passive reset by waiting more than
$5-10\times T_1$ when initializing qubits.
This results in a thermal state with a population following a
Boltzmann distribution~\cite{Riste2012}.
Therefore, the probability to find the transmon in the excited state
is $p_\text{sp}=\exp(-\hbar \omega_{01}/k_B T_\text{eff})$.
Here, $\omega_{01}$ is the energy difference between the first
excited state and the ground state, see Fig.~\ref{fig:experiment}(c).
The effective temperatures of superconducting qubits $T_\text{eff}$
are typically above the $15~\mathrm{mK}$ temperature of dilution refrigerators.
Furthermore, the population of the second excited state $\ket{2}$ is
often negligible~\cite{Jin2015}, see also Appendix~\ref{appendix:thermal}.
After a passive reset the qubit is thus in the thermal state
$\rho_\text{th}=(1-p_\text{sp})\ket{0}\!\!\bra{0}+p_\text{sp}\ket{1}\!\!\bra{1}$
with typical $p_\text{sp}$ values in the $1-10\%$ range.
Crucially, $p_\text{sp}$ can be precisely measured by driving an
oscillation in the $\ket{1}\leftrightarrow\ket{2}$ subspace with the
$R_{12}(\theta)$ gate and comparing it to a reference
oscillation~\cite{Geerlings2013, Jin2015}, see Fig.~\ref{fig:experiment}(d).
The probability $\psp$ is estimated from the amplitudes of the no-pi
($\cancel{\pi}$) signal and the reference, or pi ($\pi$), signal, see
Appendix~\ref{appendix:rabief}.
This procedure, also  known as a \emph{RabiEF}
experiment~\footnote{The E and F in RabiEF come from an alternative
  transmon level naming where the states $\ket{0}$, $\ket{1}$, and
  $\ket{2}$ are labeled by $\ket{g}$, $\ket{e}$, and $\ket{f}$,
  respectively. Here, $g$ and $e$ stand for ground and excited,
  respectively.}, provides a direct measurement of the
state-preparation error $p_\text{sp}$ of a passive reset.
Through the comparison to a reference oscillation, which is equally
affected by measurement errors, the RabiEF experiment is mostly
insensitive to measurement noise, see Appendix~\ref{appendix:meas_sensitivity}.
We can thus place ourselves in a situation where the
state-preparation error is well known and measurable with sufficient
accuracy by passively resetting the qubit.

\subsection{Split mitigation of state and measurement
  errors}\label{sec:experiment_mitig}

We can leverage our knowledge of state-preparation errors to prevent
the TREX over-corrections shown in \cref{sec:exp}.
First, we measure the fidelities $f_{\text{sp},i}=1-2p_{\text{sp},i}$
of each qubit with RabiEF following a $10~\mathrm{ms}$ passive reset.
Next, we multiply the correction factor $\langle Z^\star\rangle$ by
the product $\prod_{i=1}^{n-1} f_{\text{sp},i}$ to obtain the
mitigated expectation value
\begin{equation}\label{eq:trexpp}
  \langle O\rangle_{\text{mit}}=\frac{\langle
    O\rangle_\text{raw}\prod_{i=1}^{n-1} f_{\text{sp},i}}{\langle Z^\star\rangle}.
\end{equation}
Crucially, the circuits to compute $\langle Z^\star\rangle$ must use
the same passive reset as RabiEF so that the state-preparation error
is consistent across all experiments.
This results in physical expectation values, see the green curve in
Fig.~\ref{fig:experiment}(e).
Furthermore, these values are consistent with the noise levels of the
CNOT gates.
We learn their noise model and compute their impact on the
expectation value of $X^{\otimes{}n}$ via a Clifford simulation,
details are in Appendix~\ref{appendix:cnot_noise}.
Since we can only learn the CNOT noise up to the unlearnable degrees
of freedom, we compute upper and lower bounds for the ideal
observable with only CNOT noise, represented as the shaded area in
Fig.~\ref{fig:experiment}(e), assuming physical noise channels.
This area represents the possible range for $\langle
  O\rangle_\text{raw}$ after removing SPAM errors, and therefore
$\langle O\rangle_{\text{mit}}$ falling within this area indicates success.
The two experiments with six and eight qubits fall well within this area.
The experiment with four qubits does not.
Out-of-model errors, such as non-Markovian noise and noise generators
with a weight greater than two, may cause this~\cite{Govia2025}.

\subsection{Noise model of the measurement\label{sec:error_model}}

We now develop a framework to completely specify the noise model of
(mid-circuit) measurements
assuming projective $Z$ basis measurements and ideal single-qubit and -qutrit
gates.
As a result of appropriate twirling, state-preparation and measurement errors
are bit flips, i.e., Pauli strings containing only $I$'s or $X$'s.
We enforce this by applying random $Z$ gates to the initial state with
probability $50\%$ and twirling measurements following Ref.~\cite{Beale2023}.
This diagonalizes measurements and the initial state density matrix.
It also permutes measurement outcomes while
ensuring the post-measurement state is unchanged, see Fig.~3 of
Ref.~\cite{Beale2023}.
Note that we employ the convention that noise occurs before
measurements without loss of generality, see Fig.~\ref{fig:meas_models}.

After twirling, the Pauli Transfer
Matrices (PTM), introduced in Appendix~\ref{appendix:ptm}, of the measurement
noise channels will only contain non-zero diagonal elements in the entries
corresponding to Pauli strings containing only $I$'s and $Z$'s.
Therefore, whenever we now write the PTM of such a noise channel we
omit rows and columns containing $X$'s or $Y$'s.
Furthermore, we model the classical bit (cbit) as if it were a qubit
living in a corresponding Hilbert space with the convention
\begin{equation}
  \text{qubit}\otimes\text{cbit}.
\end{equation}
For example, a bit-flip error on the qubit is denoted by $XI$ and a
bit flip on the classical bit is $IX$.
The entries of any PTM will always appear in lexicographic order,
e.g., $II$, $IZ$, $ZI$, $ZZ$.
To construct the PTM, we represent a measurement as a CNOT gate
targeting the classical bit controlled by the qubit, and a subsequent
measurement on the qubit; see Fig.~\ref{fig:meas_models}.
The model of the noisy measurement is built using the principle of
deferred measurement~\cite{Nielsen2000}.

Our measurement noise model includes three possible error channels.
(i) A \emph{state error}, modeled as a bit flip on the qubit only,
with probability $\ps$ and noise channel
$\Lambda_\text{s}(\rho)=(1-p_\text{s})\rho + \ps XI \rho XI$.
The PTM of $\Lambda_\text{s}$ is
\begin{equation}
  \Gamma_\text{s} =
  \begin{pmatrix}
    1 & 0        \\
    0 & 1 - 2\ps \\
  \end{pmatrix} \otimes I.
\end{equation}
This may correspond to a qubit decay at the end of the readout
through a $T_1$ event. With twirling, this becomes an $X$ error
instead of an amplitude damping channel.
(ii) An \emph{assignment error}, modeled as a bit flip on the
classical bit only, with probability $\pa$ and noise channel
$\Lambda_\text{a}(\rho)=(1-\pa)\rho + \pa IX \rho IX$.
The PTM of $\Lambda_a$ is
\begin{equation}\label{eq:ptm_assign}
  \Gamma_\text{a} = I \otimes
  \begin{pmatrix}
    1 & 0        \\
    0 & 1 - 2\pa \\
  \end{pmatrix}.
\end{equation}
This may correspond to the readout misclassifying the state without a
quantum error occurring on the qubit.
Finally, (iii) a \emph{correlated error} is a bit flip on both the
classical and quantum bits with probability $\pc$ and noise channel
$\Lambda_\text{c}(\rho)=(1-\pc)\rho + \pc XX \rho XX$.
The PTM of $\Lambda_\text{c}$ is
\begin{equation}\label{eq:ptm_correlated}
  \Gamma_\text{c} =
  \begin{pmatrix}
    1 & 0        & 0        & 0 \\
    0 & 1 - 2\pc & 0        & 0 \\
    0 & 0        & 1 - 2\pc & 0 \\
    0 & 0        & 0        & 1 \\
  \end{pmatrix}.
\end{equation}
For example, this corresponds to a post-measurement $X$ error on the
qubit, which is equivalent to the correlated error $\Lambda_\text{c}$
by propagating the $X$ back through the measurement.

The PTM $\Gamma_\text{m}$ of the noise model is obtained by
multiplying the PTMs of the three error channels and the PTM of the
ideal CNOT between the classical and quantum bit such that
\begin{equation}\label{eqn:meas_ptm}
  \Gamma_{\text{m}} = \Gamma_{\text{CX}}\Gamma_{s}\Gamma_{a}\Gamma_{c} =
  \begin{pmatrix}
    1 & 0       & 0       & 0       \\
    0 & 0       & 0       & \fa \fs \\
    0 & 0       & \fc \fs & 0       \\
    0 & \fc \fa & 0       & 0       \\
  \end{pmatrix}.
\end{equation}
For readability, we write the error fidelities $f_\text{x}\coloneqq
  (1-2p_\text{x})$ instead of their probabilities $p_\text{x}$.
To learn the elements of $\Gamma_\text{m}$, we repeat the measurement
$2k$ times in a cycle benchmarking
experiment~\cite{zhang2025generalized, hines2025pauli}, resulting in
\begin{equation}
  \Gamma_\text{m}^{2k} =
  \begin{pmatrix}
    1 & 0                    & 0              & 0                    \\
    0 & \fa^{2k} (\fc \fs)^k & 0              & 0                    \\
    0 & 0                    & (\fc \fs)^{2k} & 0                    \\
    0 & 0                    & 0              & \fa^{2k} (\fc \fs)^k \\
  \end{pmatrix}.
\end{equation}

\begin{figure}
  \centering
  \includegraphics[width=\columnwidth, clip, trim=0 290 170 0]{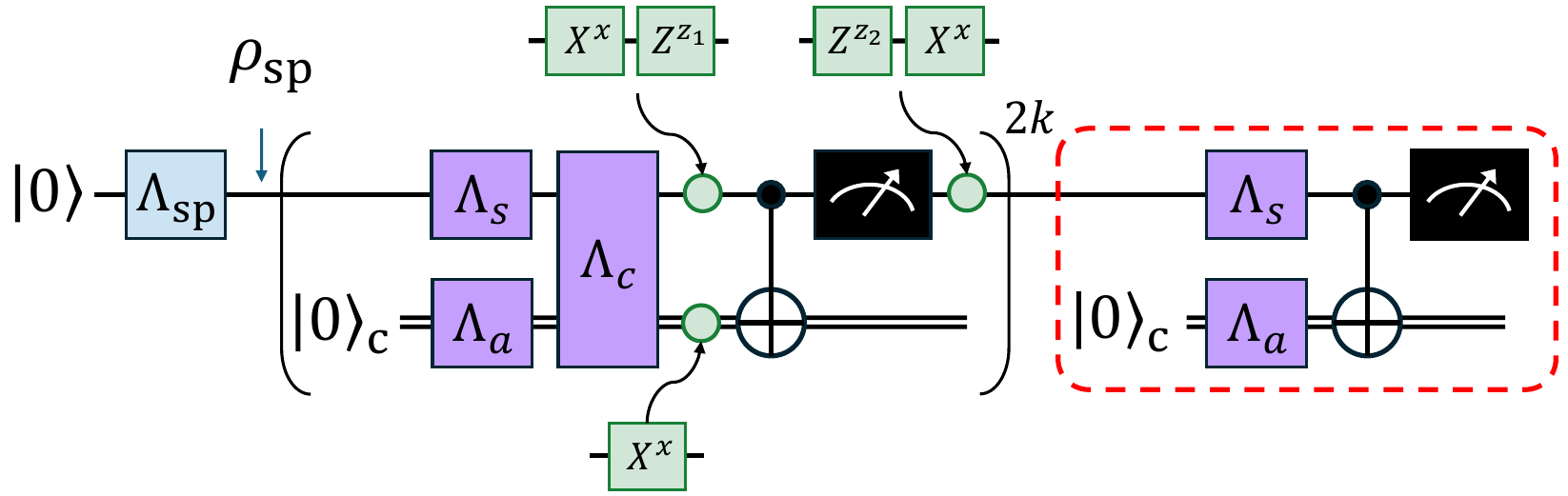}
  \caption{
    \textbf{Measurement noise model.}
    The qubit and the classical bit are indicated by the single and
    double lines, respectively.
    Cycle benchmarking learns the noise model, shown in purple, by
    repeating the measurement $2k$ times.
    The twirling gates are indicated in green; the binary variables
    $x,z_1, z_2$ are independently drawn from a uniform distribution
    on $\{0,1\}$~\cite{Beale2023}.
    The red dashed rectangle indicates the final measurement which is
    insensitive to correlated errors $\Lambda_c$.
  }
  \label{fig:meas_models}
\end{figure}

In practice, state-preparation errors may cause the qubit to start
the circuit in the $\ket{1}$ state with probability $\psp$, such that
the initial state is $\rho_\text{sp}\neq\ket{0}\!\!\bra{0}$.
The PTM that prepares $\rho_\text{sp}$ from $\ket{0}$ is
\begin{equation}\label{eq:stateprep}
  \Gamma_\text{sp} =
  \begin{pmatrix}
    1 & 0         \\
    0 & 1 - 2\psp \\
  \end{pmatrix} \otimes I.
\end{equation}
At the end of the execution, we measure the qubit and record its
outcome on the same wire, which now functions as a classical bit, see
Fig.~\ref{fig:meas_models}.
The PTM of the final measurement, derived in
Appendix~\ref{appendix:final_measurement}, is
\begin{equation}
  \Gamma_\text{fm} =
  \begin{pmatrix}
    1 & 0       \\
    0 & \fa \fs \\
  \end{pmatrix} \otimes I.
\end{equation}
Therefore, the PTM of the full measurement cycle benchmarking (MCB)
circuit, shown in Fig.~\ref{fig:meas_models}, is $\Gamma_{2k} =
  \Gamma_\text{fm}\Gamma_\text{m}^{2k}\Gamma_\text{sp}$ which equates to
\begin{equation}\label{eq:ptm_full}
  \begin{pmatrix}
    1 & 0                    & 0                           & 0                                \\
    0 & \fa^{2k} (\fc \fs)^k & 0                           & 0                                \\
    0 & 0                    & (\fc \fs)^{2k} \fa \fs \fsp & 0                                \\
    0 & 0                    & 0                           & \fa^{2k} (\fc \fs)^k\fa \fs \fsp \\
  \end{pmatrix}.
\end{equation}
We learn the entries of the PTM by measuring the corresponding noisy
expectation values $\langle{}IZ\rangle{}$, $\langle{}ZI\rangle{}$,
$\langle{}ZZ\rangle{}$ on the MCB circuit for different values of $k$.
We then fit decaying exponential curves, of the form $Af^{2k}$, to the data.
Thus, we can learn the offset $A=\fa \fs \fsp$ and the products
$f=\fa\sqrt{\fc\fs}$ and $f=\fc\fs$.
Consequently, we can learn (i) the product $\fs \fc$ of the state and
correlated fidelities,
(ii) the probability $\pa$ of a bit flip on the classical bit, and
(iii) the product $\fs \fsp$ of the state and state-preparation fidelities.
This shows that we can fully specify the noise model if we can
measure either of $\fs$, $\fc$, or $\fsp$ individually.
We therefore propose to estimate $\fsp$ with a RabiEF experiment.
Thereby splitting the fidelity products and therefore, also the SPAM
error fidelities.

Here we focus on correlated errors between qubits and their classical bits.
Our model can be extended to capture correlations between qubits by increasing
the support of the cycle benchmarking experiment, resulting in more
Pauli-Lindblad generators~\cite{van2023probabilistic, gupta2024probabilistic,
  Beale2023}.
As an illustrative example, Appendix~\ref{sec:app_correlated} shows how our
protocol learns the noise model for two correlated qubits undergoing
measurement.

\subsection{Mitigation workflow}\label{sec:general_mitig}

In \cref{sec:experiment_mitig}, we split the error mitigation of SPAM errors.
This requires a slow passive reset when learning the TREX correction
factor $\langle Z^\star \rangle$ and when running the circuit in
Fig.~\ref{fig:experiment}(b) to keep $\fsp$ consistent across all
executed circuits.
Furthermore, this correction is possible because the circuit only has
Clifford gates.
We can thus compute which qubits affect the raw observable resulting
in Eq.~(\ref{eq:trexpp}).
By contrast, an ideal error mitigation protocol
is (i) fast, and (ii) mitigates both state-preparation and
mid-circuit measurement errors for
arbitrary circuits.

We could avoid slow resets by executing a RabiEF experiment on the
initial state resulting from a fast, active reset, consisting of a
measurement and a feedforward $X$ gate conditioned on measuring a $\ket{1}$.
For a RabiEF experiment to function properly, the reset must always
prepare the same initial state of the form
$(1-p_\text{sp})\ket{0}\!\bra{0} + p_\text{sp}\ket{1}\!\bra{1}$,
i.e., with a negligible $\ket{2}$-state population.
If such a reliable fast reset is available, we propose a noise
learning and mitigation protocol that (i)~executes RabiEF with fast
resets to learn $\fsp$, (ii)~runs MCB to learn $\fa$, $\fs$, and
$\fc$, (iii)~runs gate noise learning, and (iv) mitigates
state-preparation, gate, and measurement errors in the intended
circuit, with schemes such as probabilistic error cancellation
(PEC)~\cite{van2023probabilistic} or probabilistic error
amplification (PEA)~\cite{kim2023evidence}.

\begin{figure*}
  \centering
  \tikzset{main/.style={C1},alt/.style={C0}}
  \tikzset{blocks/.style={draw=black,shape=rectangle,rounded
        corners=0mm,inner sep=2mm,align=center,text width=2.2cm}}
  \tikzset{mainblock/.style={main,blocks,black,fill=C1light},
    altblock/.style={alt,blocks,black,fill=C0light},}
  \tikzset{mainpath/.style={main,thick}}
  \tikzset{altpath/.style={alt,thick}}
  \begin{tikzpicture}[node distance=2cm and 4cm]

    \node[blocks,text width=2.5cm] (available) at (3,0) {Start: Do
      you have stable $\ket{2}$ reset with fast active resets?};
    \node[altblock,] (rabief_slow) at (6.5,1) {RabiEF with slow resets};
    \node[mainblock,] (rabief_fast) at (6.5,-1) {RabiEF with fast resets};
    \node[altblock,] (measlearn_slow) at (10,1) {MCB with slow resets};
    \node[draw=none,text width=1cm] (dummy_slow) at (14.2,1) {\vphantom{MCB}};
    \node[draw=none,text width=1cm] (dummy_fast) at (14.2,-1) {\vphantom{MCB}};
    \node[blocks,fit=(dummy_slow)(dummy_fast), text width=2cm]
    (measlearn_fast) {MCB with fast resets};
    \node[blocks, text width=1.8cm] (mitig) at (18,0) {Learn Gate
      Noise and Mitigate Circuits};

    \draw[altpath,->] (available)    |-      node[above left]{No}
    (rabief_slow);
    \draw[mainpath,->] (available)      |-     node[below left]{Yes}
    (rabief_fast);
    \draw[altpath,->] (rabief_slow) -- (measlearn_slow);
    \draw[altpath,->] (measlearn_slow) -- (measlearn_slow-|measlearn_fast.west);
    \draw[mainpath,->] (rabief_fast) -- (rabief_fast-|measlearn_fast.west);
    \draw[->] (measlearn_fast) -- (mitig);

    \node[alt,above right] at (rabief_slow.east) {$f_{sp}^\text{slow}$};
    \node[main,above right] at (rabief_fast.east) {$f_{sp}^\text{fast}$};
    \node[alt,above right] at (measlearn_slow.east) {$f_a,f_s,f_c$};
    \node[main,below right] at (measlearn_fast.east) {$f_a,f_s,f_c$};
    \node[alt,above right] at (measlearn_fast.east) {$f_{sp}^\text{fast}$};
  \end{tikzpicture}
  \caption{
    \textbf{Measurement noise learning protocol.}
    Active resets that can and cannot remove the $\ket{2}$-state
    population can both be supported.
    The learned parameters for a given step are shown on the outgoing
    arrows, e.g., $\fa$, $\fs$, and
    $\fc$ are learned by the MCB with slow resets.
    The preferred case is the lower blue path as it has fewer
    experiments, but requires active resets that can remove the
    $\ket{2}$-state population.
    In the event that they cannot, the purple path should be taken as
    this circumvents poor $\ket{2}$-state reset and its impact on the
    RabiEF experiment.
    Regardless of the path taken, the final MCB experiment completes
    the set of learned SPAM fidelities $\fa$, $\fs$, $\fc$, and $\fspfast$.
  }
  \label{fig:workflow}
\end{figure*}
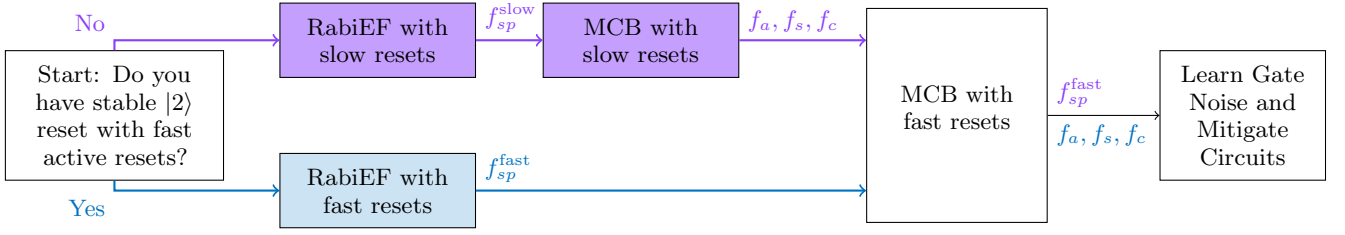

More generally, if such a fast reset --- with \textit{good}
$\ket{2}$-state reset --- is not available, we can still use MCB from
Sec.~\ref{sec:error_model} and RabiEF with both slow and fast resets
to learn all the parameters of our noise model.
First, we leverage thermal states to learn the slow reset fidelity
$\fspslow$ with a RabiEF experiment.
Second, we learn the measurement noise fidelities with a MCB
experiment and slow resets.
Next, we perform an additional MCB experiment with fast resets to
obtain $\fspfast$, where we assume the measurement noise is constant
throughout all experiments and independent of the state-preparation method.
Finally, we learn the remaining gate noise, with other noise learning
protocols, and execute the intended quantum circuits with fast resets
and an error mitigation scheme such as PEC.
Figure~\ref{fig:workflow} summarizes the proposed \textit{noise
  learning} protocol.

\section{Simulations}\label{sec:sim}

We now numerically show how to leverage a RabiEF experiment with a
fast active \emph{qutrit reset} to error mitigate measurements while
accounting for errors in the reset.
As RabiEF explicitly populates the $\ket{2}$ state, it is necessary
to remove as much of the $\ket{2}$-state population as possible in
the state-preparation step.
We therefore use \emph{qutrit} resets.
With simulations of MCB, we show correctly learned noise fidelities
and how the full MCM noise learning protocol in
\cref{sec:error_model} and \cref{fig:workflow} bypasses insufficient
$\ket{2}$-state reset.
We also simulate the same experiment in \cref{sec:exp}, without CNOT
noise, and a three-qubit noisy quantum teleportation circuit with PEC
mitigation.

\subsection{RabiEF with fast resets}\label{sec:sim_rabief}

\begin{figure}[!tb]
  \centering
  \includegraphics[width=\columnwidth, clip, trim=3 3 3 0]{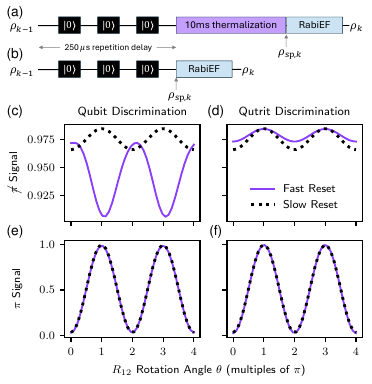}
  \caption{
    \textbf{RabiEF with different resets}.
    (a) Slow reset. A $10~\mathrm{ms}$ delay after the repetition
    delay thermalizes the transmon before the RabiEF circuits.
    (b) A fast reset followed by the RabiEF circuits.
    (c-f) Classically simulated signals with (c, e) qubit and (d, f)
    qutrit discrimination in the active resets in the repetition delay.
    Fast and slow resets are shown as solid purple and dotted black
    lines, respectively.
  }
  \label{fig:qutrit_sim_rabief}
\end{figure}

An active \emph{qubit reset} is a qubit measurement followed by a
feedforward $X$ gate conditioned on measuring the $\ket{1}$ state
that can be used for qubit initialization.
Here, the qubit measurement misclassifies the $\ket{2}$ state as
$\ket{1}$, owing to the overlap of their complex readout
signals~\cite{fischer_ancilla-free_2022,hazra2025benchmarking}.
Such active resets are ineffective at removing the $\ket{2}$-state
population as they only implement binary classification and the
feedforward $X$ gate does not interact with $\ket{2}$.
In \cref{sec:exp}, this was overcome with a long passive reset before
executing each circuit, which we call \textit{slow reset}, see
\cref{fig:qutrit_sim_rabief}(a).
\textit{Fast reset} forgoes this long delay, relying only on active
resets in the repetition delay, see \cref{fig:qutrit_sim_rabief}(b).
Qubit measurements with fast resets cause distortions in the RabiEF
signals, owing to their poor $\ket{2}$-state reset, which
contaminates the estimates of $p_\text{sp}$ in Eq.~(\ref{eq:stateprep}).
Therefore, we consider an imperfect \emph{qutrit reset} which
discriminates between $\ket{0}$, $\ket{1}$, and
$\ket{2}$~\cite{Bianchetti2010, fischer_ancilla-free_2022} and
applies the feedforward $I$, $X$, and $XX_{12}$ gates, depending on
the respective outcomes.

We simulate RabiEF experiments on qutrits with ideal qubit gates
embedded in $SU(3)$, e.g., $X=\ket{1}\!\bra{0} + \ket{0}\!\bra{1} +
  \ket{2}\!\bra{2}$.
This is justified in practice as experimental single-qubit and -qutrit
gates typically have much higher fidelities than two-qubit gates and
measurements~\cite{blok2021,abughanem2025}.
Thermal relaxation and measurement errors are included as separate
operations, attached to appropriate circuit instructions.
To investigate how state-preparation removes the $\ket{2}$-state
population we perform density matrix simulations of RabiEF with
$N=1000$ shots and $40$ rotation angles $\theta\in[0,4\pi]$ for
$R_{12}(\theta)$.
This is done for fast qubit reset, slow qubit reset, fast qutrit
reset, and slow qutrit reset.
The projected post-measurement state $\rho_{k-1}$ of the previous
RabiEF circuit and shot is the input to the repetition delay of the
next circuit, as was done in Ref.~\cite{hauptRestless2023}, see
\cref{fig:qutrit_sim_rabief}(a)-(b).
This allows us to observe how post-measurement states and imperfect
qutrit reset contaminate subsequent shots.
The measurement outcomes are saved, alongside all input states
$\rho_{\text{sp},k}$, see \cref{fig:qutrit_sim_rabief}.
See Appendix~\ref{appendix:qutrit_sim} for more details on the simulation setup.

Our simulations include three active resets interspaced in a
$250~\mathrm{\mu{}s}$ repetition delay since multiple reset
instructions increase the fidelity of initializing the
transmon~\cite{Brandhofer2023}.
The long delay for passive resets is set to $10~\mathrm{ms}$.
We simulate thermal relaxation of the qutrit during all idle times
and measurements.
Measurement errors are modeled as thermal relaxation,
measurement-induced leakage, and readout signal misclassifications.
Misclassifications are simulated with a \textit{readout assignment
  error} matrix $R$, storing probabilities $q_{jk}$ to misclassify
$\ket{k}$ as $\ket{j}$.
For a state with ideal measurement probabilities $(\alpha, \beta,
  \gamma)$, for $(\ket{0},\ket{1},\ket{2})$, the noisy final
measurement probabilities are $R(\alpha, \beta, \gamma)^T$, where
\begin{equation}\label{eq:readout_assignment_error}
  R=
  \begin{pmatrix}
    q_{00} & q_{01} & q_{02} \\
    q_{10} & q_{11} & q_{12} \\
    q_{20} & q_{21} & q_{22} \\
  \end{pmatrix}
  =
  \begin{pmatrix}
    0.991 & 0.009 & 0    \\
    0.009 & 0.931 & 0.06 \\
    0     & 0.06  & 0.94 \\
  \end{pmatrix}
  .
\end{equation}
The normalization constraint $q_{0j} + q_{1j} + q_{2j}=1$ applies to
all $j=0,1,2$.
For example, the probability of measuring a $\ket{1}$ for a given
state is $q_{10}\alpha + q_{11}\beta + q_{12}\gamma$.
The values in $R$ are chosen to match hardware experiments on
superconducting qubits~\cite{fischer_ancilla-free_2022,chen2023,kanazawa2023}.
As a result, our $R$ satisfies $q_{02}=q_{20}=0$, $q_{jk}=q_{kj}$,
and $q_{01}<q_{12}$.
Furthermore, the simulated measurement errors correspond to
measurement noise channel fidelities $\fa=0.99096$ and $\fs=0.99096$,
the associated derivation is in Appendix~\ref{app:assignment_matrix}.
The \textit{readout assignment error} matrix for qubit discrimination
and resets is obtained from $R$ with transformations $q_{10}\to
  q_{10}+q_{20}$, $q_{11}\to q_{11}+q_{21}$.
This is equivalent to qutrit discrimination followed by a
$\ket{2}$-$\ket{1}$ misclassification with $100~\%$ probability.

These simulation results are plotted in
Figs.~\ref{fig:qutrit_sim_rabief}(c)-(f), with the true and estimated
state-preparation probabilities in Tab.~\ref{tab:qutrit_sim_fits}.
Both RabiEF signals \textit{should} be of the form $a\sin^2(b\theta)
  + c$ with non-negative amplitudes $a$ being physical, and all
estimates $\hatpsp$ should be valid probabilities, i.e.,
${0\leq{}\hat{p}_\text{sp}\leq{}1}$.
However, the non-negligible $\ket{2}$-state population with
\emph{fast qubit} reset results in a negative amplitude of the no-pi
signal ($\cancel{\pi}$).
This gives an incorrect negative probability estimate, see
\cref{fig:qutrit_sim_rabief}(c) and $\hat{p}_\text{sp}$ in
Tab.~\ref{tab:qutrit_sim_fits}.
Indeed, initial states of the form
\begin{equation}
  \rho_\text{sp} = (1-p_\text{sp}-p_\text{sp}^{(2)})\ket{0}\!\bra{0}
  + p_\text{sp}\ket{1}\!\bra{1} + p_\text{sp}^{(2)}\ket{2}\!\bra{2}
\end{equation}
bias our estimate $\hat{p}_\text{sp}$ as
\begin{equation}\label{eq:rabief_bias}
  \hat{p}_\text{sp}=\frac{a_{\cancel{\pi}}}{a_{\cancel{\pi}} +
    a_{\pi}}=\frac{p_\text{sp} - p_\text{sp}^{(2)}}{1-3p_\text{sp}^{(2)}}.
\end{equation}
If $p_\text{sp}^{(2)}$ is sufficiently large, the no-pi signal has a
negative amplitude and our estimate becomes negative and unphysical
\footnote{This argumentation holds for an initial state with
  non-negligible $\ket{2}$-state population which is the same for each shot.
  However, this is not the case in our simulations as the initial
  state changes from shot to shot.
  This manifests as large standard deviations in $\psp$ and
  $p_\text{sp}^{(2)}$ with fast qubit resets, see \cref{tab:qutrit_sim_fits}.
  The large fluctuations in $\ket{1}$- and $\ket{2}$-state populations
  is the cause of distortions in the RabiEF signals, deviating from the
  form $a\sin^2(b\theta) + c$ entirely.}.
In contrast, our simulations of fast \emph{qutrit} resets and slow
resets are physical.
Furthermore, if we look at the standard deviation of the true value
$\psp$, shown in Tab.~\ref{tab:qutrit_sim_fits},
we see that the prepared state is more stable with fast qutrit resets
and slow resets than with fast resets based on qubit discrimination.
Crucially, the relative accuracy of the RabiEF with fast qutrit reset
is -5.6\% on a true $\psp$ of $0.01206\%$.
\begin{table}[!bt]
  \centering
  \caption{
    \textbf{State-preparation errors from simulated RabiEF.}
    The estimated probabilities $\hat{p}_\text{sp}$ for fast and slow
    resets are given, with the corresponding \textit{true}
    probabilities $p_\text{sp}$.
    The standard errors for $\hat{p}_\text{sp}$ are those reported by
    the curve fitting software.
    True values are averages over the ensemble of initial states
    $\rho_{\text{sp},k}$ from all shots $k$ in the qutrit
    simulations, e.g.,
    $\smash{p_\text{sp}=\operatorname{avg}_k\Tr{[\ket{1}\!\bra{1}\rho_{\text{sp},k}]}}$.
    True values $\smash{p_\text{sp}^{(2)}}$ for the $\ket{2}$ state
    populations are obtained with
    $\smash{p_\text{sp}^{(2)}=\operatorname{avg}_k\Tr{[\ket{2}\!\bra{2}\rho_{\text{sp},k}]}}$.
    Problematic cases are highlighted in bold.
    This includes fits that give unphysical results, $\psp$ and
    $\hat{p}_\text{sp}$ that fluctuate widely, and non-negligible
    $\smash{p_\text{sp}^{(2)}}$ values.
  }\label{tab:qutrit_sim_fits}
  \setlength\extrarowheight{2pt}
  \sisetup{table-format=0e-3, table-alignment-mode=format,
    table-number-alignment=center,table-fixed-width,table-column-width=22mm}
  \begin{ruledtabular}
    \begin{tabular}{lrSS}
      \multicolumn{2}{c}{}                   & \multicolumn{2}{c}{Fast Reset}
      \\\cline{3-4}
      \multicolumn{1}{c}{}                   &
                                             & {Qubit Reset}                  & {Qutrit Reset}         \\
      \hline
      \multirow{2}{1mm}{$\hatpsp$}           & Fit
                                             & \bfseries -0.07375362633110621 & 0.01138434338016134    \\
                                             & Std. Err.                      & \bfseries
      0.002365811295332557                   & 2.066484330496584e-06                                   \\
      \hline
      \multirow{2}{1mm}{$\psp$}              & Mean
                                             & 0.02842563063242803            & 0.012064411613571418   \\
                                             & Std. Dev.                      & \bfseries
      0.018402575246837197                   & 1.3869525212574677e-05                                  \\
      \hline
      \multirow{2}{1mm}{$p_\text{sp}^{(2)}$} & Mean
                                             & \bfseries 0.024294292321174377 & 0.0006517619781123185  \\
                                             & Std. Dev.                      & \bfseries
      0.026619774732092136                   & 2.393038451362641e-05                                   \\
      \hline
      \\
      \multicolumn{2}{c}{}                   & \multicolumn{2}{c}{Slow Reset}
      \\\cline{3-4}
      \multicolumn{1}{c}{}                   &
                                             & {Qubit Reset}                  & {Qutrit Reset}         \\
      \hline
      \multirow{2}{1mm}{$\hatpsp$}           & Fit
                                             & 0.018993790238475614           & 0.018993872162638512   \\
                                             & Std. Err.                      & 4.3537407080999903e-07
                                             & 3.982392987750876e-10                                   \\
      \hline
      \multirow{2}{1mm}{$\psp$}              & Mean
                                             & 0.019460143125448002           & 0.01945693629988275    \\
                                             & Std. Dev.                      & 3.913718433775682e-06
                                             & 3.2283019593791506e-09                                  \\
      \hline
      \multirow{2}{1mm}{$p_\text{sp}^{(2)}$} & Mean
                                             & 0.0004911951199155133          & 0.0004910374920579088  \\
                                             & Std. Dev.                      & 1.9237438001444048e-07
                                             & 1.5868437451608185e-10
    \end{tabular}

  \end{ruledtabular}
\end{table}

These simulations demonstrate how unstable state-preparation, where
$\ket{2}$ is not reset correctly, is insufficient for RabiEF.
Without an accurate estimate of $\psp$, separating SPAM errors in our
model to mitigate dynamic circuits is not possible.
Our full MCM noise learning protocol bypasses \textit{bad} fast
resets to obtain the estimate $\hat{p}_\text{sp}$, using slow resets.
Though slow resets can be used for mitigation, they result in
expensive circuit execution owing to their slow rates.
Fast \emph{qutrit} resets --- even if imperfect --- allow for the
state-preparation characterization required to quickly split SPAM errors.

\subsection{Cycle benchmarking for readout}\label{sec:sim_cb}

We now show that including an estimate of the state-preparation error
in a MCB experiment results in accurate estimates of the measurement
noise fidelities $\fa$, $\fs$, and $\fc$.
We run two simulations with slow and fast resets to illustrate both
noise learning paths in the full MCM noise learning protocol in
\cref{fig:workflow}.

The true measurement noise fidelities are set to $\fa=0.99096$,
$\fs=0.99096$, and $\fc=0.9950$ to match the values in
\cref{sec:sim_rabief}, see Appendix~\ref{app:assignment_matrix}.
The first simulation learns $\fa$, $\fs$, and $\fc$ with slow
state-preparation $\pspslow=0.01946$.
Then, the fast state-preparation error probability $\pspfast=0.01206$
is learned with a second MCB experiment, using the now learned
measurement fidelities $\fa$, $\fs$, and $\fc$.
The first simulation splits the products of fidelities in
Eq.~(\ref{eq:ptm_full}) with the slow reset estimate from RabiEF,
given in Tab.~\ref{tab:qutrit_sim_fits}, i.e.,
$\hat{p}_\text{sp}^\text{slow}=0.01899$.

\begin{figure}
  \centering
  \includegraphics{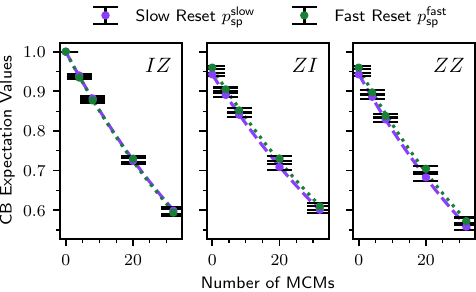}
  \caption{
    \textbf{Measurement cycle benchmarking on a qubit.}
    Fits to these data determine the noise channel fidelities.
    Error bars show the $\pm{}2\sigma$ standard errors for each
    expectation value.
    Dotted and dashed lines are obtained from non-linear least-square
    fits of the equation $Af^{2k}$ to the data, where $k$ is a
    function of the depth.
  }
  \label{fig:qubit_sim_cb}
\end{figure}
We simulated MCB with $256$ twirled-circuit randomizations for each
depth $k=0,1,2,5,8$, and $128$ shots per randomization.
The simulations were run with Qiskit~Aer~\cite{qiskitaer2025}, the
details of which are in Appendix~\ref{appendix:qubit_sim}.
The resulting six decay curves are shown in \cref{fig:qubit_sim_cb}:
three per reset type, corresponding to the observables $IZ$, $ZI$,
and $ZZ$ identifying the PTM entries in \cref{eq:ptm_full}.
We estimate the measurement noise fidelities by fitting decaying
exponentials to the slow-reset data.
The same fitting procedure is carried out for the fast reset
simulation data, except that the fast reset fidelity $\fspfast$ is
estimated using the learned measurement noise fidelities from the
first simulation.

Not only does our learning procedure effectively learn the noise
fidelities to less than $0.01~\%$ relative error, but we can also
estimate the fast reset fidelity $\fspfast$ with the second MCB
experiment, see Tab.~\ref{tab:cb_sim_estimates}.
The state-preparation method can thus be different between full MCM
noise learning and error mitigation, as long as an additional MCB
experiment is run to estimate the fast state-preparation error $\pspfast$.
This demonstrates that both paths in our full MCM noise learning
protocol result in an estimate of the full noise model in
\cref{eq:ptm_full} with split SPAM errors.
An extension of our scheme to high-rate and high fidelity state
preparation is discussed in Appendix~\ref{sec:high_rate}.

\begin{table}[!b]
  \centering
  \caption{
    \textbf{Comparison of noise fidelities from simulated MCB.}
    The estimated fidelities $\hat{f}$ are given, with their
    corresponding true fidelities $f$ and the relative error $\abs{f-\hat{f}}/f$.}
  \label{tab:cb_sim_estimates}
  \begin{ruledtabular}
    \setlength\extrarowheight{2pt}
    \sisetup{table-format=0e-3, table-alignment-mode=format,
      table-number-alignment=center}
    \begin{tabular}{lccc}
      \multicolumn{4}{c}{\bfseries MCB with slow resets}
      \\
      \hline     & {True $f$}   & {Estimated $\hat{f}$} & {Rel. Error}           \\ \hline
      $\fa$      & $0.99096$    & $0.99101$             & $5.637\times{}10^{-5}$ \\
      $\fs$      & $0.99096$    & $0.99022$             & $7.460\times{}10^{-4}$ \\
      $\fc$      & $0.99500$    & $0.99570$             & $7.055\times{}10^{-4}$ \\
      \hline                                                                     \\
      \multicolumn{4}{c}{\bfseries Second MCB experiment with fast resets}       \\
      \hline
                 & {True $f$}   & {Estimated $\hat{f}$}
                 & {Rel. Error}
      \\
      \hline
      $\fspfast$ & $0.97588$    & $0.97676$             & $9.061\times{}10^{-4}$ \\
    \end{tabular}
  \end{ruledtabular}
\end{table}

\subsection{Example mitigation of final readout errors}\label{sec:sim_mitig}

We now show classical simulations of our method on the stabilizer
circuit in \cref{fig:experiment}(a).
The simulations use the same slow reset fidelity as in
\cref{sec:sim_rabief,sec:sim_cb}, and implement both TREX and the
mitigation from \cref{eq:trexpp}.
These simulations, done with Qiskit~Aer, do not simulate leakage, as
the $\ket{2}$-state is assumed negligible for all experiments except RabiEF.

\begin{figure}
  \centering
  \includegraphics{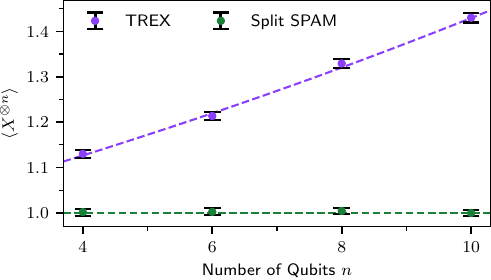}
  \caption{
    \textbf{Mitigation of the $\langle X^{\otimes n}\rangle$ expectation value.}
    The values are mitigated with TREX and the modified mitigator
    from the hardware experiments, which splits state-preparation and
    measurement errors (``Split SPAM'').
    Error bars show the $\pm{}2\sigma$ standard errors for each
    mitigated expectation value.
    Dashed lines show the expected expectation values given knowledge
    of the simulated GHZ state preparation circuit.
    The green dashed line shows the noiseless expectation value of $+1$.
  }
  \label{fig:qubit_sim_mitig}
\end{figure}

The TREX mitigator $\langle{}Z^\star\rangle$ is learned with $16$
twirling randomizations and $5000$ shots per randomization, for
$n=4,6,8,10$ qubits, i.e., the same as in \cref{sec:exp}.
The measurement and state-preparation noise fidelities were taken
from \cref{sec:sim_cb}.
The stabilizer circuit was simulated for $n=4,6,8,10$ qubits, with
$128$ shots per twirling randomization, and $2^{10}\times{}n^2$ total shots.
With the hardware experiments, we compare our mitigated results to
Clifford simulations with CNOT noise.
By contrast, here we assume ideal gates to compare the mitigated
expectation values to their ideal value of $+1$, the  dashed green
line in \cref{fig:qubit_sim_mitig}.
The expected result with TREX is $f_\text{sp}^{1-n}$, shown by the
dashed purple line in \cref{fig:qubit_sim_mitig}.

As seen in the hardware experiments, TREX overshoots the ideal
expectation values by a factor $\fsp^{1-n}$, see purple dots in
\cref{fig:qubit_sim_mitig}.
This is expected since $\langle Z^\star\rangle$ picks-up the same
factor from imperfect state-preparation, i.e., the purple dashed line.
If, however, the values are mitigated with \cref{eq:trexpp}, then the
expectation values do not overshoot, and instead lie around the ideal
value of $\langle{}O\rangle{}=+1$, see green dots in \cref{fig:qubit_sim_mitig}.
This shows a correct mitigation of SPAM errors, further validating
the error mitigation employed in \cref{sec:noise_learning}.

\subsection{Example mitigation of dynamic circuits}\label{sec:sim_dynamic}

\begin{figure}
  \centering
  \includegraphics[width=\linewidth, clip, trim=0 150 50 0]{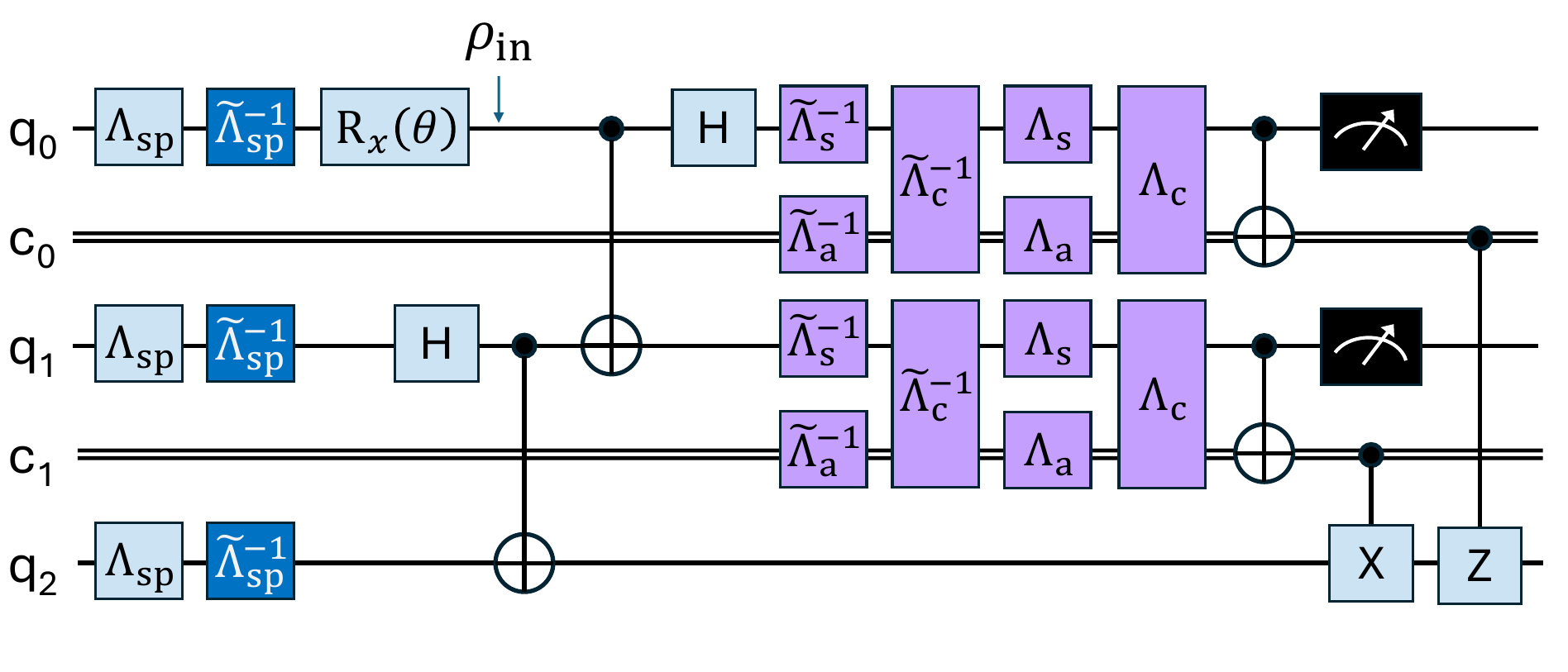}
  \caption{
    \textbf{Teleportation Circuit in the CNOT picture.}
    The circuit includes state-preparation noise $\Lambda_\text{sp}$
    and measurement noise $\Lambda_\text{s}$, $\Lambda_\text{a}$, and
    $\Lambda_\text{c}$.
    Inverse channels $\tilde{\Lambda}^{-1}$ are implemented by
    sampling their quasi-probability distributions for PEC.
    The output state $\rho_\text{out}$ is the state of the last qubit
    $q_2$, i.e., with the other qubits and classical bits traced out.
    The state to be teleported is $\rho_\text{in}$, which is affected
    by the state-preparation noise $\Lambda_\text{sp}$ of qubit zero
    and its inverse channel $\tilde{\Lambda}_{sp}^{-1}$.
  }
  \label{fig:teleportation_circuit}
\end{figure}

Splitting SPAM errors is most beneficial with mid-circuit
measurements, where measurement noise can be mitigated separately
from state-preparation and gate noise.
We demonstrate that our MCM noise learning protocol helps mitigate
errors in dynamic circuits.
In particular, we simulate a noisy three-qubit teleportation circuit
and mitigate SPAM errors with
PEC~\cite{van2023probabilistic,gupta2024probabilistic} built on our
noise learning protocol.
The circuit is constructed in the CNOT picture, i.e., including the
classical bits as is done in \cref{fig:meas_models}, and simulated
with Qiskit~Aer~\cite{qiskitaer2025}.
The teleportation circuit, shown in \cref{fig:teleportation_circuit},
includes explicit noise channels for state-preparation
$\Lambda_\text{sp}$, measurement assignment $\Lambda_\text{a}$, state
$\Lambda_\text{s}$, and correlated $\Lambda_\text{c}$ errors, as
defined in \cref{sec:noise_learning}.
The underlying fidelities $f_\text{x}$ are taken from
\cref{sec:sim_cb}, with a state-preparation fidelity $\fsp=0.96108$
equivalent to the slow resets.

The ideal teleported state
\begin{equation}\label{eq:teleportation_ideal_state}
  \rho_\text{in}^\text{ideal} = \ket{\psi}\!\!\bra{\psi} =
  R_X(\theta)\ket{0}\!\!\bra{0}R_X^\dagger{}(\theta).
\end{equation}
is prepared by an $R_X$ gate with rotation angle $\theta\in[0,2\pi]$.
However, SPAM noise introduce errors into the final teleported state
on qubit $\text{q}_2$.

\begin{figure}
  \centering
  \includegraphics{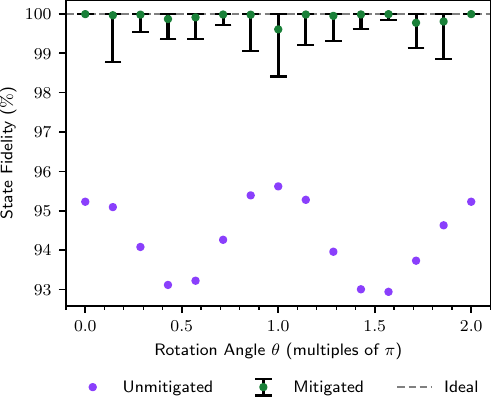}
  \caption{
    \textbf{State fidelities for unmitigated and PEC mitigated
      teleportation circuits.}
    PEC is used to mitigate Pauli expectation values which are then
    used with state-tomography to estimate the state infidelity.
    The median mitigated state infidelities are shown in green, over
    $300$ random realizations of the simulation.
    Error-bars show the interquartile range over the $300$ realizations.
    The unmitigated fidelities, in purple, are computed directly on
    the unmitigated state $\rho_{out}$ without PEC samples or state tomography.
  }
  \label{fig:sim_dynamic_fidelities}
\end{figure}

To apply PEC, we generate $128$ PEC circuit realizations, replacing
the inverse channels $\tilde{\Lambda}^{-1}$ with Paulis from their
quasi-probability distribution representations.
Each circuit is then simulated, resulting in $128$ noisy teleported
states $\rho_{\text{out},j}$ on qubit $\text{q}_2$, with $j\in[0,127]$.
As PEC mitigates expectation values and not states, we obtain
mitigated states via state tomography.
From each noisy teleported state $\rho_{\text{out},j}$, we obtain
three ideal Pauli expectation values
\begin{equation}
  \langle{}P\rangle{}_j = \Tr[P\rho_{\text{out},j}],
\end{equation}
for Paulis $P=X,Y,Z$. We emulate shot-noise by sampling $100$ times,
per circuit realization, from the corresponding Bernoulli
distributions, resulting in expectation values $E_{P,j}$ impacted by
both SPAM and shot noise.
The PEC mitigated expectation values are then
\begin{equation}\label{eq:pec}
  \langle{}P\rangle{}_\text{mit} = \gamma\sum_j(-1)^{m_j}E_{P,j}
\end{equation}
where $\gamma$ is the noise factor for the inverse noise channels and
$m_j$ is an integer from PEC controlling how the quasi-probability
inverse channels are
engineered~\cite{van2023probabilistic,gupta2024probabilistic}.
Physical states are obtained from state tomography by minimizing a
mean-squared error as follows:
\begin{equation}\label{eq:state_mse}
  \rho_\text{mit} =
  \argmin_{\rho_t}
  \sum_{P=X,Y,Z}
  \left(
  \Tr[P\rho_t]
  -
  \langle{}P\rangle{}_\text{mit}
  \right)^2.
\end{equation}
Here, the state $\rho_t$ is parameterized as
\begin{equation}
  \rho_t=\rho(t_1,t_2,t_3,t_4) = \frac{T^TT}{\Tr[T^TT]},
\end{equation}
with real parameters $t_1$, $t_2$, $t_3$, $t_4$, and triangular matrix
\begin{equation}
  T =
  \begin{pmatrix}
    t_1        & 0   \\
    t_3 + it_4 & t_2
  \end{pmatrix}.
\end{equation}
This ensures a physical mitigated state $\rho_\text{mit}$ for a given
ensemble of PEC circuit realizations.
We then compute the state fidelity
\begin{equation}\label{eq:state_fidelity}
  \mathcal{F}(\rho_\text{in}^\text{ideal},\rho_\text{mit})
  =
  \Tr\!\left[\sqrt{\sqrt{\rho_\text{in}^\text{ideal}}\rho_\text{mit}\sqrt{\rho_\text{in}^\text{ideal}}}\right]^2
\end{equation}
as a metric of success.
This is done for $300$ realizations of the simulation and $15$
rotation angles in the input state, see \cref{fig:sim_dynamic_fidelities}.
The unmitigated case does not need PEC and is thus simulated as a
single circuit directly resulting in a physical state.
Therefore, the state fidelity is computed on the output state
$\rho_\text{out}$ directly, instead of using state tomography.

The PEC mitigated state fidelities are significantly higher than the
unmitigated values, showing that we are mitigating SPAM errors, see
\cref{fig:sim_dynamic_fidelities}.
The majority of the $300$ simulation realizations result in
state-fidelities above $98~\%$, whereas the best-case unmitigated
fidelity is less than $96~\%$.
We also confirm that PEC effectively mitigates both state-preparation
and measurement errors separately by rerunning the simulations with
only one of the error sources, still observing improvements to the
state fidelities with PEC (data not shown).
Our MCM noise learning procedure thus effectively splits SPAM errors,
facilitating PEC mitigation of dynamic circuits.
Splitting SPAM errors was made possible by learning the
state-preparation fidelity $\psp$ using non-computational states with
a RabiEF experiment.

\section{Discussion and Conclusion\label{sec:conclusion}}

Our experimental data show how state-preparation errors prevent the
accurate mitigation of SPAM errors for certain expectation values,
corroborating recent results~\cite{Chen2025}.
A passive reset of the qubits into a thermal state, whose population
we measure with non-computational states, allows us to resolve this
by mitigating state-preparation errors independently of measurement errors.
This prevents the unphysical behavior seen in some TREX-mitigated
expectation values.

We introduce a noise model of (mid-)circuit measurements which fully
accounts for state-preparation errors and errors in the final measurement.
This extends recent noise learning protocols, such as
Ref.~\cite{zhang2025generalized}, by learning the state-preparation
error independently of measurement noise.
Our work also complements a recently introduced noise learning
framework that learns state-preparation, gate, and measurement noise
is a self-consistent way up to unlearnable gauge degrees of
freedom~\cite{chen2024efficient}.
While this protocol also resolves the aforementioned issues of
TREX~\cite{Chen2025}, the unconstrained degrees of freedom render it
unsuitable for dynamic circuits.

With the added resource of non-computational states, we learn the
ground truth of the state-preparation error.
Crucially, our noise learning then allows us to mitigate errors that
occur in dynamic circuits which explicitly require MCMs and feedforward gates.
We successfully demonstrate this with simulations of a teleportation
circuit, mitigating errors with PEC.
The noise model resulting from our protocol is applicable to other noise-aware
error mitigation methods which utilize twirled noise channels, such as
PEA~\cite{kim2023evidence}.

As an outlook, we propose to combine RabiEF with other gate- and
measurement-noise learning techniques and, more generally, the formalism from
Refs.~\cite{chen2024efficient, Chen2025}.
This will fully constrain the self-consistent noise models by
anchoring the gauge, thus significantly broadening their scope.

We propose to learn the state-preparation error with a RabiEF experiment which
requires either a slow passive reset or a fast active qutrit reset.
This fits well with recent developments since error correction requires a strong
suppression of leakage which can be achieved through appropriate qutrit
resets~\cite{Battistel2021, Miao2023}.
Importantly, our work requires the transmon to stay within its first
three levels.
Therefore, we do not capture effects where the transmon may escape from this
subspace~\cite{Khezri2023} or even confined states~\cite{Lescanne2019}.
Here, we believe that operating transmons in such regimes should be avoided.

Our work leverages a non-computational state of transmon qubits.
However, we expect the work should generalize to other hardware
platforms that have addressable non-computational states.
These may be found in, for instance, solid state electron spin
qutrits~\cite{Fu2022, Guo2024} and trapped ions~\cite{Klimov2003}.
An analytical treatment of the impact of all errors in the
measurements and how they propagate to the mitigated observables
would be a useful extension of the numerical analysis in
Appendix~\ref{appendix:meas_sensitivity}.
In addition, future work may also explore generalizations of the qubit readout
model to a full qutrit model, including measurement-induced leakage, and
the subsequent impact on noise
learnability~\cite{blok2021,hazra2025benchmarking}.
In our work we show noise models for up to two correlated qubits undergoing
measurement.
Generalizing our protocol and noise model to include correlations between more
qubits and with idle qubits should allow for more accurate error mitigation on
systems where strong correlations occur~\cite{gupta2024probabilistic,
  Govia2025, Beale2023, van2023probabilistic}.

In summary, non-computational states allow us to overcome no-go
theorems that hold when the learning framework is restricted to the
computational subspace.
As a result, we can fully specify degrees of freedom of noise models
that are fundamentally unlearnable when restricted to qubit circuits.
We illustrate this by separately mitigating state and preparation errors.
Crucially, this is also possible without a slow passive reset of the
quantum computer when a noisy qutrit reset is available.

\emph{Note added:} Recently we became aware of related work by S. Chen \emph{et
  al.}~\cite{Chen2025b}, in which the authors also leverage non-computational
states of the transmon.

\section{Acknowledgements}

The authors acknowledge M. Mergenthaler, A. Fuhrer, and L. C. G.
Govia, M. Takita, and A. Seif for useful discussions.
This work was supported as a part of NCCR SPIN, a National Centre of
Competence in Research, funded by the Swiss National Science
Foundation (grant number 225153).
D.J.E.~acknowledges funding within the HPQC project by the Austrian
Research Promotion Agency (FFG, project number 897481) supported by
the European Union -- NextGenerationEU.
L.E.F. acknowledges funding from the European Union's Horizon 2020
research and innovation program under the Marie Sk\l{}odowska-Curie
grant agreement No.~955479 (MOQS -- Molecular Quantum Simulations).
\appendix

\section{Measuring thermal state population\label{appendix:rabief}}

The RabiEF experiment measures the population $\alpha$ in the excited
state of the qubit for a state $\rho=(1-\alpha)\ket{0}\!\bra{0} +
  \alpha\ket{1}\!\bra{1}$~\cite{Geerlings2013}.
RabiEF drives oscillations in the $\{\ket{1}, \ket{2}\}$ subspace
with a $R_{12}(\theta)$ gate which rotates around the $X$-axis of the
corresponding Bloch sphere.
Measuring $\alpha$ requires two independent experiments called $\pi$
and no-$\pi$, indicated by $\cancel{\pi}$.
In the $\pi$-experiment we apply the gate sequence
$X-R_{12}(\theta)-X$ while the  no-$\pi$-experiment omits the first $X$ gate.
The difference between both experiments is thus the $\pi$-pulse on
the thermal state at the beginning of the gate sequence.
Following this gate sequence, which we assume is ideal, we apply a
noisy measurement with discrimination errors.
This results in two oscillating signals $s_\pi(\theta)$ and
$s_{\cancel{\pi}}(\theta)$ which we fit to two independent functions
of the form $a\sin^2(\theta + c) +b$, see Fig.~\ref{fig:experiment}(d).
Here, $a$, $b$, and $c$ are the fit parameters.
Finally, the excited state population $\alpha$ is estimated by
\begin{align}\label{eqn:thermal_meas}
  \hat{\alpha}=\frac{a_{\cancel{\pi}}}{a_{\cancel{\pi}}+a_{\pi}}
\end{align}
where $a_{\cancel{\pi}}$ and $a_{\pi}$ are the amplitudes of the
function that fit $s_{\cancel{\pi}}(\theta)$ and $s_\pi(\theta)$, respectively.
In practice, we implement $R_{12}(\theta)$ with a Rabi pulse
$R_{12}(A)$ of amplitude $A\propto{}\theta$.
The exact relationship between $A$ and $\theta$ is irrelevant, as it
only controls the frequency of the RabiEF signals, and not their amplitudes.

Superconducting qubit measurements return complex values in the IQ
plane~\cite{Krantz2019} to which we apply a principal component
analysis to forgo classification and instead use the projected
signals $s_\pi^\prime(A)\propto{}s_\pi(\theta)$ and
$s_{\cancel{\pi}}^\prime(A)\propto{}s_{\cancel{\pi}}(\theta)$.
These signals have arbitrary units and can be larger than $1$ owing
to the location of the reference points in the IQ plane, see
\cref{fig:experiment}(d).
Crucially, gate and measurement errors tend to cancel out in
\Cref{eqn:thermal_meas} as they affect $a_\pi$ and $a_{\cancel{\pi}}$
in the same way.

\subsection{Thermal state populations \label{appendix:thermal}}

Here, we discuss the experimental thermal state data presented in
Fig.~\ref{fig:experiment} which is taken on qubits 114 to 121 on the
IBM Quantum device \texttt{ibm\_pinguino3}.
The thermal populations of these qubits are measured with RabiEF
after a $10~\mathrm{ms}$ passive reset.
The resulting $\ket{1}$ state population $\alpha$ ranges from $3.06$
to $8.75~\%$, see Tab.~\ref{tab:qubit_properties}, and creates a
significant state-preparation error.
From the measured population $\alpha$ in the $\ket{1}$ state we
compute an effective qubit temperature $T_\text{eff}$ following the
relation $\alpha=\exp[-\hbar\omega_{01}/(k_bT_\text{eff})]$.
The resulting effective temperatures, shown in
Tab.~\ref{tab:qubit_properties}, are larger than
the~${\sim15~\mathrm{mK}}$ of dilution refrigerators.
We then use the above relation to estimate the population in the
second excited state by replacing $\omega_{01}$ with
$2\omega_{01}+\Delta$, where $\Delta$ is the transmon anharmonicity.
The resulting populations $\beta$ are at least an order of magnitude
lower than their corresponding populations $\alpha$ in the first excited state.
This justifies the assumption that the $\ket{2}$-state population is
negligible when slow resets are used.

\begin{table}[]
  \centering
  \caption{
    Properties of qubits used in the data presented in
    Fig.~\ref{fig:experiment} of the main text.
    The qubit frequency is $\omega_{01}$.
    The transmon anharmonicity is $\Delta$.
    The population in the $\ket{1}$ state measured with RabiEF is $\alpha$.
    The inferred population in the second excited state is $\beta$.
  }
  \begin{ruledtabular}
    \begin{tabular}{l r r r r r r r r}
      Qubit                               & 114  & 115  & 116  & 117  & 118  & 119  & 120  & 121  \\ \hline
      $\omega_{01}/(2\pi)~(\mathrm{GHz})$ & 5.14 & 5.22 & 5.08 & 5.22
                                          & 5.28 & 5.16 & 5.03 & 5.15                             \\
      $\Delta/(2\pi)~(\mathrm{MHz})$      & -303 & -303 & -304 & -302 &
      -303                                & -304 & -305 & -302                                    \\ \hline
      $\alpha~(\%)$                       & 4.99 & 8.75 & 4.01 & 6.00 & 4.17 & 3.06 & 5.42 & 2.79 \\
      $T_\text{eff}~(\mathrm{mK})$        & 82   & 103  & 76   & 89   & 80   & 71   & 83   & 69   \\
      $\beta~(\%)$                        & 0.30 & 0.88 & 0.19 & 0.42 & 0.21 & 0.11 & 0.35 &
      0.10                                                                                        \\ \hline
      $T_1~(\mu\mathrm{s})$               & 281  & 259  & 207  & 222  & 199  & 293  & 175  & 261  \\
    \end{tabular}
  \end{ruledtabular}
  \label{tab:qubit_properties}
\end{table}

\subsection{RabiEF and assignment errors\label{appendix:meas_sensitivity}}

\begin{figure*}
  \centering
  \includegraphics[width=0.95\textwidth]{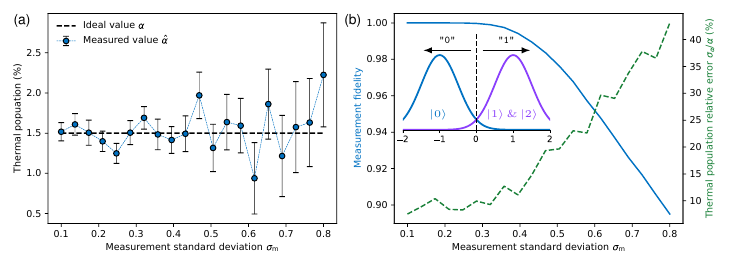}
  \caption{
    \textbf{Sensitivity of RabiEF on assignment errors.}
    (a) Simulations of estimates of $\ket{1}$ state population
    $\alpha$ with 300 shots per data point.
    The error bars are obtained by propagating fit errors on $a_\pi$
    and $a_{\cancel{\pi}}$ through Eq.~(\ref{eqn:thermal_meas}).
    The dashed horizontal line shows the ideal value $\alpha=1.5\%$.
    (b) The solid line shows how the readout fidelity $F$ varies with
    $\sigma_\text{m}$.
    The green dashed line shows how the error bars from panel (a),
    normalized to $\alpha$, grow with the strength of assignment
    errors $\sigma_\text{m}$ for a fixed 300 shots per $\hat{\alpha}$.
    The inset shows how the three states are assigned to labels
    \texttt{"0"} and \texttt{"1"}.
  }
  \label{fig:thermal_pop_vs_sigma}
\end{figure*}

We provide numerical evidence that the RabiEF measurement of $\alpha$
is accurate despite measurement assignment errors.
We construct a three level model with states $\ket{0}$, $\ket{1}$,
and $\ket{2}$.
The initial state is the thermal state
$\rho_\text{th}=(1-\alpha)\ket{0}\bra{0}+\alpha\ket{1}\bra{1}$ where
we assume that the thermal population in $\ket{2}$ is negligible.
The gates in the RabiEF circuits are assumed ideal.
Crucially, we add finite sampling effects and discrimination errors
in the readout process with a discrimination in a one-dimensional
space with overlapping Gaussians.

Here, the state $\ket{0}$ is mapped to a Gaussian distribution
$\mathcal{N}(\mu=-1,\sigma_\text{m})$ with mean $-1$ and standard
deviation $\sigma_\text{m}$.
The $\ket{1}$ and $\ket{2}$ states are mapped to a single Gaussian
distribution $\mathcal{N}(\mu=1,\sigma_\text{m})$ with mean $+1$ and
standard deviation $\sigma_\text{m}$.
To draw a shot from a state $\rho$ we first chose a random number
$\mu$ from $\{-1, 1\}$ where the probability of $-1$ is
$\Tr[\rho\ket{0}\!\!\bra{0}]$ and the probability of $+1$ is
$\Tr[\rho(\ket{1}\!\!\bra{1}+\ket{2}\!\!\bra{2}]$.
Next, we mimic the readout process by sampling from
$\mathcal{N}(\mu, \sigma_\text{m})$ and assigning a count of 1 if
the result is greater than 0.
Therefore, the overlap between the two Gaussian distributions
$\mathcal{N}(\mu=-1,\sigma_\text{m})$ and
$\mathcal{N}(\mu=1,\sigma_\text{m})$ --- controlled by
$\sigma_\text{m}$ --- creates measurement assignment errors, see
inset in Fig.~\ref{fig:thermal_pop_vs_sigma}(b).
The relation between $\sigma_\text{m}$ and the readout fidelity
$F=1-[P(1|0)+P(0|1)]/2$ is shown in Fig.~\ref{fig:thermal_pop_vs_sigma}(b).

\begin{figure}
  \centering
  \includegraphics[width=0.99\columnwidth, clip, trim=0 0 0 20]{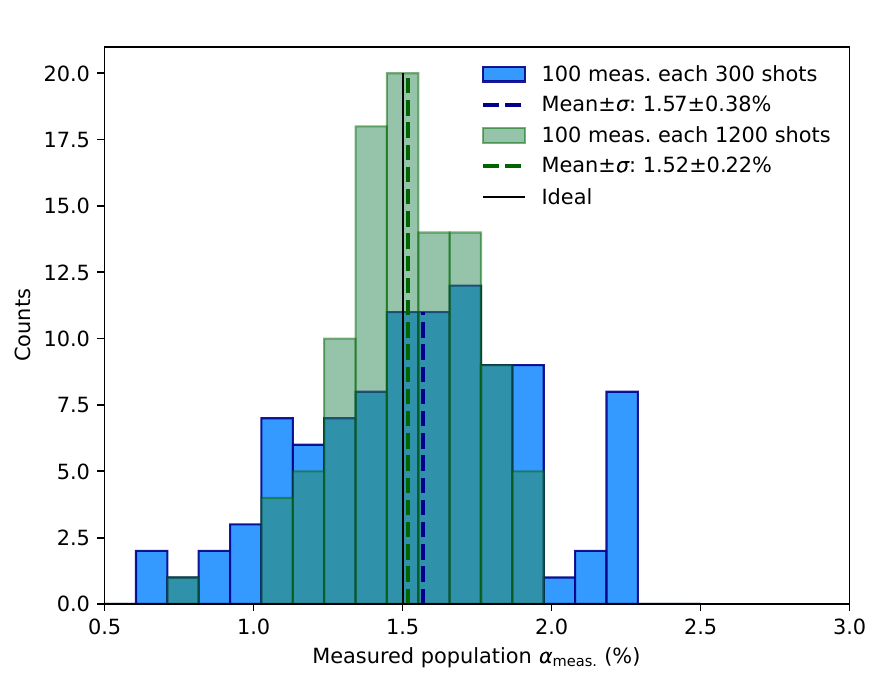}
  \caption{
    \textbf{Increasing shots improves RabiEF precision.}
    Distribution of 100 individual measurements of $\alpha$ with
    300 and 1200 shots.
    Here, $\sigma_\text{m}=0.6$ which corresponds to a readout error of 4.8\%.
    By comparison, the mean readout error on \emph{ibm\_fez} is 1.7\%.
  }
  \label{fig:thermal_pop_statistics}
\end{figure}

Even in the presence of strong assignment errors, e.g.,
$\sigma_\text{m}>0.4$ for which $F<99\%$, RabiEF accurately measures $\alpha$.
The estimates $\hat{\alpha}$ center around the ideal value set to
$1.5~\%$, see Fig.~\ref{fig:thermal_pop_vs_sigma}(a).
Furthermore, the magnitude of the error bars, correspondingly the
standard deviation of the distribution of independent measures,
increases with the assignment error, see
Fig.~\ref{fig:thermal_pop_vs_sigma}(b).
Crucially, we can reduce these errors by increasing the number of
shots, see Fig.~\ref{fig:thermal_pop_statistics}.
Here, the distribution of measurements is again centered around the
ideal value.
Importantly, increasing the shots reduces the variance of the distribution.

These results are expected given assignment errors, and potentially
other imperfections, affect both $a_\pi$ and $a_{\cancel{\pi}}$.
Therefore, as $\alpha$ is estimated with a ratio where both
numerator and denominator depend linearly on the fitted amplitudes
we can expect the measurement to be first-order insensitive to
assignment errors.

\section{Error propagation\label{appendix:error_propagation}}

Here, we give details on the computation of the error bars shown in
Fig.~\ref{fig:experiment}.
The uncertainty is calculated via standard error propagation.
For example, the propagated standard deviation on $\langle
  O\rangle_\text{mit}$ is
\begin{align}\notag
  \sigma_\text{mit}^2= &
  \left(\frac{\partial \langle O\rangle_{\text{mit}}}{\partial
    \langle O\rangle_{\text{raw}}}\sigma_{\text{raw}}\right)^2 +
  \left(\frac{\partial \langle O\rangle_{\text{mit}}}{\partial
    \langle Z^\star\rangle}\sigma_{ \langle Z^\star\rangle}\right)^2
  \\ +& \sum_{i=1}^{n-1}\left(\frac{\partial \langle
    O\rangle_{\text{mit}}}{\partial
    f_{\text{sp},i}}\sigma_{\text{sp},i}\right)^2.
\end{align}
Here, $\sigma_{ \langle Z^\star\rangle}$ denotes the standard
deviation on the TREX denominator.
The standard deviations for observables $O$, i.e.,
$\sigma_\text{raw}$ and $\sigma_{ \langle Z^\star\rangle}$, are
computed from the samples based on $\langle O^2\rangle-\langle O\rangle^2$.
The standard deviation $\sigma_{\text{sp},i}$ from the RabiEF
experiment on qubit $i$ is based on the uncertainties stemming from
the underlying curve fits.

\section{CNOT noise\label{appendix:cnot_noise}}

The shaded area in Fig.~\ref{fig:experiment} represents the range
of possible values for the observable when only CNOT noise is
present, i.e., the raw observable after removing SPAM errors and
assuming no out-of-model errors.
To obtain this shaded area, first, we twirl the two-qubit gates and
learn a sparse Pauli-Lindblad noise model for the two layers $L_1$
and $L_2$ in Fig.~\ref{fig:experiment}(b) following
Ref.~\cite{van2023probabilistic}.
Next, we use the resulting noise model to compute the noisy
observable via a classical Clifford simulation which contains only
the noise for the CNOT gates.
Crucially, the error rates of conjugate Pauli pairs, i.e., $P$ and
$U P U^\dagger$ (where $P$ is a Pauli and $U$ the unitary of the
CNOT layer) can only be inferred as a product by this noise learning protocol.
However, the simulation requires specifying individual rates which
introduces degrees of freedom.
Initially, we split the product symmetrically, as supported by
recent theoretical work up to leading order~\cite{malekakhlagh2025efficient}.
Finally, we solve two optimization problems that adjust the splits
of conjugate Pauli pairs to either maximize or minimize the
measured observable under the constraint of all Pauli fidelities
remaining physical, i.e., $\leq 1$.
This results in the shaded area in Fig.~\ref{fig:experiment}(e).
As these optimization problems are non-convex, we don't know
whether the solutions we find are the global optima.
This may also explain why the data on four qubits in
Fig.~\ref{fig:experiment}(c) lies outside of the gray area.
We have opted for this procedure out of simplicity as it suffices
to highlight the importance of separating state-preparation and
measurement noise.
The resulting spread in possible observable values is large.
With additional noise learning circuits, the noise model of the
CNOT gate layers could be further constrained by implementing
interleaved cycle benchmarking~\cite{chen2023learnability},
multi-layer cycle benchmarking~\cite{calzona2024multilayer}, or a
self-consistent gate set learning scheme~\cite{chen2024efficient}.

\section{Pauli transfer matrices\label{appendix:ptm}}

For completeness, we now provide a definition of Pauli Transfer
Matrices (PTM)~\cite{ptm}.
They provide a useful representation of quantum channels.
The elements of the PTM of an $n$-qubit quantum channel $\Lambda$ are
\begin{equation}
  \left(\Gamma_{\Lambda}\right)_{ij}=\frac{1}{2^n}\Tr[P_i \Lambda(P_j)],
\end{equation}
where $[P_0, \ldots, P_{4^n -1}]$ denotes the $n$-qubit Pauli basis
arranged in lexicographic order.
The PTM of a composite map $H\circ G$ is the matrix product of the
PTMs of the individual maps $\Gamma_H\Gamma_G$.
Therefore, the convention that noise $\Lambda_G$
occurs before a gate $G$ translates to PTMs as $\Gamma_G\Gamma_{\Lambda_G}$.

Next, the PTM of a measurement is a matrix whose only nonzero
entries are $1$’s in the top-left and bottom-right corners, with
all other entries equal to $0$.
As a result, it annihilates any prior $X$ or $Y$ components of an operation.
Consequently, our PTMs are defined as above, but with the basis
restricted to the set $\{I,Z\}^{\otimes n}$ arranged in lexicographic order.
As an example, consider the one-qubit channel
$\Lambda(\rho)=(1-p)\rho + pX\rho X$.
Its PTM in the full basis $\{I,X,Y,Z\}$ is
\begin{equation}
  \Gamma_{\Lambda} =
  \begin{pmatrix}
    1 & 0 & 0    & 0    \\
    0 & 1 & 0    & 0    \\
    0 & 0 & 1-2p & 0    \\
    0 & 0 & 0    & 1-2p \\
  \end{pmatrix}.
\end{equation}
Since we consider only the $\{I,Z\}$ part of the basis we express this PTM as
\begin{equation}
  \Gamma_{\Lambda} =
  \begin{pmatrix}
    1 & 0    \\
    0 & 1-2p \\
  \end{pmatrix}.
\end{equation}
Finally, an important property of PTMs combined with twirling is
that twirling transforms any noise channel into a Pauli noise
channel, resulting in a diagonal PTM.

\section{PTM of final measurements\label{appendix:final_measurement}}

\begin{figure}
  \centering
  \includegraphics[width=0.8\columnwidth, clip, trim=0 420 450
    0]{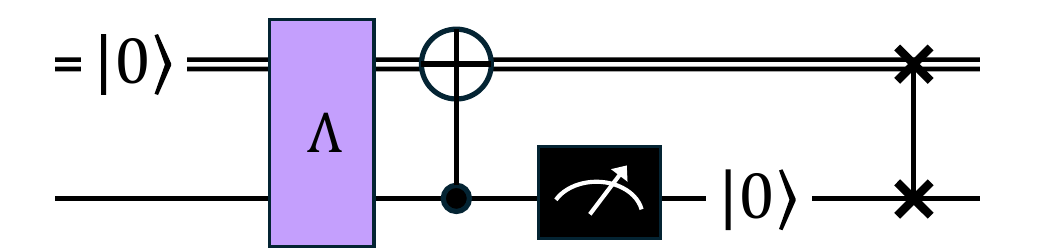}
  \caption{In the CNOT picture, the final measurement corresponds
    to measuring the control qubit. The outcome is written back onto
    the qubit (control) wire, which thereafter functions as a classical line.}
  \label{fig:final_meas}
\end{figure}

In a final measurement --- i.e., measuring the control qubit in the
CNOT picture --- the outcome is written back onto the same wire,
which then acts as a classical bit.
To derive the corresponding PTM, we temporarily introduce an
ancillary classical line and decompose the measurement into the
following sequence, also shown in Fig.~\ref{fig:final_meas}.
(i)~Apply an ideal reset to the ancillary classical line to
initialize it in the $0$ state.
(ii) Apply the noisy measurement in the CNOT picture by
transferring the control-qubit outcome onto the ancillary line.
(iii) Reset the qubit wire, which from this point on acts as a classical line.
(iv) Swap the value from the ancillary classical line back onto the
(now classical) qubit wire.
(v) Remove the ancillary classical line.
We compute the PTMs of each component in this circuit and multiply
them to obtain the PTM for the final noisy measurement.
The PTM of a reset operator $\mathcal{R}(\rho)=E_0 \rho E_0^\dagger
  + E_1 \rho E_1 ^\dagger$, where $E_0=\ket{0}\!\!\bra{0}$ and
$E_1=\ket{0}\!\!\bra{1}$, is
\begin{equation}\label{eqn:reset_ptm}
  \Gamma_{\ket{0}} =
  \begin{pmatrix}
    1 & 0 \\
    1 & 0 \\
  \end{pmatrix}.
\end{equation}
The PTM of the SWAP is
\begin{equation}\label{eqn:swap_ptm}
  \Gamma_{\text{SWAP}} =
  \begin{pmatrix}
    1 & 0 & 0 & 0 \\
    0 & 0 & 1 & 0 \\
    0 & 1 & 0 & 0 \\
    0 & 0 & 0 & 1 \\
  \end{pmatrix}.
\end{equation}
The PTM of a final measurement can then be calculated as
$\Gamma_{\text{fm}} = \Gamma_{\text{SWAP}} \left(\Gamma_{\ket{0}}
  \otimes I \right) \Gamma_\text{CX}\Gamma_\Lambda \left(I\otimes
  \Gamma_{\ket{0}}\right)$ where $\Gamma_\text{CX}\Gamma_\Lambda$ is
defined in \cref{eqn:meas_ptm}, and $\Gamma_{\ket{0}}$ and
$\Gamma_\text{SWAP}$ are defined in \cref{eqn:swap_ptm} and
\cref{eqn:reset_ptm}, respectively.
The result is
\begin{align}
  \Gamma_{\text{fm}} =
  \begin{pmatrix}
    1 & 0       \\
    0 & f_a f_s \\
  \end{pmatrix} \otimes
  \begin{pmatrix}
    1 & 0 \\
    1 & 0 \\
  \end{pmatrix}.\label{eq:final_meas_ptm}
\end{align}
Here, $\fa$ and $\fs$ are, as before, the fidelity of the
assignment and state errors, respectively.
Thus, after discarding the ancillary classical line, the PTM of a
final noisy measurement is $\text{diag}(1, \fa\fs)$.
The absence of $\fc$ in this PTM is understood as follows.
The correlated error simultaneously flips the qubit and the classical bit.
For example, a qubit, initially in state $\ket{0}$, is measured as
$1$, and the qubit ends up in $\ket{1}$.
However, in final measurements we are only concerned with the
correctness of the reported outcome.
Thus, such an event is effectively indistinguishable from a simple
bit-flip error and can be interpreted either as a readout
assignment error or a state-preparation error.

\section{Setup of simulations}\label{appendix:simulation}

Our simulations fall into two categories, qubit and qutrit simulations.
As the RabiEF experiment uses gates that interact with the
$\ket{2}$ state it is the only experiment simulated with qutrits.
Appendix~\ref{appendix:qutrit_sim} describes our qutrit simulator
and its application to the RabiEF experiment.
Appendix~\ref{app:assignment_matrix} translates the readout
assignment matrix $R$ to the measurement noise fidelities.
Appendix~\ref{appendix:qubit_sim} discusses the setup of qubit
simulations for the results shown in \cref{sec:sim_cb} and
\cref{sec:sim_mitig}.
Appendix~\ref{appendix:teleportation} gives details on the
simulation of the teleportation circuit, including PEC.

\subsection{Qutrit simulations for RabiEF}\label{appendix:qutrit_sim}

To simulate the RabiEF experiments while taking into account the
$\ket{2}$-state population, we implemented a qutrit simulator with
Qiskit~\cite{qiskit2024}.
All qubit instructions are embedded as ideal gates in $SU(3)$, and
thus do not interact with the $\ket{2}$ state.
Only the $R_{12}(\theta)$ and $X_{12}$ gates, noise channels, and
measurements interact with the $\ket{2}$-state.
The circuits are converted into a series of operators, i.e.,
\texttt{SuperOp} class instances with qutrit dimensions.
For example, a single qutrit error channel is a $9\times{}9$
superoperator matrix.
A state $\rho$ is evolved by a circuit $c$ by repeatedly calling
the \texttt{DensityMatrix.evolve(op, qargs)} Qiskit method, where
the \texttt{op} argument is the operator for each gate and
\texttt{qargs} is a list of qutrit indices identifying the support
of the operator.
A RabiEF experiment consists of three circuits: the two
$\cancel{\pi}$ and $\pi$, circuits $c_{\cancel{\pi}}$ and $c_\pi$,
respectively, and a repetition delay circuit $c_\text{rep}$
consisting of three active resets and an optional delay for slow
resets. The delay is not simulated when fast resets are used.
The simulator evolves an initial state
$\rho=(\ket{0}\!\bra{0})^{\otimes{}n}$ by simulating the circuits
in the following loop $N$ samples many times: (i) apply
$c_\text{rep}$, (ii) apply $c_{\cancel{\pi}}$, (iii) apply
$c_\text{rep}$, and (iv) apply $c_\pi$.
One loop corresponds to a single sample per $\cancel{\pi}$ and $\pi$ circuit.
Both RabiEF circuits $c_{\cancel{\pi}}$ and $c_\pi$ produce
measurement outcomes which are stored in two length-$N$ arrays.
The repetition delay also produces measurement outcomes, but they
are only used to control the feedforward instructions in the active
resets and are not stored long-term.
The initial states for each circuit, $\rho_{\text{in},k}$, over all
$k$ samples are saved to compute the \textit{true}
state-preparation probability $\psp$, shown in
Tab.~\ref{tab:qutrit_sim_fits}, as
\begin{equation}
  p_\text{sp} \coloneqq
  \underset{\substack{s=\cancel{\pi},\pi\\ k=0,\ldots{},N-1}}{\operatorname{avg}}\Tr\left[\ket{1}\!\bra{1}\rho_{\text{in},k}^{(s)}\right].
\end{equation}

We simulate thermal relaxation during slow resets, measurements,
and idle times in the repetition delay.
Noise channels for measurements are applied before the ideal measurement.
We numerically compute the thermal relaxation channel with the
Lindblad master equation and six Lindblad operators defining
relaxation, heating, and dephasing.
Each Lindblad operator $L_i$ has a corresponding rate $\gamma_i$,
controlling the strength of the operator.
The operators and rates are
\begin{equation}\label{eq:lindblad_ops}
  \begin{aligned}
    L_1 & = \ket{0}\!\bra{1},                    & L_2 & = \ket{1}\!\bra{0}, \\
    L_3 & = \ket{1}\!\bra{2},                    & L_4 & = \ket{2}\!\bra{1}, \\
    L_5 & = \ket{0}\!\bra{0} - \ket{1}\!\bra{1}, & L_6 & =
    \ket{1}\!\bra{1} - \ket{2}\!\bra{2},
  \end{aligned}
\end{equation}
\begin{equation}\label{eq:lindblad_rates}
  \begin{aligned}
    \gamma_1 & = \frac{1}{T_1},
             & \gamma_2              & = \gamma_1 \exp({-\hbar\omega_{01}/k_BT_\text{eff}}) \\
    \gamma_3 & = \frac{1}{T_1^{12}},
             & \gamma_4              & = \gamma_3
    \exp({-\hbar(\omega_{01}+\Delta)/k_BT_\text{eff}})                                      \\
    \gamma_5 & = \frac{1}{T_2},
             & \gamma_6              & = \frac{1}{T_2^{12}}.
  \end{aligned}
\end{equation}

The solution to the Lindblad master equation
$\Lambda_\text{th}(\rho,t)$ for a given evolution time $t$ is
computed with Qiskit Dynamics~\cite{qiskitdynamics2023} and
included as a \texttt{SuperOp} instruction in the circuits.
The rates $\gamma_1$, $\gamma_3$, $\gamma_5$, and $\gamma_6$ are
chosen from typical superconducting qubit decoherence
times~\cite{dane2025}, with their definitions given in
\cref{eq:lindblad_rates} and their values in Tab.~\ref{tab:qutrit_sim_params}.
The heating rates $\gamma_2$ and $\gamma_4$ are calculated using
the energy differences between $\ket{0}$, $\ket{1}$, and $\ket{2}$,
and an assumed effective qutrit temperature
$T_\text{eff}=60~\mathrm{mK}$ following the Boltzmann distribution.
We further assume that the qutrits decay sequentially, i.e.,
$\ket{2}$-$\ket{0}$ transitions are naturally suppressed by the
quantum hardware~\cite{fischer_ancilla-free_2022}.
We include measurement-induced leakage with an additional error
channel prior to the ideal measurements, after the
measurement-associated thermal relaxation.
This is modelled as a projection to the $\ket{2}\!\bra{2}$ state
with probability $p_\text{leak}=0.2~\%$.
The error channel on a qubit undergoing measurement, with duration
$t_\text{meas}=1244~\mathrm{ns}$, is thus
\begin{equation}
  \Lambda_\text{meas}(\rho)=(1-p_\text{leak})\Lambda_\text{th}(\rho,t_\text{meas})
  + p_\text{leak}\ket{2}\!\!\bra{2}.
\end{equation}
Following this noise channel, our simulator applies an ideal
projective measurement and then the readout assignment matrix $R$,
see Appendix~\ref{app:assignment_matrix} and
\cref{eq:readout_assignment_error}.
The simulator then records the noisy measurement outcome as the
integer $o=0,1,2$ and uses the post-measurement state as the input
to the next circuit.
Qutrit discrimination is only used for measurements in the
repetition delay, as RabiEF requires binary classification.

\begin{table}[!ht]
  \centering
  \caption{Parameters used in the simulations.}
  \label{tab:qutrit_sim_params}
  \begin{ruledtabular}
    \begin{tabular}{lrl}
      Symbol                & Value                 & Description                           \\
      \hline
      $T_\text{eff}$        & $60~\mathrm{mK}$      & Qubit Effective Temperature           \\
      $T_1$                 & $200~\mu{}\mathrm{s}$ & Relaxation time in the qubit subspace \\
      $T_2$                 & $100~\mu{}\mathrm{s}$ & Dephasing time in the qubit subspace  \\
      $T_1^{12}$            & $100~\mu{}\mathrm{s}$ & Relaxation time in the
      qutrit subspace                                                                       \\
      $T_2^{12}$            & $50~\mu{}\mathrm{s}$  & Dephasing time in the
      qutrit subspace                                                                       \\
      \hline
      $\omega_{01}$         & $4.9~\mathrm{GHz}$    & Qubit $0$-$1$ frequency               \\
      $\Delta$              & $-0.3~\mathrm{GHz}$   & Qubit anharmonicity                   \\
      \hline
      $t_\text{rep. delay}$ & $250~\mu{}\mathrm{s}$ & Repetition delay time                 \\
      $t_\text{meas}$       & $1244~\mathrm{ns}$    & Measurement time                      \\
      $P_\text{leak}$       & $0.2~\%$              & Prob. to leak during a measurement    \\
    \end{tabular}
  \end{ruledtabular}
\end{table}

\subsection{Assignment error fidelities for
  qutrits}\label{app:assignment_matrix}

Errors in final measurements can be specified as a \textit{readout assignment
  error} matrix where entry $j,k$ is the probability $q_{jk}$ to misclassify
state $\ket{k}$ as $\ket{j}$~\cite{heinsoo2018}.
The $3\times 3$ qutrit readout assignment matrix
\begin{equation}
  R=
  \begin{pmatrix}
    q_{00} & q_{01} & q_{02} \\
    q_{10} & q_{11} & q_{12} \\
    q_{20} & q_{21} & q_{22} \\
  \end{pmatrix}
\end{equation}
used in the RabiEF simulation is based on existing hardware
experiments~\cite{fischer_ancilla-free_2022,chen2023,kanazawa2023}.
We thus neglect the overlap between the $\ket{0}$ and $\ket{2}$
readout signals, i.e., $q_{02}=q_{20}=0$, and assume that the
qutrit misclassification probability for $\ket{1}$-$\ket{2}$ is
approximately an order of magnitude larger than for $\ket{0}$-$\ket{1}$.
Therefore, we set $q_{12}=6~\%$ and $q_{01}=0.9~\%$ resulting in
the readout assignment matrix shown in \cref{eq:readout_assignment_error}.

When qubit reset is used, we require a readout assignment matrix
that always misclassifies $\ket{2}$ as $\ket{1}$, see
\cref{eq:readout_assignment_error}.
The qubit readout assignment errors $p_{jk}$ are related to the
qutrit ones following
\begin{subequations}\label{eq:readout_matrix_mapping}
  \begin{align}
    p_{00} & = q_{00}          \\
    p_{11} & = q_{11} + q_{21} \\
    p_{01} & = q_{01}          \\
    p_{10} & = q_{10}+q_{20}
  \end{align}
\end{subequations}
with $q_{20}=0$ as discussed above.
From the values in \cref{eq:readout_assignment_error} we obtain
qubit assignment matrix probabilities $p_{00}=p_{11}=99.1~\%$ and
$p_{01}=p_{10}=0.9~\%$.

To ensure consistency between the RabiEF simulations in
\cref{sec:sim_rabief} and \cref{sec:sim_cb} we now connect the
qubit readout assignment matrix $p_{jk}$ to the measurement noise
fidelities $\fa$, $\fs$, and $\fc$ of the model in \cref{sec:error_model}.
We apply the PTM of the readout to the different qubit initial states.
In the $\text{qubit}\otimes\text{cbit}$ notation, the initial state
$\ket{00}=(II+IZ+ZI+ZZ)/4$, for example, corresponds to the vector
$\Gamma_{00}=(1, 1, 1, 1)^T$ in the basis $\{II, IZ, ZI, ZZ\}$.
The PTM of the readout $\Gamma_\text{m}$, given in
Eq.~(\ref{eqn:meas_ptm}), transforms this vector according to
\begin{align}\notag
  \Gamma_\text{m}\Gamma_{00}=
    & \frac{1}{4}\left(II+\fa \fs IZ + \fc \fs ZI + \fc\fa ZZ\right)      \\\notag
  = & \frac{1}{4}\left(1+\fa\fs+\fc\fs+\fc\fa\right)\ket{00}\!\!\bra{00}
  \\\notag
  + & \frac{1}{4}\left(1+\fa\fs-\fc\fs-\fc\fa\right)\ket{10}\!\!\bra{10}
  \\\notag
  + & \frac{1}{4}\left(1-\fa\fs+\fc\fs-\fc\fa\right)\ket{01}\!\!\bra{01}
  \\\notag
  + & \frac{1}{4}\left(1-\fa\fs-\fc\fs+\fc\fa\right)\ket{11}\!\!\bra{11}.
\end{align}
Since we consider a final measurement we only care about the state
of the classical bit and thus trace-out the qubit.
Therefore, the measurement acts on the qubit $\ket{0}$ state as
\begin{align}\notag
  \ket{0}\!\!\bra{0}\to\frac{1+\fa\fs}{2}\ket{0}\!\!\bra{0}+\frac{1-\fa\fs}{2}\ket{1}\!\!\bra{1}.
\end{align}
This implies $p_{00}=(1 + \fa\fs)/2$ and $p_{10}=(1-\fa\fs)/2$.
Similarly, for the excited qubit state we consider $\ket{10}$ in
the $\text{qubit}\otimes\text{cbit}$ notation and apply the same reasoning.
In summary, for a noisy measurement the corresponding assignment
error probabilities $p_{jk}$ are
\begin{equation}\label{eq:readout_matrix_fidelities}
  \begin{aligned}
    p_{00} & = \frac{1 + \fa\fs}{2}, & p_{10} & = \frac{1-\fa\fs}{2}  \\
    p_{10} & = \frac{1 - \fa\fs}{2}, & p_{11} & = \frac{1+\fa\fs}{2}.
  \end{aligned}
\end{equation}
This result also shows that the correlated error $\fc$ is
irrelevant when considering final measurements, see also
Appendix~\ref{appendix:final_measurement}.
The values $p_{00}=p_{11}=99.1~\%$ and $p_{01}=p_{10}=0.9~\%$ thus
require $\fa\fs=0.982$.
In our simulations we assume that assignment and state errors are
equally likely such that $
  \fa\fs$ splits equally into $\fa=\fs=0.99096$.
As the readout assignment matrix $R$ does not capture correlated
errors, this analysis cannot determine the value of $\fc$.
However, as the correlated error is of a higher weight than the
state and assignment errors, we assume it has a higher fidelity and
set $\fc=0.995$.

\subsection{Qubit simulation setup for noise learning and
  mitigation}\label{appendix:qubit_sim}

This section covers the setup for simulations in
\cref{sec:sim_cb,sec:sim_mitig,sec:sim_dynamic}.
We used Qiskit~Aer~\cite{qiskitaer2025} to simulate MCB and the
mitigation circuits.
We insert additional dummy gates to engineer the noise model in
\cref{sec:error_model}.
As Qiskit~Aer applies noise before measurements, implementing state
errors requires only an $X$ error on the measurement qubit with
probability $\ps=(1-\fs)/2$.
Qiskit~Aer has native support for readout assignment matrices which
we leverage to implement assignment errors on the classical bit.
The assignment-error PTM in \cref{eq:ptm_assign} is simulated with
the readout assignment matrix $R_\text{a}$
\begin{equation}
  R_\text{a} =
  \begin{pmatrix}
    1-\pa & \pa   \\
    \pa   & 1-\pa
  \end{pmatrix}.
\end{equation}
Qiskit~Aer does not support correlated errors between qubits and
classical bits.
Therefore, we implement $\Lambda_\text{c}$ in
\cref{fig:meas_models} with a dummy delay instruction
post-measurement to which we attach an $X$ error.
This post-measurement error is equivalent to a correlated error
pre-measurement, which can be seen by back propagating the $X$
Pauli through the measurement in the CNOT picture.
Therefore, an $X$ gate post-measurement with probability
$\pc=(1-\fc)/2$ is equivalent to a correlated error pre-measurement
with the same probability.
We set the duration of the delay to $0$ to ensure that no
additional error channels are attached to it.

To simulate a given prepared initial state, we prepend reset
instructions to the MCB and mitigation circuits, and attach an $X$
error to them with probability $\psp$.
This forces Qiskit~Aer to prepare the state
$\rho_\text{sp}=(1-\psp)\ket{0}\!\!\bra{0} + \psp\ket{1}\!\!\bra{1}$.
The measurement outcomes returned by Qiskit~Aer are post-processed
to obtain the expectation values in
\cref{fig:qubit_sim_cb,fig:qubit_sim_mitig}.

\subsection{Error mitigation of the teleportation
  circuit}\label{appendix:teleportation}

The teleportation circuit in \cref{sec:sim_dynamic} is simulated
with density matrices using Qiskit~Aer~\cite{qiskitaer2025}.
State-preparation and measurement noise are mitigated with
probabilistic error
cancellation~\cite{van2023probabilistic,gupta2024probabilistic}.
The simulation is carried out as follows.
(i) We generate $N=128$ teleportation circuit realizations where
the inverse noise channels $\tilde{\Lambda}^{-1}$ in
\cref{fig:teleportation_circuit} are replaced with Paulis sampled
from their quasi-probability distributions.
(ii) The circuits are classically simulated, obtaining density
matrices $\rho_{\text{out},j}$ on the target qubit \texttt{q2}, for
$j\in[0,N-1]$.
(iii) We obtain a mitigated physical state $\rho_\text{mit}$ with
state tomography.
Since PEC is applied to expectation values and not states, this is
done as follows.
(iii.a)~We compute the expectation values
$\langle{}P\rangle{}_{\text{raw},j}=\Tr[P\rho_{\text{out},j}]$, for
$P=X$, $Y$, $Z$, without shot noise.
However, since PEC is very sensitive to shot noise we emulate it by
sampling $100$ times from a Bernoulli distribution with probability
$(1+\langle{}P\rangle{}_{\text{raw},j})/2$ for outcome
\texttt{"0"}, approximating $\langle P\rangle_{\text{raw},j}$ as
$E_{P,j}$ with both shot noise and SPAM errors.
(iii.b)~The mitigated expectation values
$\langle{}P\rangle{}_\text{mit}$ are computed using PEC, i.e.,
\cref{eq:pec}, over all $N$ circuit realizations and $P=X,Y, Z$ Paulis.
(iii.c)~We minimize the Mean-Squared Error (MSE) in
\cref{eq:state_mse} over all expectation values to obtain a
physical state $\rho_\text{mit}$ based on the mitigated expectation values.
(iv) We compute the state fidelity between the mitigated state
$\rho_\text{mit}$ and the ideal one with \cref{eq:state_fidelity}.
Steps (i)-(iv) are repeated for $15$ equidistant rotation angles
$\theta\in[0,2\pi]$ to study different states to teleport.
The mitigated fidelities in \cref{sec:sim_dynamic} display the
interquartile range and the median obtained from $300$ realizations
of the above simulation.
To avoid simulating $300\times 128$ circuits we simulate $1000$ PEC
circuit realizations and compute the raw expectation values
$\langle{}P\rangle_{\text{raw},j}$ on all of them.
Next, we bootstrap these results into $300$ datasets by sampling
$N=128$ circuit realizations from the $1000$ circuit executions.

The fifteen unmitigated fidelities, shown as purple dots in
\cref{fig:teleportation_circuit}, are obtained using the same
procedure with $N=1$ circuit and no PEC sampling.
Furthermore, since there is only one circuit, and thus only one
density matrix $\rho_{\text{out},0}$, the MSE minimization is not necessary.

\section{Extension of the Noise Model to Correlated Qubits Undergoing Measurement}\label{sec:app_correlated}
\begin{figure*}[!th]
  \centering
  \includegraphics[width=0.9\linewidth]{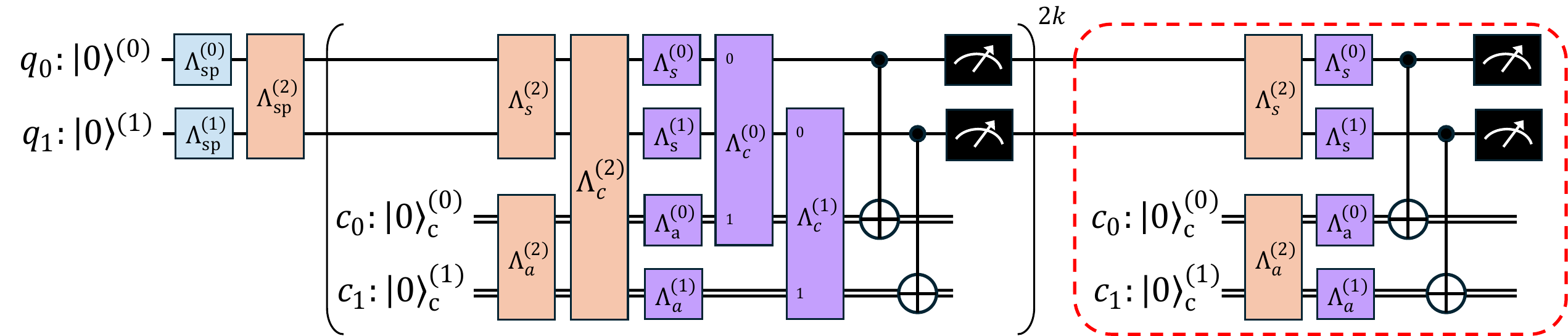}
  \caption{
    \textbf{Measurement noise model for two correlated qubits.}
    Qubits and classical bits are indicated by the single and double lines,
    respectively.
    Noise channels from the original noise model in the main text are shown
    in purple and blue.
    New noise channels are in light orange.
    The support of some channels differ from the wires they cross.
    For these, indices are used to denote the qubits and classical bits on
    which they are applied.
    For example, $\Lambda_c^{(0)}$ is supported by $q_0$ and $c_0$, but not
    $q_1$.
    Superscripts $(0)$ and $(1)$ denote errors that arise from the original
    noise model on qubit-classical bit pairs $(q_0,c_0)$ and $(q_1,c_1)$,
    respectively.
    New errors which model correlations between the qubit-classical bit
    pairs are denoted with superscript $(2)$.
    The red dashed rectangle indicates the final measurement which is
    insensitive to post-measurement qubit errors, equivalent to
    $\Lambda_c^{(j)}$ for $j=0,1,2$.
  }
  \label{fig:meas_model_correlated}
\end{figure*}

Our proposed noise model in \cref{sec:error_model} assumes that noise is
local to the qubit and associated classical bit.
Accounting for correlations between neighboring qubits requires increasing
the support of the Pauli-Lindblad noise model and measurement cycle
benchmarking experiments.
The same noise learning procedure as before can be employed, now with more
MCB decay curves and model fidelities to learn.
This is similar to how gate-noise learning with cycle benchmarking is
extended to include higher-weight noise
generators~\cite{gupta2024probabilistic, van2023probabilistic}.
The chosen support determines which correlations are assumed as
negligible.

As an example, we derive the twirled-noise PTM for two neighboring qubits
undergoing measurement and describe the noise learning procedure, as per our
protocol in \cref{fig:workflow}.
We assume that correlations between qubits arise from measurements, and that
thermal states are uncorrelated.
Given fast resets use measurements, this model accounts for correlations in
prepared states using fast resets.

We model correlations between two subsystems $j=0,1$, each composed of a
qubit-classical bit pair $(q_j, c_j)$.
In addition to the errors possible in our original noise model in
\cref{sec:error_model}, we include correlated errors between the qubits and
correlated errors between the classical bits.
The resulting error model is shown in \cref{fig:meas_model_correlated} for
the measurement cycle benchmarking circuit.
Employing the same twirling as before~\cite{Beale2023, van2023probabilistic,
  gupta2024probabilistic}, all errors are bitflips, i.e., Pauli strings
composed of $I$ and $X$.
We order the subsystems so that qubits $q_0$ and $q_1$ are before classical
bits $c_0$ and $c_1$, i.e., $q_0q_1c_0c_1$.
Therefore, Pauli string $IXIX$ corresponds to the original noise-model's
correlated error between qubit $q_1$ and classical bit $c_1$; and Pauli
string $XXII$ corresponds to a weight-two error between qubits $q_0$ and
$q_1$.
Errors local to subsystems $j=0,1$ are denoted with superscripts $(0)$ and
$(1)$, respectively.
New errors between subsystems are denoted with superscript $(2)$, and are
described by the following channels:
\begin{enumerate}
  \item Weight-two correlated state-errors between qubits:
        \begin{equation}
          \Lambda_\text{s}^{(2)}
          =
          (1-\ps^{(2)})\rho + \ps^{(2)}XXII\rho XXII
        \end{equation}
        where $\fs^{(2)}=1-2\ps^{(2)}$.
        These errors can be caused by cross-talk between qubits, potentially
        induced by measurements~\cite{Hashim2025}.
  \item Weight-four correlated errors between both qubits and both classical
        bits:
        \begin{equation}
          \Lambda_\text{c}^{(2)}
          =
          (1-\pc^{(2)})\rho + \pc^{(2)}XXXX\rho XXXX
        \end{equation}
        where $\fc^{(2)}=1-2\pc^{(2)}$.
        This arises from a weight-two post-measurement error on both qubits
        $\tilde{\Lambda}_\text{c}^{(2)}$, commuted backwards through the
        measurements:
        \begin{equation}
          \tilde{\Lambda}_\text{c}^{(2)}
          =
          (1-\pc^{(2)})\rho + \pc^{(2)}XXII\rho XXII.
        \end{equation}
        This is similar to how the correlated error $\Lambda_c$ in
        \cref{fig:meas_models} and \cref{eq:ptm_correlated} is derived from
        a post-measurement qubit-error.
        Similarly to $\Lambda_\text{s}^{(2)}$, these errors can be caused by
        cross-talk between qubits, only post measurement.
  \item Weight-two correlated assignment-errors between classical bits:
        \begin{equation}
          \Lambda_\text{a}^{(2)}
          =
          (1-\pa^{(2)})\rho + \pa^{(2)}IIXX\rho IIXX
        \end{equation}
        where $\fa^{(2)}=1-2\pa^{(2)}$.
        These errors can occur because of frequency collisions in readout
        resonators.
  \item Weight-two correlated state-preparation errors between qubits:
        \begin{equation}
          \Lambda_\text{sp}^{(2)}
          =
          (1-\psp^{(2)})\rho + \psp^{(2)}XXII\rho XXII
        \end{equation}
        where $\fsp^{(2)}=1-2\psp^{(2)}$.
        As active resets utilize multiple measurements, correlated errors
        can manifest as a result of measurement-induced correlations.
\end{enumerate}

The full PTM $\Gamma_{2k}^{(2q)}$ for this noise model in a MCB circuit is
diagonal and has twelve fidelities to learn: four per qubit ($\fsp^{(j)}$,
$\fs^{(j)}$, $\fa^{(j)}$, and $\fc^{(j)}$ for $j=0,1$), from our original
noise model, and four from the new correlated errors ($\fsp^{(2)}$,
$\fs^{(2)}$, $\fa^{(2)}$, and $\fc^{(2)}$).
The entries $y_{\scriptscriptstyle P}$ of the diagonal, for $P\in\{I,Z\}^4$,
are given in
Eqs.~(\ref{eq:ptm_2q_diagonal_first})-(\ref{eq:ptm_2q_diagonal_last}).
\begin{widetext}
  \begin{subequations}
    \newcommand{\faalpha}{f_{\text{a}}^{(0)}}
    \newcommand{\fabeta}{f_{\text{a}}^{(1)}}
    \newcommand{\fagamma}{f_{\text{a}}^{(2)}}
    \newcommand{\fsalpha}{f_{\text{s}}^{(0)}}
    \newcommand{\fsbeta}{f_{\text{s}}^{(1)}}
    \newcommand{\fsgamma}{f_{\text{s}}^{(2)}}
    \newcommand{\fspalpha}{f_{\text{sp}}^{(0)}}
    \newcommand{\fspbeta}{f_{\text{sp}}^{(1)}}
    \newcommand{\fspgamma}{f_{\text{sp}}^{(2)}}
    \newcommand{\fcalpha}{f_{\text{c}}^{(0)}}
    \newcommand{\fcbeta}{f_{\text{c}}^{(1)}}
    \newcommand{\fcgamma}{f_{\text{c}}^{(2)}}
    \begin{align}
      y_{\scriptscriptstyle IIII}
       & =
      1
      \label{eq:ptm_2q_diagonal_first}
      \\
      y_{\scriptscriptstyle IIIZ}
       & =
      \left(\fabeta \fagamma\right)^{2 k} \left(\fcbeta \fcgamma \fsbeta \fsgamma\right)^{k}
      \\
      y_{\scriptscriptstyle IIZI}
       & =
      \left(\faalpha \fagamma\right)^{2 k} \left(\fcalpha \fcgamma \fsalpha \fsgamma\right)^{k}
      \\
      y_{\scriptscriptstyle IIZZ}
       & =
      \left(\faalpha \fabeta\right)^{2 k} \left(\fcalpha \fcbeta \fsalpha \fsbeta\right)^{k}
      \\
      y_{\scriptscriptstyle IZII}
       & =
      \fabeta \fagamma \fspbeta \fspgamma \left(\fsbeta \fsgamma\right)^{2 k + 1} \left(\fcbeta \fcgamma\right)^{2 k}
      \\
      y_{\scriptscriptstyle IZIZ}
       & =
      \fspbeta \fspgamma \left(\fabeta \fagamma\right)^{2 k + 1} \left(\fcbeta \fcgamma\right)^{k} \left(\fsbeta \fsgamma\right)^{k + 1}
      \\
      y_{\scriptscriptstyle IZZI}
       & =
      \fabeta \fspbeta \fspgamma \left(\fagamma \fsbeta\right)^{2 k + 1} \left(\faalpha \fcbeta\right)^{2 k} \left( \fcalpha \fcgamma \fsalpha\right)^{k} \left(\fsgamma\right)^{k + 1}
      \\
      y_{\scriptscriptstyle IZZZ}
       & =
      \fagamma \fspbeta \fspgamma \left(\fabeta \fsgamma\right)^{2 k + 1} \left(\faalpha \fcgamma\right)^{2 k} \left(\fcalpha \fcbeta \fsalpha\right)^{k} \left(\fsbeta\right)^{k + 1}
      \\
      y_{\scriptscriptstyle ZIII}
       & =
      \faalpha \fagamma \fspalpha \fspgamma \left(\fsalpha \fsgamma\right)^{2 k + 1} \left(\fcalpha \fcgamma\right)^{2 k}
      \\
      y_{\scriptscriptstyle ZIIZ}
       & =
      \faalpha \fspalpha \fspgamma \left(\fagamma \fsalpha\right)^{2 k + 1} \left(\fabeta \fcalpha\right)^{2 k} \left( \fcbeta \fcgamma \fsbeta\right)^{k} \left(\fsgamma\right)^{k + 1}
      \\
      y_{\scriptscriptstyle ZIZI}
       & =
      \fspalpha \fspgamma \left(\faalpha \fagamma\right)^{2 k + 1} \left(\fcalpha \fcgamma\right)^{k} \left(\fsalpha \fsgamma\right)^{k + 1}
      \\
      y_{\scriptscriptstyle ZIZZ}
       & =
      \fagamma \fspalpha \fspgamma \left(\faalpha \fsgamma\right)^{2 k + 1} \left(\fabeta \fcgamma\right)^{2 k}\left(\fcalpha \fcbeta \fsbeta\right)^{k} \left(\fsalpha\right)^{k + 1}
      \\
      y_{\scriptscriptstyle ZZII}
       & =
      \faalpha \fabeta \fspalpha \fspbeta \left(\fsalpha \fsbeta\right)^{2 k + 1} \left(\fcalpha \fcbeta\right)^{2 k}
      \\
      y_{\scriptscriptstyle ZZIZ}
       & =
      \faalpha \fspalpha \fspbeta \left(\fabeta \fsalpha\right)^{2 k + 1} \left(\fagamma \fcalpha\right)^{2 k} \left(\fcbeta \fcgamma \fsgamma\right)^{k} \left(\fsbeta\right)^{k + 1}
      \\
      y_{\scriptscriptstyle ZZZI}
       & =
      \fabeta \fspalpha \fspbeta \left(\faalpha \fsbeta\right)^{2 k + 1} \left(\fagamma \fcbeta\right)^{2 k} \left( \fcalpha \fcgamma \fsgamma\right)^{k} \left(\fsalpha\right)^{k + 1}
      \\
      y_{\scriptscriptstyle ZZZZ}
       & =
      \fspalpha \fspbeta \left(\faalpha \fabeta\right)^{2 k + 1} \left(\fcalpha \fcbeta\right)^{k} \left(\fsalpha \fsbeta\right)^{k + 1}
      \label{eq:ptm_2q_diagonal_last}
    \end{align}
  \end{subequations}
\end{widetext}

To identify any non-learnable degrees of freedom in $\Gamma_{2k}^{(2q)}$, we
compute the observability matrix with respect to the logarithm of
$y_{\scriptscriptstyle P}$ and all fidelities~\cite{burns2001}.
The observability matrix is rank $9$, which, with twelve unknown fidelities,
indicates that a single MCB experiment is insufficient to learn the full noise
model.
Learning $\fsp^{(0)}$ and $\fsp^{(1)}$ with RabiEF on thermal states,
assuming $\fsp^{(2)}=1$, reduces the number of unknown fidelities to nine.
This indicates that one RabiEF experiment and one MCB experiment is
sufficient to learn $\Gamma_{2k}^{(2q)}$ with slow resets.
To illustrate the strength of our full noise learning protocol, we now show
how all twelve fidelities can be learned when fast resets introduce
non-negligible correlations during state-preparation.

First, (i) learn $\fsp^{(0),\text{slow}}$ and $\fsp^{(1),\text{slow}}$ with
separate RabiEF experiments, using slow resets.
As we assume no correlations during state-preparation with slow resets, we
set $\fsp^{(2),\text{slow}}=1$.
This reduces the number of unknown fidelities to nine.
(ii) Run a MCB experiment with slow resets to learn the remaining nine
unknown fidelities $\fs^{(j)}$, $\fa^{(j)}$, and $\fc^{(j)}$ for $j=0,1,2$.
(iii) The last step, as per our protocol, is a MCB experiment with fast
resets.
Assuming all fidelities remain the same, other than those for
state-preparation, we need only learn $\fsp^{(0),\text{fast}}$,
$\fsp^{(1),\text{fast}}$, and $\fsp^{(2),\text{fast}}$.
The PTM $\Gamma_{2k}^{(2q)}$ for fast resets has a rank $3$ observability
matrix, indicating that the last MCB experiment allows us to complete the
full noise model on both subsystems.
This shows that our full noise learning protocol with RabiEF can also fix
the gauge in the noise model of two correlated qubits undergoing
measurement.

\section{Extension for high-fidelity resets and high-rate
  learning}\label{sec:high_rate}

As coherence times and state-preparation fidelities improve, it is
expected that $\psp$ should decrease.
This poses a problem for RabiEF as the error in the estimate of
$\psp$ is additive, owing to sampling error on the $\cancel{\pi}$
and $\pi$ RabiEF signals.
To reduce the impact of this error on the learned measurement noise
fidelities, we can
amplify $\psp$ by inserting random local $X$ gates on the initial
state with probability
$p_\text{amp}$.
The effective state-preparation error probability
$\tilde{p}_\text{sp}$ is then
\begin{equation}
  \tilde{p}_\text{sp}=p_\text{amp} +
  \psp(1-2p_\text{amp})=p_\text{amp}(1-2\psp) + \psp
\end{equation}
which is greater than $\psp$ for $\psp<1/2$.
As a result, one can utilize our MCM noise learning protocol,
specifically the purple path in \cref{fig:workflow}, to
(i) learn $\tilde{p}_\text{sp}$ with a RabiEF experiment, (ii)
learn the measurement fidelities with the amplified
state-preparation error, and then finally, (iii) learn the
high-fidelity reset probability $\psp$ with a second MCB experiment.
For the RabiEF experiment, the random $X$ gate is equivalent to a relabelling
of $\cancel{\pi}$ and $\pi$ shots with probability $p_\text{amp}$,
which can be done
in post-processing.
For the measurement cycle benchmarking and mitigation experiments,
the circuits are
already twirled; therefore, the random $X$ gates can be included in the normal
twirling samples by applying a bias to the $X$ component in the
initial state twirling.


\bibliography{references}

\begin{thebibliography}{71}%
\makeatletter
\providecommand \@ifxundefined [1]{%
 \@ifx{#1\undefined}
}%
\providecommand \@ifnum [1]{%
 \ifnum #1\expandafter \@firstoftwo
 \else \expandafter \@secondoftwo
 \fi
}%
\providecommand \@ifx [1]{%
 \ifx #1\expandafter \@firstoftwo
 \else \expandafter \@secondoftwo
 \fi
}%
\providecommand \natexlab [1]{#1}%
\providecommand \enquote  [1]{``#1''}%
\providecommand \bibnamefont  [1]{#1}%
\providecommand \bibfnamefont [1]{#1}%
\providecommand \citenamefont [1]{#1}%
\providecommand \href@noop [0]{\@secondoftwo}%
\providecommand \href [0]{\begingroup \@sanitize@url \@href}%
\providecommand \@href[1]{\@@startlink{#1}\@@href}%
\providecommand \@@href[1]{\endgroup#1\@@endlink}%
\providecommand \@sanitize@url [0]{\catcode `\\12\catcode `\$12\catcode
  `\&12\catcode `\#12\catcode `\^12\catcode `\_12\catcode `\%12\relax}%
\providecommand \@@startlink[1]{}%
\providecommand \@@endlink[0]{}%
\providecommand \url  [0]{\begingroup\@sanitize@url \@url }%
\providecommand \@url [1]{\endgroup\@href {#1}{\urlprefix }}%
\providecommand \urlprefix  [0]{URL }%
\providecommand \Eprint [0]{\href }%
\providecommand \doibase [0]{https://doi.org/}%
\providecommand \selectlanguage [0]{\@gobble}%
\providecommand \bibinfo  [0]{\@secondoftwo}%
\providecommand \bibfield  [0]{\@secondoftwo}%
\providecommand \translation [1]{[#1]}%
\providecommand \BibitemOpen [0]{}%
\providecommand \bibitemStop [0]{}%
\providecommand \bibitemNoStop [0]{.\EOS\space}%
\providecommand \EOS [0]{\spacefactor3000\relax}%
\providecommand \BibitemShut  [1]{\csname bibitem#1\endcsname}%
\let\auto@bib@innerbib\@empty
\bibitem [{\citenamefont {Kim}\ \emph {et~al.}(2023)\citenamefont {Kim},
  \citenamefont {Eddins}, \citenamefont {Anand}, \citenamefont {Wei},
  \citenamefont {{van den Berg}}, \citenamefont {Rosenblatt}, \citenamefont
  {Nayfeh}, \citenamefont {Wu}, \citenamefont {Zaletel}, \citenamefont
  {Temme},\ and\ \citenamefont {Kandala}}]{kim2023evidence}%
  \BibitemOpen
  \bibfield  {author} {\bibinfo {author} {\bibfnamefont {Y.}~\bibnamefont
  {Kim}}, \bibinfo {author} {\bibfnamefont {A.}~\bibnamefont {Eddins}},
  \bibinfo {author} {\bibfnamefont {S.}~\bibnamefont {Anand}}, \bibinfo
  {author} {\bibfnamefont {K.~X.}\ \bibnamefont {Wei}}, \bibinfo {author}
  {\bibfnamefont {E.}~\bibnamefont {{van den Berg}}}, \bibinfo {author}
  {\bibfnamefont {S.}~\bibnamefont {Rosenblatt}}, \bibinfo {author}
  {\bibfnamefont {H.}~\bibnamefont {Nayfeh}}, \bibinfo {author} {\bibfnamefont
  {Y.}~\bibnamefont {Wu}}, \bibinfo {author} {\bibfnamefont {M.}~\bibnamefont
  {Zaletel}}, \bibinfo {author} {\bibfnamefont {K.}~\bibnamefont {Temme}},\
  and\ \bibinfo {author} {\bibfnamefont {A.}~\bibnamefont {Kandala}},\
  }\bibfield  {title} {\bibinfo {title} {Evidence for the utility of quantum
  computing before fault tolerance},\ }\href
  {https://doi.org/10.1038/s41586-023-06096-3} {\bibfield  {journal} {\bibinfo
  {journal} {Nature}\ }\textbf {\bibinfo {volume} {618}},\ \bibinfo {pages}
  {500} (\bibinfo {year} {2023})}\BibitemShut {NoStop}%
\bibitem [{\citenamefont {Fischer}\ \emph {et~al.}(2026)\citenamefont
  {Fischer}, \citenamefont {Leahy}, \citenamefont {Eddins}, \citenamefont
  {Keenan}, \citenamefont {Ferracin}, \citenamefont {Rossi}, \citenamefont
  {Kim}, \citenamefont {He}, \citenamefont {Pietracaprina}, \citenamefont
  {Sokolov}, \citenamefont {Dooley}, \citenamefont {Zimbor{\'a}s},
  \citenamefont {Tacchino}, \citenamefont {Maniscalco}, \citenamefont {Goold},
  \citenamefont {{Garc{\'i}a-P{\'e}rez}}, \citenamefont {Tavernelli},
  \citenamefont {Kandala},\ and\ \citenamefont
  {Filippov}}]{fischer2024dynamical}%
  \BibitemOpen
  \bibfield  {author} {\bibinfo {author} {\bibfnamefont {L.~E.}\ \bibnamefont
  {Fischer}}, \bibinfo {author} {\bibfnamefont {M.}~\bibnamefont {Leahy}},
  \bibinfo {author} {\bibfnamefont {A.}~\bibnamefont {Eddins}}, \bibinfo
  {author} {\bibfnamefont {N.}~\bibnamefont {Keenan}}, \bibinfo {author}
  {\bibfnamefont {D.}~\bibnamefont {Ferracin}}, \bibinfo {author}
  {\bibfnamefont {M.~A.~C.}\ \bibnamefont {Rossi}}, \bibinfo {author}
  {\bibfnamefont {Y.}~\bibnamefont {Kim}}, \bibinfo {author} {\bibfnamefont
  {A.}~\bibnamefont {He}}, \bibinfo {author} {\bibfnamefont {F.}~\bibnamefont
  {Pietracaprina}}, \bibinfo {author} {\bibfnamefont {B.}~\bibnamefont
  {Sokolov}}, \bibinfo {author} {\bibfnamefont {S.}~\bibnamefont {Dooley}},
  \bibinfo {author} {\bibfnamefont {Z.}~\bibnamefont {Zimbor{\'a}s}}, \bibinfo
  {author} {\bibfnamefont {F.}~\bibnamefont {Tacchino}}, \bibinfo {author}
  {\bibfnamefont {S.}~\bibnamefont {Maniscalco}}, \bibinfo {author}
  {\bibfnamefont {J.}~\bibnamefont {Goold}}, \bibinfo {author} {\bibfnamefont
  {G.}~\bibnamefont {{Garc{\'i}a-P{\'e}rez}}}, \bibinfo {author} {\bibfnamefont
  {I.}~\bibnamefont {Tavernelli}}, \bibinfo {author} {\bibfnamefont
  {A.}~\bibnamefont {Kandala}},\ and\ \bibinfo {author} {\bibfnamefont {S.~N.}\
  \bibnamefont {Filippov}},\ }\bibfield  {title} {\bibinfo {title} {Dynamical
  simulations of many-body quantum chaos on a quantum computer},\ }\href
  {https://doi.org/10.1038/s41567-025-03144-9} {\bibfield  {journal} {\bibinfo
  {journal} {Nature Physics}\ ,\ \bibinfo {pages} {1}} (\bibinfo {year}
  {2026})}\BibitemShut {NoStop}%
\bibitem [{\citenamefont {Fuller}\ \emph {et~al.}(2025)\citenamefont {Fuller},
  \citenamefont {Tran}, \citenamefont {Lykov}, \citenamefont {Johnson},
  \citenamefont {Rossmannek}, \citenamefont {Wei}, \citenamefont {He},
  \citenamefont {Kim}, \citenamefont {Vu}, \citenamefont {Sharma},
  \citenamefont {Alexeev}, \citenamefont {Kandala},\ and\ \citenamefont
  {Mezzacapo}}]{fuller2025improved}%
  \BibitemOpen
  \bibfield  {author} {\bibinfo {author} {\bibfnamefont {B.}~\bibnamefont
  {Fuller}}, \bibinfo {author} {\bibfnamefont {M.~C.}\ \bibnamefont {Tran}},
  \bibinfo {author} {\bibfnamefont {D.}~\bibnamefont {Lykov}}, \bibinfo
  {author} {\bibfnamefont {C.}~\bibnamefont {Johnson}}, \bibinfo {author}
  {\bibfnamefont {M.}~\bibnamefont {Rossmannek}}, \bibinfo {author}
  {\bibfnamefont {K.~X.}\ \bibnamefont {Wei}}, \bibinfo {author} {\bibfnamefont
  {A.}~\bibnamefont {He}}, \bibinfo {author} {\bibfnamefont {Y.}~\bibnamefont
  {Kim}}, \bibinfo {author} {\bibfnamefont {D.}~\bibnamefont {Vu}}, \bibinfo
  {author} {\bibfnamefont {K.}~\bibnamefont {Sharma}}, \bibinfo {author}
  {\bibfnamefont {Y.}~\bibnamefont {Alexeev}}, \bibinfo {author} {\bibfnamefont
  {A.}~\bibnamefont {Kandala}},\ and\ \bibinfo {author} {\bibfnamefont
  {A.}~\bibnamefont {Mezzacapo}},\ }\href
  {https://doi.org/10.48550/arXiv.2502.01897} {\bibinfo {title} {Improved
  quantum computation using operator backpropagation}} (\bibinfo {year}
  {2025}),\ \Eprint {https://arxiv.org/abs/2502.01897} {arXiv:2502.01897
  [quant-ph]} \BibitemShut {NoStop}%
\bibitem [{\citenamefont {Zimbor{\'a}s}\ \emph {et~al.}(2025)\citenamefont
  {Zimbor{\'a}s} \emph {et~al.}}]{zimboras2025myths}%
  \BibitemOpen
  \bibfield  {author} {\bibinfo {author} {\bibfnamefont {Z.}~\bibnamefont
  {Zimbor{\'a}s}} \emph {et~al.},\ }\href
  {https://doi.org/10.48550/arXiv.2501.05694} {\bibinfo {title} {Myths around
  quantum computation before full fault tolerance: {{What}} no-go theorems rule
  out and what they don't}} (\bibinfo {year} {2025}),\ \Eprint
  {https://arxiv.org/abs/2501.05694} {arXiv:2501.05694 [quant-ph]} \BibitemShut
  {NoStop}%
\bibitem [{\citenamefont {Piveteau}\ \emph {et~al.}(2021)\citenamefont
  {Piveteau}, \citenamefont {Sutter}, \citenamefont {Bravyi}, \citenamefont
  {Gambetta},\ and\ \citenamefont {Temme}}]{Piveteau2021}%
  \BibitemOpen
  \bibfield  {author} {\bibinfo {author} {\bibfnamefont {C.}~\bibnamefont
  {Piveteau}}, \bibinfo {author} {\bibfnamefont {D.}~\bibnamefont {Sutter}},
  \bibinfo {author} {\bibfnamefont {S.}~\bibnamefont {Bravyi}}, \bibinfo
  {author} {\bibfnamefont {J.~M.}\ \bibnamefont {Gambetta}},\ and\ \bibinfo
  {author} {\bibfnamefont {K.}~\bibnamefont {Temme}},\ }\bibfield  {title}
  {\bibinfo {title} {Error mitigation for universal gates on encoded qubits},\
  }\href {https://doi.org/10.1103/PhysRevLett.127.200505} {\bibfield  {journal}
  {\bibinfo  {journal} {Phys. Rev. Lett.}\ }\textbf {\bibinfo {volume} {127}},\
  \bibinfo {pages} {200505} (\bibinfo {year} {2021})}\BibitemShut {NoStop}%
\bibitem [{\citenamefont {Aharonov}\ \emph {et~al.}(2025)\citenamefont
  {Aharonov}, \citenamefont {Alberton}, \citenamefont {Arad}, \citenamefont
  {Atia}, \citenamefont {Bairey}, \citenamefont {Brakerski}, \citenamefont
  {Cohen}, \citenamefont {Golan}, \citenamefont {Gurwich}, \citenamefont
  {Kenneth}, \citenamefont {Leviatan}, \citenamefont {Lindner}, \citenamefont
  {Melcer}, \citenamefont {Meyer}, \citenamefont {Schul},\ and\ \citenamefont
  {Shutman}}]{aharonov2025importance}%
  \BibitemOpen
  \bibfield  {author} {\bibinfo {author} {\bibfnamefont {D.}~\bibnamefont
  {Aharonov}}, \bibinfo {author} {\bibfnamefont {O.}~\bibnamefont {Alberton}},
  \bibinfo {author} {\bibfnamefont {I.}~\bibnamefont {Arad}}, \bibinfo {author}
  {\bibfnamefont {Y.}~\bibnamefont {Atia}}, \bibinfo {author} {\bibfnamefont
  {E.}~\bibnamefont {Bairey}}, \bibinfo {author} {\bibfnamefont
  {Z.}~\bibnamefont {Brakerski}}, \bibinfo {author} {\bibfnamefont
  {I.}~\bibnamefont {Cohen}}, \bibinfo {author} {\bibfnamefont
  {O.}~\bibnamefont {Golan}}, \bibinfo {author} {\bibfnamefont
  {I.}~\bibnamefont {Gurwich}}, \bibinfo {author} {\bibfnamefont
  {O.}~\bibnamefont {Kenneth}}, \bibinfo {author} {\bibfnamefont
  {E.}~\bibnamefont {Leviatan}}, \bibinfo {author} {\bibfnamefont {N.~H.}\
  \bibnamefont {Lindner}}, \bibinfo {author} {\bibfnamefont {R.~A.}\
  \bibnamefont {Melcer}}, \bibinfo {author} {\bibfnamefont {A.}~\bibnamefont
  {Meyer}}, \bibinfo {author} {\bibfnamefont {G.}~\bibnamefont {Schul}},\ and\
  \bibinfo {author} {\bibfnamefont {M.}~\bibnamefont {Shutman}},\ }\href
  {https://doi.org/10.48550/arXiv.2503.17243} {\bibinfo {title} {On the
  importance of error mitigation for quantum computation}} (\bibinfo {year}
  {2025}),\ \Eprint {https://arxiv.org/abs/2503.17243} {arXiv:2503.17243
  [quant-ph]} \BibitemShut {NoStop}%
\bibitem [{\citenamefont {B{\"a}umer}\ \emph
  {et~al.}(2024{\natexlab{a}})\citenamefont {B{\"a}umer}, \citenamefont
  {Tripathi}, \citenamefont {Wang}, \citenamefont {Rall}, \citenamefont {Chen},
  \citenamefont {Majumder}, \citenamefont {Seif},\ and\ \citenamefont
  {Minev}}]{Baumer2024a}%
  \BibitemOpen
  \bibfield  {author} {\bibinfo {author} {\bibfnamefont {E.}~\bibnamefont
  {B{\"a}umer}}, \bibinfo {author} {\bibfnamefont {V.}~\bibnamefont
  {Tripathi}}, \bibinfo {author} {\bibfnamefont {D.~S.}\ \bibnamefont {Wang}},
  \bibinfo {author} {\bibfnamefont {P.}~\bibnamefont {Rall}}, \bibinfo {author}
  {\bibfnamefont {E.~H.}\ \bibnamefont {Chen}}, \bibinfo {author}
  {\bibfnamefont {S.}~\bibnamefont {Majumder}}, \bibinfo {author}
  {\bibfnamefont {A.}~\bibnamefont {Seif}},\ and\ \bibinfo {author}
  {\bibfnamefont {Z.~K.}\ \bibnamefont {Minev}},\ }\bibfield  {title} {\bibinfo
  {title} {Efficient long-range entanglement using dynamic circuits},\ }\href
  {https://doi.org/10.1103/PRXQuantum.5.030339} {\bibfield  {journal} {\bibinfo
   {journal} {PRX Quantum}\ }\textbf {\bibinfo {volume} {5}},\ \bibinfo {pages}
  {030339} (\bibinfo {year} {2024}{\natexlab{a}})}\BibitemShut {NoStop}%
\bibitem [{\citenamefont {C{\'o}rcoles}\ \emph {et~al.}(2021)\citenamefont
  {C{\'o}rcoles}, \citenamefont {Takita}, \citenamefont {Inoue}, \citenamefont
  {Lekuch}, \citenamefont {Minev}, \citenamefont {Chow},\ and\ \citenamefont
  {Gambetta}}]{Corcoles2021}%
  \BibitemOpen
  \bibfield  {author} {\bibinfo {author} {\bibfnamefont {A.~D.}\ \bibnamefont
  {C{\'o}rcoles}}, \bibinfo {author} {\bibfnamefont {M.}~\bibnamefont
  {Takita}}, \bibinfo {author} {\bibfnamefont {K.}~\bibnamefont {Inoue}},
  \bibinfo {author} {\bibfnamefont {S.}~\bibnamefont {Lekuch}}, \bibinfo
  {author} {\bibfnamefont {Z.~K.}\ \bibnamefont {Minev}}, \bibinfo {author}
  {\bibfnamefont {J.~M.}\ \bibnamefont {Chow}},\ and\ \bibinfo {author}
  {\bibfnamefont {J.~M.}\ \bibnamefont {Gambetta}},\ }\bibfield  {title}
  {\bibinfo {title} {Exploiting dynamic quantum circuits in a quantum algorithm
  with superconducting qubits},\ }\href
  {https://doi.org/10.1103/PhysRevLett.127.100501} {\bibfield  {journal}
  {\bibinfo  {journal} {Phys. Rev. Lett.}\ }\textbf {\bibinfo {volume} {127}},\
  \bibinfo {pages} {100501} (\bibinfo {year} {2021})}\BibitemShut {NoStop}%
\bibitem [{\citenamefont {B{\"a}umer}\ \emph
  {et~al.}(2024{\natexlab{b}})\citenamefont {B{\"a}umer}, \citenamefont
  {Tripathi}, \citenamefont {Seif}, \citenamefont {Lidar},\ and\ \citenamefont
  {Wang}}]{Baumer2024}%
  \BibitemOpen
  \bibfield  {author} {\bibinfo {author} {\bibfnamefont {E.}~\bibnamefont
  {B{\"a}umer}}, \bibinfo {author} {\bibfnamefont {V.}~\bibnamefont
  {Tripathi}}, \bibinfo {author} {\bibfnamefont {A.}~\bibnamefont {Seif}},
  \bibinfo {author} {\bibfnamefont {D.}~\bibnamefont {Lidar}},\ and\ \bibinfo
  {author} {\bibfnamefont {D.~S.}\ \bibnamefont {Wang}},\ }\bibfield  {title}
  {\bibinfo {title} {Quantum fourier transform using dynamic circuits},\ }\href
  {https://doi.org/10.1103/PhysRevLett.133.150602} {\bibfield  {journal}
  {\bibinfo  {journal} {Phys. Rev. Lett.}\ }\textbf {\bibinfo {volume} {133}},\
  \bibinfo {pages} {150602} (\bibinfo {year} {2024}{\natexlab{b}})}\BibitemShut
  {NoStop}%
\bibitem [{\citenamefont {B{\"a}umer}\ \emph {et~al.}(2025)\citenamefont
  {B{\"a}umer}, \citenamefont {Sutter},\ and\ \citenamefont
  {Woerner}}]{Baumer2025a}%
  \BibitemOpen
  \bibfield  {author} {\bibinfo {author} {\bibfnamefont {E.}~\bibnamefont
  {B{\"a}umer}}, \bibinfo {author} {\bibfnamefont {D.}~\bibnamefont {Sutter}},\
  and\ \bibinfo {author} {\bibfnamefont {S.}~\bibnamefont {Woerner}},\ }\href
  {https://doi.org/10.48550/arXiv.2504.20832} {\bibinfo {title} {Approximate
  quantum fourier transform in logarithmic depth on a line}} (\bibinfo {year}
  {2025}),\ \Eprint {https://arxiv.org/abs/2504.20832} {arXiv:2504.20832
  [quant-ph]} \BibitemShut {NoStop}%
\bibitem [{\citenamefont {Piveteau}\ and\ \citenamefont
  {Sutter}(2024)}]{Piveteau2024}%
  \BibitemOpen
  \bibfield  {author} {\bibinfo {author} {\bibfnamefont {C.}~\bibnamefont
  {Piveteau}}\ and\ \bibinfo {author} {\bibfnamefont {D.}~\bibnamefont
  {Sutter}},\ }\bibfield  {title} {\bibinfo {title} {Circuit knitting with
  classical communication},\ }\href {https://doi.org/10.1109/TIT.2023.3310797}
  {\bibfield  {journal} {\bibinfo  {journal} {IEEE Trans. Inf. Theory}\
  }\textbf {\bibinfo {volume} {70}},\ \bibinfo {pages} {2734} (\bibinfo {year}
  {2024})}\BibitemShut {NoStop}%
\bibitem [{\citenamefont {Brenner}\ \emph {et~al.}(2023)\citenamefont
  {Brenner}, \citenamefont {Piveteau},\ and\ \citenamefont
  {Sutter}}]{Brenner2023}%
  \BibitemOpen
  \bibfield  {author} {\bibinfo {author} {\bibfnamefont {L.}~\bibnamefont
  {Brenner}}, \bibinfo {author} {\bibfnamefont {C.}~\bibnamefont {Piveteau}},\
  and\ \bibinfo {author} {\bibfnamefont {D.}~\bibnamefont {Sutter}},\ }\href
  {https://doi.org/10.48550/arXiv.2302.03366} {\bibinfo {title} {Optimal wire
  cutting with classical communication}} (\bibinfo {year} {2023}),\ \Eprint
  {https://arxiv.org/abs/2302.03366} {arXiv:2302.03366 [quant-ph]} \BibitemShut
  {NoStop}%
\bibitem [{\citenamefont {Carrera~Vazquez}\ \emph {et~al.}(2024)\citenamefont
  {Carrera~Vazquez}, \citenamefont {Tornow}, \citenamefont {Rist{\`e}},
  \citenamefont {Woerner}, \citenamefont {Takita},\ and\ \citenamefont
  {Egger}}]{Carrera2024}%
  \BibitemOpen
  \bibfield  {author} {\bibinfo {author} {\bibfnamefont {A.}~\bibnamefont
  {Carrera~Vazquez}}, \bibinfo {author} {\bibfnamefont {C.}~\bibnamefont
  {Tornow}}, \bibinfo {author} {\bibfnamefont {D.}~\bibnamefont {Rist{\`e}}},
  \bibinfo {author} {\bibfnamefont {S.}~\bibnamefont {Woerner}}, \bibinfo
  {author} {\bibfnamefont {M.}~\bibnamefont {Takita}},\ and\ \bibinfo {author}
  {\bibfnamefont {D.~J.}\ \bibnamefont {Egger}},\ }\bibfield  {title} {\bibinfo
  {title} {Combining quantum processors with real-time classical
  communication},\ }\href {https://doi.org/10.1038/s41586-024-08178-2}
  {\bibfield  {journal} {\bibinfo  {journal} {Nature}\ }\textbf {\bibinfo
  {volume} {636}},\ \bibinfo {pages} {75} (\bibinfo {year} {2024})}\BibitemShut
  {NoStop}%
\bibitem [{\citenamefont {Mitarai}\ and\ \citenamefont
  {Fujii}(2021)}]{Mitarai2021}%
  \BibitemOpen
  \bibfield  {author} {\bibinfo {author} {\bibfnamefont {K.}~\bibnamefont
  {Mitarai}}\ and\ \bibinfo {author} {\bibfnamefont {K.}~\bibnamefont
  {Fujii}},\ }\bibfield  {title} {\bibinfo {title} {Constructing a virtual
  two-qubit gate by sampling single-qubit operations},\ }\href
  {https://doi.org/10.1088/1367-2630/abd7bc} {\bibfield  {journal} {\bibinfo
  {journal} {New J. Phys.}\ }\textbf {\bibinfo {volume} {23}},\ \bibinfo
  {pages} {023021} (\bibinfo {year} {2021})}\BibitemShut {NoStop}%
\bibitem [{\citenamefont {Singh}\ \emph {et~al.}(2024)\citenamefont {Singh},
  \citenamefont {Mitarai}, \citenamefont {Suzuki}, \citenamefont {Heya},
  \citenamefont {Tabuchi}, \citenamefont {Fujii},\ and\ \citenamefont
  {Nakamura}}]{Singh2024}%
  \BibitemOpen
  \bibfield  {author} {\bibinfo {author} {\bibfnamefont {A.~P.}\ \bibnamefont
  {Singh}}, \bibinfo {author} {\bibfnamefont {K.}~\bibnamefont {Mitarai}},
  \bibinfo {author} {\bibfnamefont {Y.}~\bibnamefont {Suzuki}}, \bibinfo
  {author} {\bibfnamefont {K.}~\bibnamefont {Heya}}, \bibinfo {author}
  {\bibfnamefont {Y.}~\bibnamefont {Tabuchi}}, \bibinfo {author} {\bibfnamefont
  {K.}~\bibnamefont {Fujii}},\ and\ \bibinfo {author} {\bibfnamefont
  {Y.}~\bibnamefont {Nakamura}},\ }\bibfield  {title} {\bibinfo {title}
  {Experimental demonstration of a high-fidelity virtual two-qubit gate},\
  }\href {https://doi.org/10.1103/PhysRevResearch.6.013235} {\bibfield
  {journal} {\bibinfo  {journal} {Phys. Rev. Res.}\ }\textbf {\bibinfo {volume}
  {6}},\ \bibinfo {pages} {013235} (\bibinfo {year} {2024})}\BibitemShut
  {NoStop}%
\bibitem [{\citenamefont {{van den Berg}}\ \emph {et~al.}(2023)\citenamefont
  {{van den Berg}}, \citenamefont {Minev}, \citenamefont {Kandala},\ and\
  \citenamefont {Temme}}]{van2023probabilistic}%
  \BibitemOpen
  \bibfield  {author} {\bibinfo {author} {\bibfnamefont {E.}~\bibnamefont {{van
  den Berg}}}, \bibinfo {author} {\bibfnamefont {Z.~K.}\ \bibnamefont {Minev}},
  \bibinfo {author} {\bibfnamefont {A.}~\bibnamefont {Kandala}},\ and\ \bibinfo
  {author} {\bibfnamefont {K.}~\bibnamefont {Temme}},\ }\bibfield  {title}
  {\bibinfo {title} {Probabilistic error cancellation with sparse
  {{Pauli}}--{{Lindblad}} models on noisy quantum processors},\ }\href
  {https://doi.org/10.1038/s41567-023-02042-2} {\bibfield  {journal} {\bibinfo
  {journal} {Nat. Phys.}\ }\textbf {\bibinfo {volume} {19}},\ \bibinfo {pages}
  {1116} (\bibinfo {year} {2023})}\BibitemShut {NoStop}%
\bibitem [{\citenamefont {McKay}\ \emph {et~al.}(2023)\citenamefont {McKay},
  \citenamefont {Hincks}, \citenamefont {Pritchett}, \citenamefont {Carroll},
  \citenamefont {Govia},\ and\ \citenamefont {Merkel}}]{Mckay2023}%
  \BibitemOpen
  \bibfield  {author} {\bibinfo {author} {\bibfnamefont {D.~C.}\ \bibnamefont
  {McKay}}, \bibinfo {author} {\bibfnamefont {I.}~\bibnamefont {Hincks}},
  \bibinfo {author} {\bibfnamefont {E.~J.}\ \bibnamefont {Pritchett}}, \bibinfo
  {author} {\bibfnamefont {M.}~\bibnamefont {Carroll}}, \bibinfo {author}
  {\bibfnamefont {L.~C.~G.}\ \bibnamefont {Govia}},\ and\ \bibinfo {author}
  {\bibfnamefont {S.~T.}\ \bibnamefont {Merkel}},\ }\href
  {https://doi.org/10.48550/arXiv.2311.05933} {\bibinfo {title} {Benchmarking
  quantum processor performance at scale}} (\bibinfo {year} {2023}),\ \Eprint
  {https://arxiv.org/abs/2311.05933} {arXiv:2311.05933 [quant-ph]} \BibitemShut
  {NoStop}%
\bibitem [{\citenamefont {Erhard}\ \emph {et~al.}(2019)\citenamefont {Erhard},
  \citenamefont {Wallman}, \citenamefont {Postler}, \citenamefont {Meth},
  \citenamefont {Stricker}, \citenamefont {Martinez}, \citenamefont
  {Schindler}, \citenamefont {Monz}, \citenamefont {Emerson},\ and\
  \citenamefont {Blatt}}]{erhard2019characterizing}%
  \BibitemOpen
  \bibfield  {author} {\bibinfo {author} {\bibfnamefont {A.}~\bibnamefont
  {Erhard}}, \bibinfo {author} {\bibfnamefont {J.~J.}\ \bibnamefont {Wallman}},
  \bibinfo {author} {\bibfnamefont {L.}~\bibnamefont {Postler}}, \bibinfo
  {author} {\bibfnamefont {M.}~\bibnamefont {Meth}}, \bibinfo {author}
  {\bibfnamefont {R.}~\bibnamefont {Stricker}}, \bibinfo {author}
  {\bibfnamefont {E.~A.}\ \bibnamefont {Martinez}}, \bibinfo {author}
  {\bibfnamefont {P.}~\bibnamefont {Schindler}}, \bibinfo {author}
  {\bibfnamefont {T.}~\bibnamefont {Monz}}, \bibinfo {author} {\bibfnamefont
  {J.}~\bibnamefont {Emerson}},\ and\ \bibinfo {author} {\bibfnamefont
  {R.}~\bibnamefont {Blatt}},\ }\bibfield  {title} {\bibinfo {title}
  {Characterizing large-scale quantum computers via cycle benchmarking},\
  }\href {https://doi.org/10.1038/s41467-019-13068-7} {\bibfield  {journal}
  {\bibinfo  {journal} {Nat. Commun.}\ }\textbf {\bibinfo {volume} {10}},\
  \bibinfo {pages} {5347} (\bibinfo {year} {2019})}\BibitemShut {NoStop}%
\bibitem [{\citenamefont {{Carignan-Dugas}}\ \emph {et~al.}(2023)\citenamefont
  {{Carignan-Dugas}}, \citenamefont {Dahlen}, \citenamefont {Hincks},
  \citenamefont {Ospadov}, \citenamefont {Beale}, \citenamefont {Ferracin},
  \citenamefont {{Skanes-Norman}}, \citenamefont {Emerson},\ and\ \citenamefont
  {Wallman}}]{carignan-dugas2023}%
  \BibitemOpen
  \bibfield  {author} {\bibinfo {author} {\bibfnamefont {A.}~\bibnamefont
  {{Carignan-Dugas}}}, \bibinfo {author} {\bibfnamefont {D.}~\bibnamefont
  {Dahlen}}, \bibinfo {author} {\bibfnamefont {I.}~\bibnamefont {Hincks}},
  \bibinfo {author} {\bibfnamefont {E.}~\bibnamefont {Ospadov}}, \bibinfo
  {author} {\bibfnamefont {S.~J.}\ \bibnamefont {Beale}}, \bibinfo {author}
  {\bibfnamefont {S.}~\bibnamefont {Ferracin}}, \bibinfo {author}
  {\bibfnamefont {J.}~\bibnamefont {{Skanes-Norman}}}, \bibinfo {author}
  {\bibfnamefont {J.}~\bibnamefont {Emerson}},\ and\ \bibinfo {author}
  {\bibfnamefont {J.~J.}\ \bibnamefont {Wallman}},\ }\href
  {https://doi.org/10.48550/arXiv.2303.17714} {\bibinfo {title} {The error
  reconstruction and compiled calibration of quantum computing cycles}}
  (\bibinfo {year} {2023}),\ \Eprint {https://arxiv.org/abs/2303.17714v1}
  {arXiv:2303.17714v1 [quant-ph]} \BibitemShut {NoStop}%
\bibitem [{\citenamefont {Bennett}\ \emph {et~al.}(1996)\citenamefont
  {Bennett}, \citenamefont {Brassard}, \citenamefont {Popescu}, \citenamefont
  {Schumacher}, \citenamefont {Smolin},\ and\ \citenamefont
  {Wootters}}]{twirling_bennet}%
  \BibitemOpen
  \bibfield  {author} {\bibinfo {author} {\bibfnamefont {C.~H.}\ \bibnamefont
  {Bennett}}, \bibinfo {author} {\bibfnamefont {G.}~\bibnamefont {Brassard}},
  \bibinfo {author} {\bibfnamefont {S.}~\bibnamefont {Popescu}}, \bibinfo
  {author} {\bibfnamefont {B.}~\bibnamefont {Schumacher}}, \bibinfo {author}
  {\bibfnamefont {J.~A.}\ \bibnamefont {Smolin}},\ and\ \bibinfo {author}
  {\bibfnamefont {W.~K.}\ \bibnamefont {Wootters}},\ }\bibfield  {title}
  {\bibinfo {title} {Purification of noisy entanglement and faithful
  teleportation via noisy channels},\ }\href
  {https://doi.org/10.1103/PhysRevLett.76.722} {\bibfield  {journal} {\bibinfo
  {journal} {Phys. Rev. Lett.}\ }\textbf {\bibinfo {volume} {76}},\ \bibinfo
  {pages} {722} (\bibinfo {year} {1996})}\BibitemShut {NoStop}%
\bibitem [{\citenamefont {Knill}(2004)}]{twirling_knill}%
  \BibitemOpen
  \bibfield  {author} {\bibinfo {author} {\bibfnamefont {E.}~\bibnamefont
  {Knill}},\ }\href {https://doi.org/10.48550/arXiv.quant-ph/0404104} {\bibinfo
  {title} {Fault-tolerant postselected quantum computation: Threshold
  analysis}} (\bibinfo {year} {2004}),\ \Eprint
  {https://arxiv.org/abs/quant-ph/0404104} {arXiv:quant-ph/0404104}
  \BibitemShut {NoStop}%
\bibitem [{\citenamefont {Wallman}\ and\ \citenamefont
  {Emerson}(2016)}]{wallman2016noise}%
  \BibitemOpen
  \bibfield  {author} {\bibinfo {author} {\bibfnamefont {J.~J.}\ \bibnamefont
  {Wallman}}\ and\ \bibinfo {author} {\bibfnamefont {J.}~\bibnamefont
  {Emerson}},\ }\bibfield  {title} {\bibinfo {title} {Noise tailoring for
  scalable quantum computation via randomized compiling},\ }\href
  {https://doi.org/10.1103/PhysRevA.94.052325} {\bibfield  {journal} {\bibinfo
  {journal} {Phys. Rev. A}\ }\textbf {\bibinfo {volume} {94}},\ \bibinfo
  {pages} {052325} (\bibinfo {year} {2016})}\BibitemShut {NoStop}%
\bibitem [{\citenamefont {Van Den~Berg}\ \emph {et~al.}(2022)\citenamefont {Van
  Den~Berg}, \citenamefont {Minev},\ and\ \citenamefont
  {Temme}}]{VanDenBerg2022}%
  \BibitemOpen
  \bibfield  {author} {\bibinfo {author} {\bibfnamefont {E.}~\bibnamefont {Van
  Den~Berg}}, \bibinfo {author} {\bibfnamefont {Z.~K.}\ \bibnamefont {Minev}},\
  and\ \bibinfo {author} {\bibfnamefont {K.}~\bibnamefont {Temme}},\ }\bibfield
   {title} {\bibinfo {title} {Model-free readout-error mitigation for quantum
  expectation values},\ }\href {https://doi.org/10.1103/PhysRevA.105.032620}
  {\bibfield  {journal} {\bibinfo  {journal} {Phys. Rev. A}\ }\textbf {\bibinfo
  {volume} {105}},\ \bibinfo {pages} {032620} (\bibinfo {year}
  {2022})}\BibitemShut {NoStop}%
\bibitem [{\citenamefont {Hines}\ and\ \citenamefont
  {Proctor}(2025)}]{hines2025pauli}%
  \BibitemOpen
  \bibfield  {author} {\bibinfo {author} {\bibfnamefont {J.}~\bibnamefont
  {Hines}}\ and\ \bibinfo {author} {\bibfnamefont {T.}~\bibnamefont
  {Proctor}},\ }\bibfield  {title} {\bibinfo {title} {Pauli noise learning for
  mid-circuit measurements},\ }\href
  {https://doi.org/10.1103/PhysRevLett.134.020602} {\bibfield  {journal}
  {\bibinfo  {journal} {Phys. Rev. Lett.}\ }\textbf {\bibinfo {volume} {134}},\
  \bibinfo {pages} {020602} (\bibinfo {year} {2025})}\BibitemShut {NoStop}%
\bibitem [{\citenamefont {Zhang}\ \emph {et~al.}(2025)\citenamefont {Zhang},
  \citenamefont {Chen}, \citenamefont {Liu},\ and\ \citenamefont
  {Jiang}}]{zhang2025generalized}%
  \BibitemOpen
  \bibfield  {author} {\bibinfo {author} {\bibfnamefont {Z.}~\bibnamefont
  {Zhang}}, \bibinfo {author} {\bibfnamefont {S.}~\bibnamefont {Chen}},
  \bibinfo {author} {\bibfnamefont {Y.}~\bibnamefont {Liu}},\ and\ \bibinfo
  {author} {\bibfnamefont {L.}~\bibnamefont {Jiang}},\ }\bibfield  {title}
  {\bibinfo {title} {Generalized cycle benchmarking algorithm for
  characterizing midcircuit measurements},\ }\href
  {https://doi.org/10.1103/PRXQuantum.6.010310} {\bibfield  {journal} {\bibinfo
   {journal} {PRX Quantum}\ }\textbf {\bibinfo {volume} {6}},\ \bibinfo {pages}
  {010310} (\bibinfo {year} {2025})}\BibitemShut {NoStop}%
\bibitem [{\citenamefont {Hazra}\ \emph {et~al.}(2025)\citenamefont {Hazra},
  \citenamefont {Dai}, \citenamefont {Connolly}, \citenamefont {Kurilovich},
  \citenamefont {Wang}, \citenamefont {Frunzio},\ and\ \citenamefont
  {Devoret}}]{hazra2025benchmarking}%
  \BibitemOpen
  \bibfield  {author} {\bibinfo {author} {\bibfnamefont {S.}~\bibnamefont
  {Hazra}}, \bibinfo {author} {\bibfnamefont {W.}~\bibnamefont {Dai}}, \bibinfo
  {author} {\bibfnamefont {T.}~\bibnamefont {Connolly}}, \bibinfo {author}
  {\bibfnamefont {P.~D.}\ \bibnamefont {Kurilovich}}, \bibinfo {author}
  {\bibfnamefont {Z.}~\bibnamefont {Wang}}, \bibinfo {author} {\bibfnamefont
  {L.}~\bibnamefont {Frunzio}},\ and\ \bibinfo {author} {\bibfnamefont {M.~H.}\
  \bibnamefont {Devoret}},\ }\bibfield  {title} {\bibinfo {title} {Benchmarking
  the readout of a superconducting qubit for repeated measurements},\ }\href
  {https://doi.org/10.1103/PhysRevLett.134.100601} {\bibfield  {journal}
  {\bibinfo  {journal} {Phys. Rev. Lett.}\ }\textbf {\bibinfo {volume} {134}},\
  \bibinfo {pages} {100601} (\bibinfo {year} {2025})}\BibitemShut {NoStop}%
\bibitem [{\citenamefont {Gupta}\ \emph {et~al.}(2024)\citenamefont {Gupta},
  \citenamefont {{van den Berg}}, \citenamefont {Takita}, \citenamefont
  {Rist{\`e}}, \citenamefont {Temme},\ and\ \citenamefont
  {Kandala}}]{gupta2024probabilistic}%
  \BibitemOpen
  \bibfield  {author} {\bibinfo {author} {\bibfnamefont {R.~S.}\ \bibnamefont
  {Gupta}}, \bibinfo {author} {\bibfnamefont {E.}~\bibnamefont {{van den
  Berg}}}, \bibinfo {author} {\bibfnamefont {M.}~\bibnamefont {Takita}},
  \bibinfo {author} {\bibfnamefont {D.}~\bibnamefont {Rist{\`e}}}, \bibinfo
  {author} {\bibfnamefont {K.}~\bibnamefont {Temme}},\ and\ \bibinfo {author}
  {\bibfnamefont {A.}~\bibnamefont {Kandala}},\ }\bibfield  {title} {\bibinfo
  {title} {Probabilistic error cancellation for dynamic quantum circuits},\
  }\href {https://doi.org/10.1103/PhysRevA.109.062617} {\bibfield  {journal}
  {\bibinfo  {journal} {Phys. Rev. A}\ }\textbf {\bibinfo {volume} {109}},\
  \bibinfo {pages} {062617} (\bibinfo {year} {2024})}\BibitemShut {NoStop}%
\bibitem [{\citenamefont {Koh}\ \emph {et~al.}(2026)\citenamefont {Koh},
  \citenamefont {Koh},\ and\ \citenamefont {Thompson}}]{koh2024readout}%
  \BibitemOpen
  \bibfield  {author} {\bibinfo {author} {\bibfnamefont {J.~M.}\ \bibnamefont
  {Koh}}, \bibinfo {author} {\bibfnamefont {D.~E.}\ \bibnamefont {Koh}},\ and\
  \bibinfo {author} {\bibfnamefont {J.}~\bibnamefont {Thompson}},\ }\bibfield
  {title} {\bibinfo {title} {Readout error mitigation for mid-circuit
  measurements and feedforward},\ }\href {https://doi.org/10.1103/cj89-4h5t}
  {\bibfield  {journal} {\bibinfo  {journal} {PRX Quantum}\ }\textbf {\bibinfo
  {volume} {7}},\ \bibinfo {pages} {10317} (\bibinfo {year}
  {2026})}\BibitemShut {NoStop}%
\bibitem [{\citenamefont {Hashim}\ \emph {et~al.}(2025)\citenamefont {Hashim},
  \citenamefont {{Carignan-Dugas}}, \citenamefont {Chen}, \citenamefont
  {J{\"u}nger}, \citenamefont {Fruitwala}, \citenamefont {Xu}, \citenamefont
  {Huang}, \citenamefont {Wallman},\ and\ \citenamefont
  {Siddiqi}}]{Hashim2025}%
  \BibitemOpen
  \bibfield  {author} {\bibinfo {author} {\bibfnamefont {A.}~\bibnamefont
  {Hashim}}, \bibinfo {author} {\bibfnamefont {A.}~\bibnamefont
  {{Carignan-Dugas}}}, \bibinfo {author} {\bibfnamefont {L.}~\bibnamefont
  {Chen}}, \bibinfo {author} {\bibfnamefont {C.}~\bibnamefont {J{\"u}nger}},
  \bibinfo {author} {\bibfnamefont {N.}~\bibnamefont {Fruitwala}}, \bibinfo
  {author} {\bibfnamefont {Y.}~\bibnamefont {Xu}}, \bibinfo {author}
  {\bibfnamefont {G.}~\bibnamefont {Huang}}, \bibinfo {author} {\bibfnamefont
  {J.~J.}\ \bibnamefont {Wallman}},\ and\ \bibinfo {author} {\bibfnamefont
  {I.}~\bibnamefont {Siddiqi}},\ }\bibfield  {title} {\bibinfo {title}
  {Quasiprobabilistic readout correction of midcircuit measurements for
  adaptive feedback via measurement randomized compiling},\ }\href
  {https://doi.org/10.1103/PRXQuantum.6.010307} {\bibfield  {journal} {\bibinfo
   {journal} {PRX Quantum}\ }\textbf {\bibinfo {volume} {6}},\ \bibinfo {pages}
  {010307} (\bibinfo {year} {2025})}\BibitemShut {NoStop}%
\bibitem [{\citenamefont {Chen}\ \emph
  {et~al.}(2023{\natexlab{a}})\citenamefont {Chen}, \citenamefont {Liu},
  \citenamefont {Otten}, \citenamefont {Seif}, \citenamefont {Fefferman},\ and\
  \citenamefont {Jiang}}]{chen2023learnability}%
  \BibitemOpen
  \bibfield  {author} {\bibinfo {author} {\bibfnamefont {S.}~\bibnamefont
  {Chen}}, \bibinfo {author} {\bibfnamefont {Y.}~\bibnamefont {Liu}}, \bibinfo
  {author} {\bibfnamefont {M.}~\bibnamefont {Otten}}, \bibinfo {author}
  {\bibfnamefont {A.}~\bibnamefont {Seif}}, \bibinfo {author} {\bibfnamefont
  {B.}~\bibnamefont {Fefferman}},\ and\ \bibinfo {author} {\bibfnamefont
  {L.}~\bibnamefont {Jiang}},\ }\bibfield  {title} {\bibinfo {title} {The
  learnability of {{Pauli}} noise},\ }\href
  {https://doi.org/10.1038/s41467-022-35759-4} {\bibfield  {journal} {\bibinfo
  {journal} {Nat. Commun.}\ }\textbf {\bibinfo {volume} {14}},\ \bibinfo
  {pages} {52} (\bibinfo {year} {2023}{\natexlab{a}})}\BibitemShut {NoStop}%
\bibitem [{\citenamefont {Yu}\ and\ \citenamefont {Wei}(2025)}]{Yu2025}%
  \BibitemOpen
  \bibfield  {author} {\bibinfo {author} {\bibfnamefont {H.}~\bibnamefont
  {Yu}}\ and\ \bibinfo {author} {\bibfnamefont {T.-C.}\ \bibnamefont {Wei}},\
  }\bibfield  {title} {\bibinfo {title} {Efficient separate quantification of
  state preparation errors and measurement errors on quantum computers and
  their mitigation},\ }\href {https://doi.org/10.22331/q-2025-05-05-1724}
  {\bibfield  {journal} {\bibinfo  {journal} {Quantum}\ }\textbf {\bibinfo
  {volume} {9}},\ \bibinfo {pages} {1724} (\bibinfo {year} {2025})}\BibitemShut
  {NoStop}%
\bibitem [{\citenamefont {Chen}\ \emph {et~al.}(2026)\citenamefont {Chen},
  \citenamefont {Zhang}, \citenamefont {Jiang},\ and\ \citenamefont
  {Flammia}}]{chen2024efficient}%
  \BibitemOpen
  \bibfield  {author} {\bibinfo {author} {\bibfnamefont {S.}~\bibnamefont
  {Chen}}, \bibinfo {author} {\bibfnamefont {Z.}~\bibnamefont {Zhang}},
  \bibinfo {author} {\bibfnamefont {L.}~\bibnamefont {Jiang}},\ and\ \bibinfo
  {author} {\bibfnamefont {S.~T.}\ \bibnamefont {Flammia}},\ }\bibfield
  {title} {\bibinfo {title} {Efficient self-consistent learning of gate set
  pauli noise},\ }\href {https://doi.org/10.1103/1pnv-t9px} {\bibfield
  {journal} {\bibinfo  {journal} {PRX Quantum}\ }\textbf {\bibinfo {volume}
  {7}},\ \bibinfo {pages} {10305} (\bibinfo {year} {2026})}\BibitemShut
  {NoStop}%
\bibitem [{\citenamefont {Chen}\ \emph
  {et~al.}(2025{\natexlab{a}})\citenamefont {Chen}, \citenamefont {Chen},
  \citenamefont {Fischer}, \citenamefont {Eddins}, \citenamefont {Govia},
  \citenamefont {Mitchell}, \citenamefont {He}, \citenamefont {Kim},
  \citenamefont {Jiang},\ and\ \citenamefont {Seif}}]{Chen2025}%
  \BibitemOpen
  \bibfield  {author} {\bibinfo {author} {\bibfnamefont {E.~H.}\ \bibnamefont
  {Chen}}, \bibinfo {author} {\bibfnamefont {S.}~\bibnamefont {Chen}}, \bibinfo
  {author} {\bibfnamefont {L.~E.}\ \bibnamefont {Fischer}}, \bibinfo {author}
  {\bibfnamefont {A.}~\bibnamefont {Eddins}}, \bibinfo {author} {\bibfnamefont
  {L.~C.~G.}\ \bibnamefont {Govia}}, \bibinfo {author} {\bibfnamefont
  {B.}~\bibnamefont {Mitchell}}, \bibinfo {author} {\bibfnamefont
  {A.}~\bibnamefont {He}}, \bibinfo {author} {\bibfnamefont {Y.}~\bibnamefont
  {Kim}}, \bibinfo {author} {\bibfnamefont {L.}~\bibnamefont {Jiang}},\ and\
  \bibinfo {author} {\bibfnamefont {A.}~\bibnamefont {Seif}},\ }\href
  {https://doi.org/10.48550/arXiv.2505.22629} {\bibinfo {title} {Disambiguating
  {{Pauli}} noise in quantum computers}} (\bibinfo {year}
  {2025}{\natexlab{a}}),\ \Eprint {https://arxiv.org/abs/2505.22629}
  {arXiv:2505.22629 [quant-ph]} \BibitemShut {NoStop}%
\bibitem [{\citenamefont {Huggins}\ \emph {et~al.}(2021)\citenamefont
  {Huggins}, \citenamefont {McArdle}, \citenamefont {O'Brien}, \citenamefont
  {Lee}, \citenamefont {Rubin}, \citenamefont {Boixo}, \citenamefont {Whaley},
  \citenamefont {Babbush},\ and\ \citenamefont {McClean}}]{huggins2021}%
  \BibitemOpen
  \bibfield  {author} {\bibinfo {author} {\bibfnamefont {W.~J.}\ \bibnamefont
  {Huggins}}, \bibinfo {author} {\bibfnamefont {S.}~\bibnamefont {McArdle}},
  \bibinfo {author} {\bibfnamefont {T.~E.}\ \bibnamefont {O'Brien}}, \bibinfo
  {author} {\bibfnamefont {J.}~\bibnamefont {Lee}}, \bibinfo {author}
  {\bibfnamefont {N.~C.}\ \bibnamefont {Rubin}}, \bibinfo {author}
  {\bibfnamefont {S.}~\bibnamefont {Boixo}}, \bibinfo {author} {\bibfnamefont
  {K.~B.}\ \bibnamefont {Whaley}}, \bibinfo {author} {\bibfnamefont
  {R.}~\bibnamefont {Babbush}},\ and\ \bibinfo {author} {\bibfnamefont {J.~R.}\
  \bibnamefont {McClean}},\ }\bibfield  {title} {\bibinfo {title} {Virtual
  distillation for quantum error mitigation},\ }\href
  {https://doi.org/10.1103/PhysRevX.11.041036} {\bibfield  {journal} {\bibinfo
  {journal} {Phys. Rev. X}\ }\textbf {\bibinfo {volume} {11}},\ \bibinfo
  {pages} {41036} (\bibinfo {year} {2021})}\BibitemShut {NoStop}%
\bibitem [{\citenamefont {Koczor}(2021)}]{koczor2021}%
  \BibitemOpen
  \bibfield  {author} {\bibinfo {author} {\bibfnamefont {B.}~\bibnamefont
  {Koczor}},\ }\bibfield  {title} {\bibinfo {title} {Exponential error
  suppression for near-term quantum devices},\ }\href
  {https://doi.org/10.1103/PhysRevX.11.031057} {\bibfield  {journal} {\bibinfo
  {journal} {Phys. Rev. X}\ }\textbf {\bibinfo {volume} {11}},\ \bibinfo
  {pages} {31057} (\bibinfo {year} {2021})}\BibitemShut {NoStop}%
\bibitem [{\citenamefont {Laflamme}\ \emph {et~al.}(2022)\citenamefont
  {Laflamme}, \citenamefont {Lin},\ and\ \citenamefont {Mor}}]{laflamme2022}%
  \BibitemOpen
  \bibfield  {author} {\bibinfo {author} {\bibfnamefont {R.}~\bibnamefont
  {Laflamme}}, \bibinfo {author} {\bibfnamefont {J.}~\bibnamefont {Lin}},\ and\
  \bibinfo {author} {\bibfnamefont {T.}~\bibnamefont {Mor}},\ }\bibfield
  {title} {\bibinfo {title} {Algorithmic cooling for resolving state
  preparation and measurement errors in quantum computing},\ }\href
  {https://doi.org/10.1103/PhysRevA.106.012439} {\bibfield  {journal} {\bibinfo
   {journal} {Phys. Rev. A}\ }\textbf {\bibinfo {volume} {106}},\ \bibinfo
  {pages} {12439} (\bibinfo {year} {2022})}\BibitemShut {NoStop}%
\bibitem [{\citenamefont {Santos}\ and\ \citenamefont
  {Uzdin}(2025)}]{santos2025}%
  \BibitemOpen
  \bibfield  {author} {\bibinfo {author} {\bibfnamefont {J.~P.}\ \bibnamefont
  {Santos}}\ and\ \bibinfo {author} {\bibfnamefont {R.}~\bibnamefont {Uzdin}},\
  }\href {https://doi.org/10.48550/arXiv.2506.11270} {\bibinfo {title}
  {Drift-resilient mid-circuit measurement and state preparation error
  mitigation for dynamic circuits}} (\bibinfo {year} {2025}),\ \Eprint
  {https://arxiv.org/abs/2506.11270} {arXiv:2506.11270 [quant-ph]} \BibitemShut
  {NoStop}%
\bibitem [{\citenamefont {Krantz}\ \emph {et~al.}(2019)\citenamefont {Krantz},
  \citenamefont {Kjaergaard}, \citenamefont {Yan}, \citenamefont {Orlando},
  \citenamefont {Gustavsson},\ and\ \citenamefont {Oliver}}]{Krantz2019}%
  \BibitemOpen
  \bibfield  {author} {\bibinfo {author} {\bibfnamefont {P.}~\bibnamefont
  {Krantz}}, \bibinfo {author} {\bibfnamefont {M.}~\bibnamefont {Kjaergaard}},
  \bibinfo {author} {\bibfnamefont {F.}~\bibnamefont {Yan}}, \bibinfo {author}
  {\bibfnamefont {T.~P.}\ \bibnamefont {Orlando}}, \bibinfo {author}
  {\bibfnamefont {S.}~\bibnamefont {Gustavsson}},\ and\ \bibinfo {author}
  {\bibfnamefont {W.~D.}\ \bibnamefont {Oliver}},\ }\bibfield  {title}
  {\bibinfo {title} {A quantum engineer's guide to superconducting qubits},\
  }\href {https://doi.org/10.1063/1.5089550} {\bibfield  {journal} {\bibinfo
  {journal} {Appl. Phys. Rev.}\ }\textbf {\bibinfo {volume} {6}},\ \bibinfo
  {pages} {021318} (\bibinfo {year} {2019})}\BibitemShut {NoStop}%
\bibitem [{\citenamefont {Rist{\`e}}\ \emph {et~al.}(2012)\citenamefont
  {Rist{\`e}}, \citenamefont {Bultink}, \citenamefont {Lehnert},\ and\
  \citenamefont {DiCarlo}}]{Riste2012}%
  \BibitemOpen
  \bibfield  {author} {\bibinfo {author} {\bibfnamefont {D.}~\bibnamefont
  {Rist{\`e}}}, \bibinfo {author} {\bibfnamefont {C.~C.}\ \bibnamefont
  {Bultink}}, \bibinfo {author} {\bibfnamefont {K.~W.}\ \bibnamefont
  {Lehnert}},\ and\ \bibinfo {author} {\bibfnamefont {L.}~\bibnamefont
  {DiCarlo}},\ }\bibfield  {title} {\bibinfo {title} {Feedback control of a
  solid-state qubit using high-fidelity projective measurement},\ }\href
  {https://doi.org/10.1103/PhysRevLett.109.240502} {\bibfield  {journal}
  {\bibinfo  {journal} {Phys. Rev. Lett.}\ }\textbf {\bibinfo {volume} {109}},\
  \bibinfo {pages} {240502} (\bibinfo {year} {2012})}\BibitemShut {NoStop}%
\bibitem [{\citenamefont {Jin}\ \emph {et~al.}(2015)\citenamefont {Jin},
  \citenamefont {Kamal}, \citenamefont {Sears}, \citenamefont {Gudmundsen},
  \citenamefont {Hover}, \citenamefont {Miloshi}, \citenamefont {Slattery},
  \citenamefont {Yan}, \citenamefont {Yoder}, \citenamefont {Orlando},
  \citenamefont {Gustavsson},\ and\ \citenamefont {Oliver}}]{Jin2015}%
  \BibitemOpen
  \bibfield  {author} {\bibinfo {author} {\bibfnamefont {X.~Y.}\ \bibnamefont
  {Jin}}, \bibinfo {author} {\bibfnamefont {A.}~\bibnamefont {Kamal}}, \bibinfo
  {author} {\bibfnamefont {A.~P.}\ \bibnamefont {Sears}}, \bibinfo {author}
  {\bibfnamefont {T.}~\bibnamefont {Gudmundsen}}, \bibinfo {author}
  {\bibfnamefont {D.}~\bibnamefont {Hover}}, \bibinfo {author} {\bibfnamefont
  {J.}~\bibnamefont {Miloshi}}, \bibinfo {author} {\bibfnamefont
  {R.}~\bibnamefont {Slattery}}, \bibinfo {author} {\bibfnamefont
  {F.}~\bibnamefont {Yan}}, \bibinfo {author} {\bibfnamefont {J.}~\bibnamefont
  {Yoder}}, \bibinfo {author} {\bibfnamefont {T.~P.}\ \bibnamefont {Orlando}},
  \bibinfo {author} {\bibfnamefont {S.}~\bibnamefont {Gustavsson}},\ and\
  \bibinfo {author} {\bibfnamefont {W.~D.}\ \bibnamefont {Oliver}},\ }\bibfield
   {title} {\bibinfo {title} {Thermal and residual excited-state population in
  a {{3D}} transmon qubit},\ }\href
  {https://doi.org/10.1103/PhysRevLett.114.240501} {\bibfield  {journal}
  {\bibinfo  {journal} {Phys. Rev. Lett.}\ }\textbf {\bibinfo {volume} {114}},\
  \bibinfo {pages} {240501} (\bibinfo {year} {2015})}\BibitemShut {NoStop}%
\bibitem [{\citenamefont {Geerlings}\ \emph {et~al.}(2013)\citenamefont
  {Geerlings}, \citenamefont {Leghtas}, \citenamefont {Pop}, \citenamefont
  {Shankar}, \citenamefont {Frunzio}, \citenamefont {Schoelkopf}, \citenamefont
  {Mirrahimi},\ and\ \citenamefont {Devoret}}]{Geerlings2013}%
  \BibitemOpen
  \bibfield  {author} {\bibinfo {author} {\bibfnamefont {K.}~\bibnamefont
  {Geerlings}}, \bibinfo {author} {\bibfnamefont {Z.}~\bibnamefont {Leghtas}},
  \bibinfo {author} {\bibfnamefont {I.~M.}\ \bibnamefont {Pop}}, \bibinfo
  {author} {\bibfnamefont {S.}~\bibnamefont {Shankar}}, \bibinfo {author}
  {\bibfnamefont {L.}~\bibnamefont {Frunzio}}, \bibinfo {author} {\bibfnamefont
  {R.~J.}\ \bibnamefont {Schoelkopf}}, \bibinfo {author} {\bibfnamefont
  {M.}~\bibnamefont {Mirrahimi}},\ and\ \bibinfo {author} {\bibfnamefont
  {M.~H.}\ \bibnamefont {Devoret}},\ }\bibfield  {title} {\bibinfo {title}
  {Demonstrating a driven reset protocol for a superconducting qubit},\ }\href
  {https://doi.org/10.1103/PhysRevLett.110.120501} {\bibfield  {journal}
  {\bibinfo  {journal} {Phys. Rev. Lett.}\ }\textbf {\bibinfo {volume} {110}},\
  \bibinfo {pages} {120501} (\bibinfo {year} {2013})}\BibitemShut {NoStop}%
\bibitem [{Note1()}]{Note1}%
  \BibitemOpen
  \bibinfo {note} {The E and F in RabiEF come from an alternative transmon
  level naming where the states $\mathinner {|{0}\rangle }$, $\mathinner
  {|{1}\rangle }$, and $\mathinner {|{2}\rangle }$ are labeled by $\mathinner
  {|{g}\rangle }$, $\mathinner {|{e}\rangle }$, and $\mathinner {|{f}\rangle
  }$, respectively. Here, $g$ and $e$ stand for ground and excited,
  respectively.}\BibitemShut {Stop}%
\bibitem [{\citenamefont {Govia}\ \emph {et~al.}(2025)\citenamefont {Govia},
  \citenamefont {Majumder}, \citenamefont {Barron}, \citenamefont {Mitchell},
  \citenamefont {Seif}, \citenamefont {Kim}, \citenamefont {Wood},
  \citenamefont {Pritchett}, \citenamefont {Merkel},\ and\ \citenamefont
  {McKay}}]{Govia2025}%
  \BibitemOpen
  \bibfield  {author} {\bibinfo {author} {\bibfnamefont {L.}~\bibnamefont
  {Govia}}, \bibinfo {author} {\bibfnamefont {S.}~\bibnamefont {Majumder}},
  \bibinfo {author} {\bibfnamefont {S.}~\bibnamefont {Barron}}, \bibinfo
  {author} {\bibfnamefont {B.}~\bibnamefont {Mitchell}}, \bibinfo {author}
  {\bibfnamefont {A.}~\bibnamefont {Seif}}, \bibinfo {author} {\bibfnamefont
  {Y.}~\bibnamefont {Kim}}, \bibinfo {author} {\bibfnamefont {C.}~\bibnamefont
  {Wood}}, \bibinfo {author} {\bibfnamefont {E.}~\bibnamefont {Pritchett}},
  \bibinfo {author} {\bibfnamefont {S.}~\bibnamefont {Merkel}},\ and\ \bibinfo
  {author} {\bibfnamefont {D.}~\bibnamefont {McKay}},\ }\bibfield  {title}
  {\bibinfo {title} {Bounding the systematic error in quantum error mitigation
  due to model violation},\ }\href
  {https://doi.org/10.1103/PRXQuantum.6.010354} {\bibfield  {journal} {\bibinfo
   {journal} {PRX Quantum}\ }\textbf {\bibinfo {volume} {6}},\ \bibinfo {pages}
  {010354} (\bibinfo {year} {2025})}\BibitemShut {NoStop}%
\bibitem [{\citenamefont {Beale}\ and\ \citenamefont
  {Wallman}(2023)}]{Beale2023}%
  \BibitemOpen
  \bibfield  {author} {\bibinfo {author} {\bibfnamefont {S.~J.}\ \bibnamefont
  {Beale}}\ and\ \bibinfo {author} {\bibfnamefont {J.~J.}\ \bibnamefont
  {Wallman}},\ }\href {https://doi.org/10.48550/arXiv.2304.06599} {\bibinfo
  {title} {Randomized compiling for subsystem measurements}} (\bibinfo {year}
  {2023}),\ \Eprint {https://arxiv.org/abs/2304.06599} {arXiv:2304.06599
  [quant-ph]} \BibitemShut {NoStop}%
\bibitem [{\citenamefont {Nielsen}\ and\ \citenamefont
  {Chuang}(2010)}]{Nielsen2000}%
  \BibitemOpen
  \bibfield  {author} {\bibinfo {author} {\bibfnamefont {M.~A.}\ \bibnamefont
  {Nielsen}}\ and\ \bibinfo {author} {\bibfnamefont {I.~L.}\ \bibnamefont
  {Chuang}},\ }\href@noop {} {\emph {\bibinfo {title} {Quantum Computation and
  Quantum Information}}},\ \bibinfo {edition} {10th}\ ed.\ (\bibinfo
  {publisher} {Cambridge University Press},\ \bibinfo {address} {Cambridge ;
  New York},\ \bibinfo {year} {2010})\BibitemShut {NoStop}%
\bibitem [{\citenamefont {Fischer}\ \emph {et~al.}(2022)\citenamefont
  {Fischer}, \citenamefont {Miller}, \citenamefont {Tacchino}, \citenamefont
  {Barkoutsos}, \citenamefont {Egger},\ and\ \citenamefont
  {Tavernelli}}]{fischer_ancilla-free_2022}%
  \BibitemOpen
  \bibfield  {author} {\bibinfo {author} {\bibfnamefont {L.~E.}\ \bibnamefont
  {Fischer}}, \bibinfo {author} {\bibfnamefont {D.}~\bibnamefont {Miller}},
  \bibinfo {author} {\bibfnamefont {F.}~\bibnamefont {Tacchino}}, \bibinfo
  {author} {\bibfnamefont {P.~K.}\ \bibnamefont {Barkoutsos}}, \bibinfo
  {author} {\bibfnamefont {D.~J.}\ \bibnamefont {Egger}},\ and\ \bibinfo
  {author} {\bibfnamefont {I.}~\bibnamefont {Tavernelli}},\ }\bibfield  {title}
  {\bibinfo {title} {Ancilla-free implementation of generalized measurements
  for qubits embedded in a qudit space},\ }\href
  {https://doi.org/10.1103/PhysRevResearch.4.033027} {\bibfield  {journal}
  {\bibinfo  {journal} {Phys. Rev. Res.}\ }\textbf {\bibinfo {volume} {4}},\
  \bibinfo {pages} {033027} (\bibinfo {year} {2022})}\BibitemShut {NoStop}%
\bibitem [{\citenamefont {Bianchetti}\ \emph {et~al.}(2010)\citenamefont
  {Bianchetti}, \citenamefont {Filipp}, \citenamefont {Baur}, \citenamefont
  {Fink}, \citenamefont {Lang}, \citenamefont {Steffen}, \citenamefont
  {Boissonneault}, \citenamefont {Blais},\ and\ \citenamefont
  {Wallraff}}]{Bianchetti2010}%
  \BibitemOpen
  \bibfield  {author} {\bibinfo {author} {\bibfnamefont {R.}~\bibnamefont
  {Bianchetti}}, \bibinfo {author} {\bibfnamefont {S.}~\bibnamefont {Filipp}},
  \bibinfo {author} {\bibfnamefont {M.}~\bibnamefont {Baur}}, \bibinfo {author}
  {\bibfnamefont {J.~M.}\ \bibnamefont {Fink}}, \bibinfo {author}
  {\bibfnamefont {C.}~\bibnamefont {Lang}}, \bibinfo {author} {\bibfnamefont
  {L.}~\bibnamefont {Steffen}}, \bibinfo {author} {\bibfnamefont
  {M.}~\bibnamefont {Boissonneault}}, \bibinfo {author} {\bibfnamefont
  {A.}~\bibnamefont {Blais}},\ and\ \bibinfo {author} {\bibfnamefont
  {A.}~\bibnamefont {Wallraff}},\ }\bibfield  {title} {\bibinfo {title}
  {Control and tomography of a three level superconducting artificial atom},\
  }\href {https://doi.org/10.1103/PhysRevLett.105.223601} {\bibfield  {journal}
  {\bibinfo  {journal} {Phys. Rev. Lett.}\ }\textbf {\bibinfo {volume} {105}},\
  \bibinfo {pages} {223601} (\bibinfo {year} {2010})}\BibitemShut {NoStop}%
\bibitem [{\citenamefont {Blok}\ \emph {et~al.}(2021)\citenamefont {Blok},
  \citenamefont {Ramasesh}, \citenamefont {Schuster}, \citenamefont {O'Brien},
  \citenamefont {Kreikebaum}, \citenamefont {Dahlen}, \citenamefont {Morvan},
  \citenamefont {Yoshida}, \citenamefont {Yao},\ and\ \citenamefont
  {Siddiqi}}]{blok2021}%
  \BibitemOpen
  \bibfield  {author} {\bibinfo {author} {\bibfnamefont {M.~S.}\ \bibnamefont
  {Blok}}, \bibinfo {author} {\bibfnamefont {V.~V.}\ \bibnamefont {Ramasesh}},
  \bibinfo {author} {\bibfnamefont {T.}~\bibnamefont {Schuster}}, \bibinfo
  {author} {\bibfnamefont {K.}~\bibnamefont {O'Brien}}, \bibinfo {author}
  {\bibfnamefont {J.~M.}\ \bibnamefont {Kreikebaum}}, \bibinfo {author}
  {\bibfnamefont {D.}~\bibnamefont {Dahlen}}, \bibinfo {author} {\bibfnamefont
  {A.}~\bibnamefont {Morvan}}, \bibinfo {author} {\bibfnamefont
  {B.}~\bibnamefont {Yoshida}}, \bibinfo {author} {\bibfnamefont {N.~Y.}\
  \bibnamefont {Yao}},\ and\ \bibinfo {author} {\bibfnamefont {I.}~\bibnamefont
  {Siddiqi}},\ }\bibfield  {title} {\bibinfo {title} {Quantum information
  scrambling on a superconducting qutrit processor},\ }\href
  {https://doi.org/10.1103/PhysRevX.11.021010} {\bibfield  {journal} {\bibinfo
  {journal} {Phys. Rev. X}\ }\textbf {\bibinfo {volume} {11}},\ \bibinfo
  {pages} {21010} (\bibinfo {year} {2021})}\BibitemShut {NoStop}%
\bibitem [{\citenamefont {AbuGhanem}(2025)}]{abughanem2025}%
  \BibitemOpen
  \bibfield  {author} {\bibinfo {author} {\bibfnamefont {M.}~\bibnamefont
  {AbuGhanem}},\ }\bibfield  {title} {\bibinfo {title} {{{IBM}} quantum
  computers: Evolution, performance, and future directions},\ }\href
  {https://doi.org/10.1007/s11227-025-07047-7} {\bibfield  {journal} {\bibinfo
  {journal} {J. Supercomput.}\ }\textbf {\bibinfo {volume} {81}},\ \bibinfo
  {pages} {687} (\bibinfo {year} {2025})}\BibitemShut {NoStop}%
\bibitem [{\citenamefont {Haupt}\ and\ \citenamefont
  {Egger}(2023)}]{hauptRestless2023}%
  \BibitemOpen
  \bibfield  {author} {\bibinfo {author} {\bibfnamefont {C.~J.}\ \bibnamefont
  {Haupt}}\ and\ \bibinfo {author} {\bibfnamefont {D.~J.}\ \bibnamefont
  {Egger}},\ }\bibfield  {title} {\bibinfo {title} {Leakage in restless quantum
  gate calibration},\ }\href {https://doi.org/10.1103/PhysRevA.108.022614}
  {\bibfield  {journal} {\bibinfo  {journal} {Phys. Rev. A}\ }\textbf {\bibinfo
  {volume} {108}},\ \bibinfo {pages} {022614} (\bibinfo {year}
  {2023})}\BibitemShut {NoStop}%
\bibitem [{\citenamefont {Brandhofer}\ \emph {et~al.}(2023)\citenamefont
  {Brandhofer}, \citenamefont {Polian},\ and\ \citenamefont
  {Krsulich}}]{Brandhofer2023}%
  \BibitemOpen
  \bibfield  {author} {\bibinfo {author} {\bibfnamefont {S.}~\bibnamefont
  {Brandhofer}}, \bibinfo {author} {\bibfnamefont {I.}~\bibnamefont {Polian}},\
  and\ \bibinfo {author} {\bibfnamefont {K.}~\bibnamefont {Krsulich}},\
  }\bibfield  {title} {\bibinfo {title} {Optimal qubit reuse for near-term
  quantum computers},\ }in\ \href {https://doi.org/10.1109/QCE57702.2023.00100}
  {\emph {\bibinfo {booktitle} {2023 {{IEEE International Conference}} on
  {{Quantum Computing}} and {{Engineering}} ({{QCE}})}}},\ Vol.~\bibinfo
  {volume} {01}\ (\bibinfo {year} {2023})\ pp.\ \bibinfo {pages}
  {859--869}\BibitemShut {NoStop}%
\bibitem [{\citenamefont {Chen}\ \emph
  {et~al.}(2023{\natexlab{b}})\citenamefont {Chen}, \citenamefont {Li},
  \citenamefont {Lu}, \citenamefont {Warren}, \citenamefont {Kri{\v z}an},
  \citenamefont {Kosen}, \citenamefont {Rommel}, \citenamefont {Ahmed},
  \citenamefont {Osman}, \citenamefont {Bizn{\'a}rov{\'a}}, \citenamefont
  {Fadavi~Roudsari}, \citenamefont {Lienhard}, \citenamefont {Caputo},
  \citenamefont {Grigoras}, \citenamefont {Gr{\"o}nberg}, \citenamefont
  {Govenius}, \citenamefont {Kockum}, \citenamefont {Delsing}, \citenamefont
  {Bylander},\ and\ \citenamefont {Tancredi}}]{chen2023}%
  \BibitemOpen
  \bibfield  {author} {\bibinfo {author} {\bibfnamefont {L.}~\bibnamefont
  {Chen}}, \bibinfo {author} {\bibfnamefont {H.-X.}\ \bibnamefont {Li}},
  \bibinfo {author} {\bibfnamefont {Y.}~\bibnamefont {Lu}}, \bibinfo {author}
  {\bibfnamefont {C.~W.}\ \bibnamefont {Warren}}, \bibinfo {author}
  {\bibfnamefont {C.~J.}\ \bibnamefont {Kri{\v z}an}}, \bibinfo {author}
  {\bibfnamefont {S.}~\bibnamefont {Kosen}}, \bibinfo {author} {\bibfnamefont
  {M.}~\bibnamefont {Rommel}}, \bibinfo {author} {\bibfnamefont
  {S.}~\bibnamefont {Ahmed}}, \bibinfo {author} {\bibfnamefont
  {A.}~\bibnamefont {Osman}}, \bibinfo {author} {\bibfnamefont
  {J.}~\bibnamefont {Bizn{\'a}rov{\'a}}}, \bibinfo {author} {\bibfnamefont
  {A.}~\bibnamefont {Fadavi~Roudsari}}, \bibinfo {author} {\bibfnamefont
  {B.}~\bibnamefont {Lienhard}}, \bibinfo {author} {\bibfnamefont
  {M.}~\bibnamefont {Caputo}}, \bibinfo {author} {\bibfnamefont
  {K.}~\bibnamefont {Grigoras}}, \bibinfo {author} {\bibfnamefont
  {L.}~\bibnamefont {Gr{\"o}nberg}}, \bibinfo {author} {\bibfnamefont
  {J.}~\bibnamefont {Govenius}}, \bibinfo {author} {\bibfnamefont {A.~F.}\
  \bibnamefont {Kockum}}, \bibinfo {author} {\bibfnamefont {P.}~\bibnamefont
  {Delsing}}, \bibinfo {author} {\bibfnamefont {J.}~\bibnamefont {Bylander}},\
  and\ \bibinfo {author} {\bibfnamefont {G.}~\bibnamefont {Tancredi}},\
  }\bibfield  {title} {\bibinfo {title} {Transmon qubit readout fidelity at the
  threshold for quantum error correction without a quantum-limited amplifier},\
  }\href {https://doi.org/10.1038/s41534-023-00689-6} {\bibfield  {journal}
  {\bibinfo  {journal} {npj Quantum Inf.}\ }\textbf {\bibinfo {volume} {9}},\
  \bibinfo {pages} {1} (\bibinfo {year} {2023}{\natexlab{b}})}\BibitemShut
  {NoStop}%
\bibitem [{\citenamefont {Kanazawa}\ \emph {et~al.}(2023)\citenamefont
  {Kanazawa}, \citenamefont {Emori},\ and\ \citenamefont
  {McKay}}]{kanazawa2023}%
  \BibitemOpen
  \bibfield  {author} {\bibinfo {author} {\bibfnamefont {N.}~\bibnamefont
  {Kanazawa}}, \bibinfo {author} {\bibfnamefont {H.}~\bibnamefont {Emori}},\
  and\ \bibinfo {author} {\bibfnamefont {D.~C.}\ \bibnamefont {McKay}},\ }\href
  {https://doi.org/10.48550/arXiv.2309.11303} {\bibinfo {title} {Qutrit state
  discrimination with mid-circuit measurements}} (\bibinfo {year} {2023}),\
  \Eprint {https://arxiv.org/abs/2309.11303} {arXiv:2309.11303 [quant-ph]}
  \BibitemShut {NoStop}%
\bibitem [{Note2()}]{Note2}%
  \BibitemOpen
  \bibinfo {note} {This argumentation holds for an initial state with
  non-negligible $\mathinner {|{2}\rangle }$-state population which is the same
  for each shot. However, this is not the case in our simulations as the
  initial state changes from shot to shot. This manifests as large standard
  deviations in $p_\protect \text {sp}$ and $p_\protect \text {sp}^{(2)}$ with
  fast qubit resets, see \protect \cref {tab:qutrit_sim_fits}. The large
  fluctuations in $\mathinner {|{1}\rangle }$- and $\mathinner {|{2}\rangle
  }$-state populations is the cause of distortions in the RabiEF signals,
  deviating from the form $a\sin ^2(b\theta ) + c$ entirely.}\BibitemShut
  {Stop}%
\bibitem [{\citenamefont {{Qiskit Aer Contributors}}(2025)}]{qiskitaer2025}%
  \BibitemOpen
  \bibfield  {author} {\bibinfo {author} {\bibnamefont {{Qiskit Aer
  Contributors}}},\ }\href {https://github.com/Qiskit/qiskit-aer} {\bibinfo
  {title} {Qiskit/qiskit-aer: {{Qiskit Aer}} 0.16.0}} (\bibinfo {year}
  {2025}),\ \bibinfo {note}
  {\url{https://github.com/Qiskit/qiskit-aer}}\BibitemShut {NoStop}%
\bibitem [{\citenamefont {Battistel}\ \emph {et~al.}(2021)\citenamefont
  {Battistel}, \citenamefont {Varbanov},\ and\ \citenamefont
  {Terhal}}]{Battistel2021}%
  \BibitemOpen
  \bibfield  {author} {\bibinfo {author} {\bibfnamefont {F.}~\bibnamefont
  {Battistel}}, \bibinfo {author} {\bibfnamefont {B.}~\bibnamefont
  {Varbanov}},\ and\ \bibinfo {author} {\bibfnamefont {B.}~\bibnamefont
  {Terhal}},\ }\bibfield  {title} {\bibinfo {title} {Hardware-efficient
  leakage-reduction scheme for quantum error correction with superconducting
  transmon qubits},\ }\href {https://doi.org/10.1103/PRXQuantum.2.030314}
  {\bibfield  {journal} {\bibinfo  {journal} {PRX Quantum}\ }\textbf {\bibinfo
  {volume} {2}},\ \bibinfo {pages} {030314} (\bibinfo {year}
  {2021})}\BibitemShut {NoStop}%
\bibitem [{\citenamefont {Miao}\ \emph {et~al.}(2023)\citenamefont {Miao} \emph
  {et~al.}}]{Miao2023}%
  \BibitemOpen
  \bibfield  {author} {\bibinfo {author} {\bibfnamefont {K.~C.}\ \bibnamefont
  {Miao}} \emph {et~al.},\ }\bibfield  {title} {\bibinfo {title} {Overcoming
  leakage in quantum error correction},\ }\href
  {https://doi.org/10.1038/s41567-023-02226-w} {\bibfield  {journal} {\bibinfo
  {journal} {Nat. Phys.}\ }\textbf {\bibinfo {volume} {19}},\ \bibinfo {pages}
  {1780} (\bibinfo {year} {2023})}\BibitemShut {NoStop}%
\bibitem [{\citenamefont {Khezri}\ \emph {et~al.}(2023)\citenamefont {Khezri},
  \citenamefont {Opremcak}, \citenamefont {Chen}, \citenamefont {Miao},
  \citenamefont {McEwen}, \citenamefont {Bengtsson}, \citenamefont {White},
  \citenamefont {Naaman}, \citenamefont {Sank}, \citenamefont {Korotkov},
  \citenamefont {Chen},\ and\ \citenamefont {Smelyanskiy}}]{Khezri2023}%
  \BibitemOpen
  \bibfield  {author} {\bibinfo {author} {\bibfnamefont {M.}~\bibnamefont
  {Khezri}}, \bibinfo {author} {\bibfnamefont {A.}~\bibnamefont {Opremcak}},
  \bibinfo {author} {\bibfnamefont {Z.}~\bibnamefont {Chen}}, \bibinfo {author}
  {\bibfnamefont {K.~C.}\ \bibnamefont {Miao}}, \bibinfo {author}
  {\bibfnamefont {M.}~\bibnamefont {McEwen}}, \bibinfo {author} {\bibfnamefont
  {A.}~\bibnamefont {Bengtsson}}, \bibinfo {author} {\bibfnamefont
  {T.}~\bibnamefont {White}}, \bibinfo {author} {\bibfnamefont
  {O.}~\bibnamefont {Naaman}}, \bibinfo {author} {\bibfnamefont
  {D.}~\bibnamefont {Sank}}, \bibinfo {author} {\bibfnamefont {A.~N.}\
  \bibnamefont {Korotkov}}, \bibinfo {author} {\bibfnamefont {Y.}~\bibnamefont
  {Chen}},\ and\ \bibinfo {author} {\bibfnamefont {V.}~\bibnamefont
  {Smelyanskiy}},\ }\bibfield  {title} {\bibinfo {title} {Measurement-induced
  state transitions in a superconducting qubit: {{Within}} the rotating-wave
  approximation},\ }\href {https://doi.org/10.1103/PhysRevApplied.20.054008}
  {\bibfield  {journal} {\bibinfo  {journal} {Phys. Rev. Appl.}\ }\textbf
  {\bibinfo {volume} {20}},\ \bibinfo {pages} {054008} (\bibinfo {year}
  {2023})}\BibitemShut {NoStop}%
\bibitem [{\citenamefont {Lescanne}\ \emph {et~al.}(2019)\citenamefont
  {Lescanne}, \citenamefont {Verney}, \citenamefont {Ficheux}, \citenamefont
  {Devoret}, \citenamefont {Huard}, \citenamefont {Mirrahimi},\ and\
  \citenamefont {Leghtas}}]{Lescanne2019}%
  \BibitemOpen
  \bibfield  {author} {\bibinfo {author} {\bibfnamefont {R.}~\bibnamefont
  {Lescanne}}, \bibinfo {author} {\bibfnamefont {L.}~\bibnamefont {Verney}},
  \bibinfo {author} {\bibfnamefont {Q.}~\bibnamefont {Ficheux}}, \bibinfo
  {author} {\bibfnamefont {M.~H.}\ \bibnamefont {Devoret}}, \bibinfo {author}
  {\bibfnamefont {B.}~\bibnamefont {Huard}}, \bibinfo {author} {\bibfnamefont
  {M.}~\bibnamefont {Mirrahimi}},\ and\ \bibinfo {author} {\bibfnamefont
  {Z.}~\bibnamefont {Leghtas}},\ }\bibfield  {title} {\bibinfo {title} {Escape
  of a driven quantum {{Josephson}} circuit into unconfined states},\ }\href
  {https://doi.org/10.1103/PhysRevApplied.11.014030} {\bibfield  {journal}
  {\bibinfo  {journal} {Phys. Rev. Appl.}\ }\textbf {\bibinfo {volume} {11}},\
  \bibinfo {pages} {014030} (\bibinfo {year} {2019})}\BibitemShut {NoStop}%
\bibitem [{\citenamefont {Fu}\ \emph {et~al.}(2022)\citenamefont {Fu},
  \citenamefont {Liu}, \citenamefont {Ye}, \citenamefont {Wang}, \citenamefont
  {Zhang}, \citenamefont {Duan}, \citenamefont {Rong},\ and\ \citenamefont
  {Du}}]{Fu2022}%
  \BibitemOpen
  \bibfield  {author} {\bibinfo {author} {\bibfnamefont {Y.}~\bibnamefont
  {Fu}}, \bibinfo {author} {\bibfnamefont {W.}~\bibnamefont {Liu}}, \bibinfo
  {author} {\bibfnamefont {X.}~\bibnamefont {Ye}}, \bibinfo {author}
  {\bibfnamefont {Y.}~\bibnamefont {Wang}}, \bibinfo {author} {\bibfnamefont
  {C.}~\bibnamefont {Zhang}}, \bibinfo {author} {\bibfnamefont {C.-K.}\
  \bibnamefont {Duan}}, \bibinfo {author} {\bibfnamefont {X.}~\bibnamefont
  {Rong}},\ and\ \bibinfo {author} {\bibfnamefont {J.}~\bibnamefont {Du}},\
  }\bibfield  {title} {\bibinfo {title} {Experimental investigation of quantum
  correlations in a two-qutrit spin system},\ }\href
  {https://doi.org/10.1103/PhysRevLett.129.100501} {\bibfield  {journal}
  {\bibinfo  {journal} {Phys. Rev. Lett.}\ }\textbf {\bibinfo {volume} {129}},\
  \bibinfo {pages} {100501} (\bibinfo {year} {2022})}\BibitemShut {NoStop}%
\bibitem [{\citenamefont {Guo}\ \emph {et~al.}(2024)\citenamefont {Guo},
  \citenamefont {Ji}, \citenamefont {Kong}, \citenamefont {Wang}, \citenamefont
  {Sun}, \citenamefont {Zhou}, \citenamefont {Chai}, \citenamefont {Rong},
  \citenamefont {Shi}, \citenamefont {Wang},\ and\ \citenamefont
  {Du}}]{Guo2024}%
  \BibitemOpen
  \bibfield  {author} {\bibinfo {author} {\bibfnamefont {Y.}~\bibnamefont
  {Guo}}, \bibinfo {author} {\bibfnamefont {W.}~\bibnamefont {Ji}}, \bibinfo
  {author} {\bibfnamefont {X.}~\bibnamefont {Kong}}, \bibinfo {author}
  {\bibfnamefont {M.}~\bibnamefont {Wang}}, \bibinfo {author} {\bibfnamefont
  {H.}~\bibnamefont {Sun}}, \bibinfo {author} {\bibfnamefont {J.}~\bibnamefont
  {Zhou}}, \bibinfo {author} {\bibfnamefont {Z.}~\bibnamefont {Chai}}, \bibinfo
  {author} {\bibfnamefont {X.}~\bibnamefont {Rong}}, \bibinfo {author}
  {\bibfnamefont {F.}~\bibnamefont {Shi}}, \bibinfo {author} {\bibfnamefont
  {Y.}~\bibnamefont {Wang}},\ and\ \bibinfo {author} {\bibfnamefont
  {J.}~\bibnamefont {Du}},\ }\bibfield  {title} {\bibinfo {title} {Single-shot
  readout of a solid-state electron spin qutrit},\ }\href
  {https://doi.org/10.1103/PhysRevLett.132.060601} {\bibfield  {journal}
  {\bibinfo  {journal} {Phys. Rev. Lett.}\ }\textbf {\bibinfo {volume} {132}},\
  \bibinfo {pages} {060601} (\bibinfo {year} {2024})}\BibitemShut {NoStop}%
\bibitem [{\citenamefont {Klimov}\ \emph {et~al.}(2003)\citenamefont {Klimov},
  \citenamefont {Guzm{\'a}n}, \citenamefont {Retamal},\ and\ \citenamefont
  {Saavedra}}]{Klimov2003}%
  \BibitemOpen
  \bibfield  {author} {\bibinfo {author} {\bibfnamefont {A.~B.}\ \bibnamefont
  {Klimov}}, \bibinfo {author} {\bibfnamefont {R.}~\bibnamefont {Guzm{\'a}n}},
  \bibinfo {author} {\bibfnamefont {J.~C.}\ \bibnamefont {Retamal}},\ and\
  \bibinfo {author} {\bibfnamefont {C.}~\bibnamefont {Saavedra}},\ }\bibfield
  {title} {\bibinfo {title} {Qutrit quantum computer with trapped ions},\
  }\href {https://doi.org/10.1103/PhysRevA.67.062313} {\bibfield  {journal}
  {\bibinfo  {journal} {Phys. Rev. A}\ }\textbf {\bibinfo {volume} {67}},\
  \bibinfo {pages} {062313} (\bibinfo {year} {2003})}\BibitemShut {NoStop}%
\bibitem [{\citenamefont {Chen}\ \emph
  {et~al.}(2025{\natexlab{b}})\citenamefont {Chen}, \citenamefont {Hashim},
  \citenamefont {Goss}, \citenamefont {Seif}, \citenamefont {Siddiqi},\ and\
  \citenamefont {Jiang}}]{Chen2025b}%
  \BibitemOpen
  \bibfield  {author} {\bibinfo {author} {\bibfnamefont {S.}~\bibnamefont
  {Chen}}, \bibinfo {author} {\bibfnamefont {A.}~\bibnamefont {Hashim}},
  \bibinfo {author} {\bibfnamefont {N.}~\bibnamefont {Goss}}, \bibinfo {author}
  {\bibfnamefont {A.}~\bibnamefont {Seif}}, \bibinfo {author} {\bibfnamefont
  {I.}~\bibnamefont {Siddiqi}},\ and\ \bibinfo {author} {\bibfnamefont
  {L.}~\bibnamefont {Jiang}},\ }\href
  {https://doi.org/10.48550/arXiv.2506.09131} {\bibinfo {title} {Enhancing
  quantum noise characterization via extra energy levels}} (\bibinfo {year}
  {2025}{\natexlab{b}}),\ \Eprint {https://arxiv.org/abs/2506.09131}
  {arXiv:2506.09131 [quant-ph]} \BibitemShut {NoStop}%
\bibitem [{\citenamefont {Malekakhlagh}\ \emph {et~al.}(2025)\citenamefont
  {Malekakhlagh}, \citenamefont {Seif}, \citenamefont {Puzzuoli}, \citenamefont
  {Govia},\ and\ \citenamefont {{van den Berg}}}]{malekakhlagh2025efficient}%
  \BibitemOpen
  \bibfield  {author} {\bibinfo {author} {\bibfnamefont {M.}~\bibnamefont
  {Malekakhlagh}}, \bibinfo {author} {\bibfnamefont {A.}~\bibnamefont {Seif}},
  \bibinfo {author} {\bibfnamefont {D.}~\bibnamefont {Puzzuoli}}, \bibinfo
  {author} {\bibfnamefont {L.~C.~G.}\ \bibnamefont {Govia}},\ and\ \bibinfo
  {author} {\bibfnamefont {E.}~\bibnamefont {{van den Berg}}},\ }\bibfield
  {title} {\bibinfo {title} {Efficient lindblad synthesis for noise model
  construction},\ }\href {https://doi.org/10.1038/s41534-025-01139-1}
  {\bibfield  {journal} {\bibinfo  {journal} {npj Quantum Information}\
  }\textbf {\bibinfo {volume} {11}},\ \bibinfo {pages} {191} (\bibinfo {year}
  {2025})}\BibitemShut {NoStop}%
\bibitem [{\citenamefont {Calzona}\ \emph {et~al.}(2024)\citenamefont
  {Calzona}, \citenamefont {Papi{\v c}}, \citenamefont {{Figueroa-Romero}},\
  and\ \citenamefont {Auer}}]{calzona2024multilayer}%
  \BibitemOpen
  \bibfield  {author} {\bibinfo {author} {\bibfnamefont {A.}~\bibnamefont
  {Calzona}}, \bibinfo {author} {\bibfnamefont {M.}~\bibnamefont {Papi{\v c}}},
  \bibinfo {author} {\bibfnamefont {P.}~\bibnamefont {{Figueroa-Romero}}},\
  and\ \bibinfo {author} {\bibfnamefont {A.}~\bibnamefont {Auer}},\ }\href
  {https://doi.org/10.48550/arXiv.2412.09332} {\bibinfo {title} {Multi-layer
  cycle benchmarking for high-accuracy error characterization}} (\bibinfo
  {year} {2024}),\ \Eprint {https://arxiv.org/abs/2412.09332} {arXiv:2412.09332
  [quant-ph]} \BibitemShut {NoStop}%
\bibitem [{\citenamefont {Greenbaum}(2015)}]{ptm}%
  \BibitemOpen
  \bibfield  {author} {\bibinfo {author} {\bibfnamefont {D.}~\bibnamefont
  {Greenbaum}},\ }\href {https://doi.org/10.48550/arXiv.1509.02921} {\bibinfo
  {title} {Introduction to quantum gate set tomography}} (\bibinfo {year}
  {2015}),\ \Eprint {https://arxiv.org/abs/1509.02921} {arXiv:1509.02921
  [quant-ph]} \BibitemShut {NoStop}%
\bibitem [{\citenamefont {{Javadi-Abhari}}\ \emph {et~al.}(2024)\citenamefont
  {{Javadi-Abhari}}, \citenamefont {Treinish}, \citenamefont {Krsulich},
  \citenamefont {Wood}, \citenamefont {Lishman}, \citenamefont {Gacon},
  \citenamefont {Martiel}, \citenamefont {Nation}, \citenamefont {Bishop},
  \citenamefont {Cross}, \citenamefont {Johnson},\ and\ \citenamefont
  {Gambetta}}]{qiskit2024}%
  \BibitemOpen
  \bibfield  {author} {\bibinfo {author} {\bibfnamefont {A.}~\bibnamefont
  {{Javadi-Abhari}}}, \bibinfo {author} {\bibfnamefont {M.}~\bibnamefont
  {Treinish}}, \bibinfo {author} {\bibfnamefont {K.}~\bibnamefont {Krsulich}},
  \bibinfo {author} {\bibfnamefont {C.~J.}\ \bibnamefont {Wood}}, \bibinfo
  {author} {\bibfnamefont {J.}~\bibnamefont {Lishman}}, \bibinfo {author}
  {\bibfnamefont {J.}~\bibnamefont {Gacon}}, \bibinfo {author} {\bibfnamefont
  {S.}~\bibnamefont {Martiel}}, \bibinfo {author} {\bibfnamefont {P.~D.}\
  \bibnamefont {Nation}}, \bibinfo {author} {\bibfnamefont {L.~S.}\
  \bibnamefont {Bishop}}, \bibinfo {author} {\bibfnamefont {A.~W.}\
  \bibnamefont {Cross}}, \bibinfo {author} {\bibfnamefont {B.~R.}\ \bibnamefont
  {Johnson}},\ and\ \bibinfo {author} {\bibfnamefont {J.~M.}\ \bibnamefont
  {Gambetta}},\ }\href {https://doi.org/10.48550/arXiv.2405.08810} {\bibinfo
  {title} {Quantum computing with {{Qiskit}}}} (\bibinfo {year} {2024}),\
  \Eprint {https://arxiv.org/abs/2405.08810} {arXiv:2405.08810 [quant-ph]}
  \BibitemShut {NoStop}%
\bibitem [{\citenamefont {Puzzuoli}\ \emph {et~al.}(2023)\citenamefont
  {Puzzuoli}, \citenamefont {Wood}, \citenamefont {Egger}, \citenamefont
  {Rosand},\ and\ \citenamefont {Ueda}}]{qiskitdynamics2023}%
  \BibitemOpen
  \bibfield  {author} {\bibinfo {author} {\bibfnamefont {D.}~\bibnamefont
  {Puzzuoli}}, \bibinfo {author} {\bibfnamefont {C.~J.}\ \bibnamefont {Wood}},
  \bibinfo {author} {\bibfnamefont {D.~J.}\ \bibnamefont {Egger}}, \bibinfo
  {author} {\bibfnamefont {B.}~\bibnamefont {Rosand}},\ and\ \bibinfo {author}
  {\bibfnamefont {K.}~\bibnamefont {Ueda}},\ }\bibfield  {title} {\bibinfo
  {title} {Qiskit {{Dynamics}}: {{A Python}} package for simulating the time
  dynamics of quantum systems},\ }\href {https://doi.org/10.21105/joss.05853}
  {\bibfield  {journal} {\bibinfo  {journal} {J. Open Source Softw.}\ }\textbf
  {\bibinfo {volume} {8}},\ \bibinfo {pages} {5853} (\bibinfo {year}
  {2023})}\BibitemShut {NoStop}%
\bibitem [{\citenamefont {Dane}\ \emph {et~al.}(2025)\citenamefont {Dane},
  \citenamefont {Balakrishnan}, \citenamefont {Wacaser}, \citenamefont {Hung},
  \citenamefont {Mamin}, \citenamefont {Rugar}, \citenamefont {Shelby},
  \citenamefont {Murray}, \citenamefont {Rodbell},\ and\ \citenamefont
  {Sleight}}]{dane2025}%
  \BibitemOpen
  \bibfield  {author} {\bibinfo {author} {\bibfnamefont {A.}~\bibnamefont
  {Dane}}, \bibinfo {author} {\bibfnamefont {K.}~\bibnamefont {Balakrishnan}},
  \bibinfo {author} {\bibfnamefont {B.}~\bibnamefont {Wacaser}}, \bibinfo
  {author} {\bibfnamefont {L.-W.}\ \bibnamefont {Hung}}, \bibinfo {author}
  {\bibfnamefont {H.~J.}\ \bibnamefont {Mamin}}, \bibinfo {author}
  {\bibfnamefont {D.}~\bibnamefont {Rugar}}, \bibinfo {author} {\bibfnamefont
  {R.~M.}\ \bibnamefont {Shelby}}, \bibinfo {author} {\bibfnamefont
  {C.}~\bibnamefont {Murray}}, \bibinfo {author} {\bibfnamefont
  {K.}~\bibnamefont {Rodbell}},\ and\ \bibinfo {author} {\bibfnamefont
  {J.}~\bibnamefont {Sleight}},\ }\href
  {https://doi.org/10.48550/arXiv.2503.12514} {\bibinfo {title} {Performance
  stabilization of high-coherence superconducting qubits}} (\bibinfo {year}
  {2025}),\ \Eprint {https://arxiv.org/abs/2503.12514} {arXiv:2503.12514
  [quant-ph]} \BibitemShut {NoStop}%
\bibitem [{\citenamefont {Heinsoo}\ \emph {et~al.}(2018)\citenamefont
  {Heinsoo}, \citenamefont {Andersen}, \citenamefont {Remm}, \citenamefont
  {Krinner}, \citenamefont {Walter}, \citenamefont {Salath{\'e}}, \citenamefont
  {Gasparinetti}, \citenamefont {Besse}, \citenamefont {Poto{\v c}nik},
  \citenamefont {Wallraff},\ and\ \citenamefont {Eichler}}]{heinsoo2018}%
  \BibitemOpen
  \bibfield  {author} {\bibinfo {author} {\bibfnamefont {J.}~\bibnamefont
  {Heinsoo}}, \bibinfo {author} {\bibfnamefont {C.~K.}\ \bibnamefont
  {Andersen}}, \bibinfo {author} {\bibfnamefont {A.}~\bibnamefont {Remm}},
  \bibinfo {author} {\bibfnamefont {S.}~\bibnamefont {Krinner}}, \bibinfo
  {author} {\bibfnamefont {T.}~\bibnamefont {Walter}}, \bibinfo {author}
  {\bibfnamefont {Y.}~\bibnamefont {Salath{\'e}}}, \bibinfo {author}
  {\bibfnamefont {S.}~\bibnamefont {Gasparinetti}}, \bibinfo {author}
  {\bibfnamefont {J.-C.}\ \bibnamefont {Besse}}, \bibinfo {author}
  {\bibfnamefont {A.}~\bibnamefont {Poto{\v c}nik}}, \bibinfo {author}
  {\bibfnamefont {A.}~\bibnamefont {Wallraff}},\ and\ \bibinfo {author}
  {\bibfnamefont {C.}~\bibnamefont {Eichler}},\ }\bibfield  {title} {\bibinfo
  {title} {Rapid high-fidelity multiplexed readout of superconducting qubits},\
  }\href {https://doi.org/10.1103/PhysRevApplied.10.034040} {\bibfield
  {journal} {\bibinfo  {journal} {Phys. Rev. Appl.}\ }\textbf {\bibinfo
  {volume} {10}},\ \bibinfo {pages} {34040} (\bibinfo {year}
  {2018})}\BibitemShut {NoStop}%
\bibitem [{\citenamefont {Burns}(2001)}]{burns2001}%
  \BibitemOpen
  \bibfield  {author} {\bibinfo {author} {\bibfnamefont {R.~S.}\ \bibnamefont
  {Burns}},\ }\href@noop {} {\emph {\bibinfo {title} {Advanced Control
  Engineering}}}\ (\bibinfo  {publisher} {Butterworth-Heinemann},\ \bibinfo
  {address} {Oxford},\ \bibinfo {year} {2001})\BibitemShut {NoStop}%
\end{thebibliography}%

\end{document}